\documentclass[twocolumn]{aastex631}

\usepackage{amsmath}
\usepackage{amsfonts,amssymb}
\usepackage{txfonts}
\usepackage{gensymb}
\usepackage{natbib}

\begin{document}
	
	\title{Spatial Scales and Time Variation of Solar Subsurface Convection}
	
	\shorttitle{Scales and Time Variation of Solar Subsurface Convection}
	\shortauthors{Getling \& Kosovichev}
	
	\correspondingauthor{A. V. Getling}
	\email{A.Getling@mail.ru}
	
\author{Alexander V. Getling}
\affil{Skobeltsyn Institute of Nuclear Physics, Lomonosov Moscow State
	University, Moscow, 119991 Russia}

\author{Alexander G. Kosovichev}
\affiliation{New Jersey Institute of Technology, NJ 07102, USA}

	\begin{abstract}
Spectral analysis of the spatial structure of solar subphotospheric convection is carried out for subsurface flow maps constructed using the time--distance helioseismological technique. The source data are obtained from the Helioseismic and Magnetic Imager (HMI) onboard Solar Dynamics Observatory (SDO) from 2010 May to 2020 September. A spherical-harmonic transform is applied to the horizontal-velocity-divergence field at depths from 0 to 19~Mm. The range of flow scales is fairly broad in shallow layers and narrows as the depth increases. The horizontal flow scales rapidly increase with depth, from supergranulation to giant-cell values, and indicate the existence of large-scale convective motions in the near-surface shear layer. The results can naturally be interpreted in terms of a superposition of differently scaled flows localized in different depth intervals. There is some tendency toward the emergence of meridionally elongated (banana-shaped) convection structures in the deep layers. The total power of convective flows is anticorrelated with the sunspot-number variation over the solar activity cycle in shallow subsurface layers and positively correlated at larger depths, which is suggestive of the depth redistribution of the convective-flow energy due to the action of magnetic fields.
    \end{abstract}
	
	\keywords{Sun: helioseismology --- Sun: granulation --- Sun: solar activity}
	
\section{Introduction}\label{sec:intro}
	
As is known, plasma flows in the solar convection zone form variously scaled structures resembling convection cells. The smallest of them, granules, have been known since the advent of high-resolution telescopic observations of the Sun \citep{Herschel1800}. \citet{Frenkiel1952} performed the first analysis of the turbulence spectrum of solar convection and, in addition to the primary maximum corresponding to granulation, found a secondary maximum at long wavelengths corresponding to 15~Mm. Based on Doppler measurements of horizontal velocities (away from the disk center), \citet{Hart_1954} discovered signs of a pattern of cells with sizes an order of magnitude larger and living much longer than granules. \citet{Leighton_etal_1962} described them in greater detail and designated them as supergranules. For a recent review of studies of supergranulation, see \citet{Rincon2018}. Further, \citet{November_etal_1981}, using Doppler measurements of vertical velocities, detected mesogranulation---a system of cells intermediate between granules and supergranules in their sizes. The existence of the largest velocity-field structures, giant cells, was suggested long ago by \citet{Simon_Weiss_1968}. Even before that, \citet{Bumba_etal_1964} noted indications of the presence of giant structures observing magnetic fields. However, giant cells were qualified to be hypothetical for over three decades.

\citet{Glatzmaier_Gilman_II_1981} performed a linear analysis of the onset of convection in a rotating spherical shell using the anelastic approximation. They found that giant cells extending from the bottom to the top of the convection zone are the most easily excitable instability mode. As the density stratification and rotation rate increase, these cells become meridionally elongated, or banana-shaped. Possibilities of banana-shaped cells have been repeatedly noted since the early 1970-s \citep{Busse_1970}; see also a survey by \citet{Busse_2002}. In particular, such cells were demonstrated in laboratory experiments \citep{Busse_Carrigan_1974}. Giant cellular structures were also found in global magnetohydrodynamic simulations of the solar-convection-zone dynamo action \citep{Ghizaru_etal_2010}. \citet{Featherstone_Hindman_2016} simulated convection in the anelastic approximation using a spectral technique. They claimed that ``the supergranular scale emerges due to a suppression of power on larger spatial scales owing to the presence of deep, rotationally constrained convection'' and that ``giant cells in the traditional sense do not exist.''

The earliest direct Doppler observations of giant cells were done by \citet{Beck_1998}. Later, \citet{Hathaway_etal_2013} revealed them using supergranules as tracers of the material flow.

Lastly, \citet{Abramenko_etal_2012} reported the detection of mini-granules, whose sizes vary extremely widely. These features yet remain very poorly studied.
	
The multiscale structure of solar convection raises some questions that can be resolved only using the information on the velocity field in the subphotospheric convection zone. In particular, hydrodynamic considerations \citep[see, e.g.,][]{Shcheritsa_etal_2018} along with the fact that smaller-scale convection cells are advected at the photospheric surface by the flows in larger-scale cells and can be considered tracers of the large-scale velocity field \citep[see, e.g.,][]{Muller_etal_1992,Rieutord_etal_2001, GetlingBuchnev2010,Hathaway_2021} suggest that the convective velocity field represents a superposition of differently scaled flows \citep{GetlingBuchnev2010}, in contrast to the idea of the mixing-length theory that the flows have unique characteristic scales gradually increasing with depth. The progress of helioseismological research has made it possible to trace the structure and temporal evolution of the subsurface flow field.

The flow scales are characterized by the spatial spectra of the velocity field. In particular, \citet{Hathaway_1987} applied a spherical-harmonic transform to the photospheric Doppler velocity signal and investigated the spatial scales of supergranules and giant cells. \citet{Hathaway_etal_2000,Hathaway_etal_2015} continued this line of research. They noted a broadband nature of the convection spectrum and employed a data-filtering technique to isolate the granular and the supergranular scale. \citet{Greer_etal_2015} employed ring-diagram techniques to study the strength and spatial scale of convective flows in the near-surface shear layer. In particular, they found that the peak of the horizontal velocity spectrum shifts with depth from higher to smaller values of the spherical-harmonic degree.

The spectral composition of the velocity spectrum at various depths below the photosphere has not yet received sufficient attention. We aim to study the spatiotemporal structure of subphotospheric convection using the full-disk horizontal-velocity maps obtained from the Helioseismic and Magnetic Imager \citep[HMI,][]{Scherrer_2012} onboard the \emph{Solar Dynamics Observatory (SDO)} and available from the Joint Science Operations Center (JSOC, \url{http://jsoc.stanford.edu/data/timed/}). In particular, we construct the spatial spectra of the horizontal-velocity fields in the depth range from 0 to 19~Mm, obtained from the time--distance helioseismology pipeline, and analyze the variations of the spectra with depth and time in the course of the solar activity cycle. These data resolve supergranulation and convective flows of larger scales.

Attempts to investigate variations in the convection patterns over the activity cycle are not numerous \citep[see, in particular, references in][]{Roudier_Reardon_1998, Muller_etal_2018}. \citet{Lefebvre_etal_2008} found that the granulation evolves with height in the photosphere but does not exhibit considerable variations in the activity cycle. In addition, \citet{McIntosh_etal_2011} studied the variation of the supergranular length scale over multiple solar minima. \citet{Muller_etal_2018} detected no significant variations in the granulation scale with the activity cycle. \citet{Ballot_etal_2021} have shown that the density and the mean area of granules experience an approximately 2\% variation in the course of the solar cycle, the density of granules being greater and the area being smaller at the solar maximum.

In this paper, we show that the subsurface flow represents a superposition of multiscale convective structures traced at various levels in the upper convection zone. In particular, the scale of giant cells is present in the velocity spectrum, along with the supergranulation scales. In addition, we demonstrate variations of the integrated power of the velocity field in the course of the solar cycle.
	
\section{The Data and the Processing Techniques Used}\label{obs}
	
The original subsurface-flow maps for the central $123\degree \times 123\degree$ area of the visible hemisphere of the Sun are routinely produced every 8 hours by the time--distance helioseismology pipeline \citep{Zhao_etal_2012} from the HMI Dopplergrams. They are represented on a grid of $1026\times 1026$ points with a spatial sampling interval of 0\fdg 12 in both longitude and latitude. We use data for the horizontal velocities at the following eight characteristic levels below the photosphere (the corresponding depth ranges for which the inversions were done are parenthesized): $d=0.5$ (0--1)~Mm, 2.0 (1--3)~Mm, 4.0 (3--5)~Mm, 6.0 (5--7)~Mm, 8.5 (7--10)~Mm, 11.5 (10--13)~Mm, 15.0 (13--17)~Mm, 19.0 (17--21)~Mm. The travel-time measurements are described by \citet{Couvidat_etal_2012}. The travel-time inversion procedure employed in the HMI pipeline uses Born-approximation sensitivity kernels and provides a good localization of the averaging kernels at the target depth. However, the vertical width of the averaging kernels increases with depth, from $\sim 2$~Mm near the surface to $\sim 10$~Mm at the bottom layer \citep[][Figures 10 and 11]{Couvidat_etal_2005}. The horizontal width of the averaging kernels also increases with depth, from $\sim 16$~Mm to $\sim 40$~Mm \citep[][Figures 11 and 12]{Couvidat_etal_2005}. Therefore, the flow maps represent the velocities convolved with the averaging kernels, and this should be taken into account in the interpretation of the presented results. To assess the possible effect of the averaging-kernel variation with depth, we present an illustrative example in Subsection~\ref{scales}.
		
\begin{figure} 
	\centering
    \includegraphics[width=0.45\textwidth]{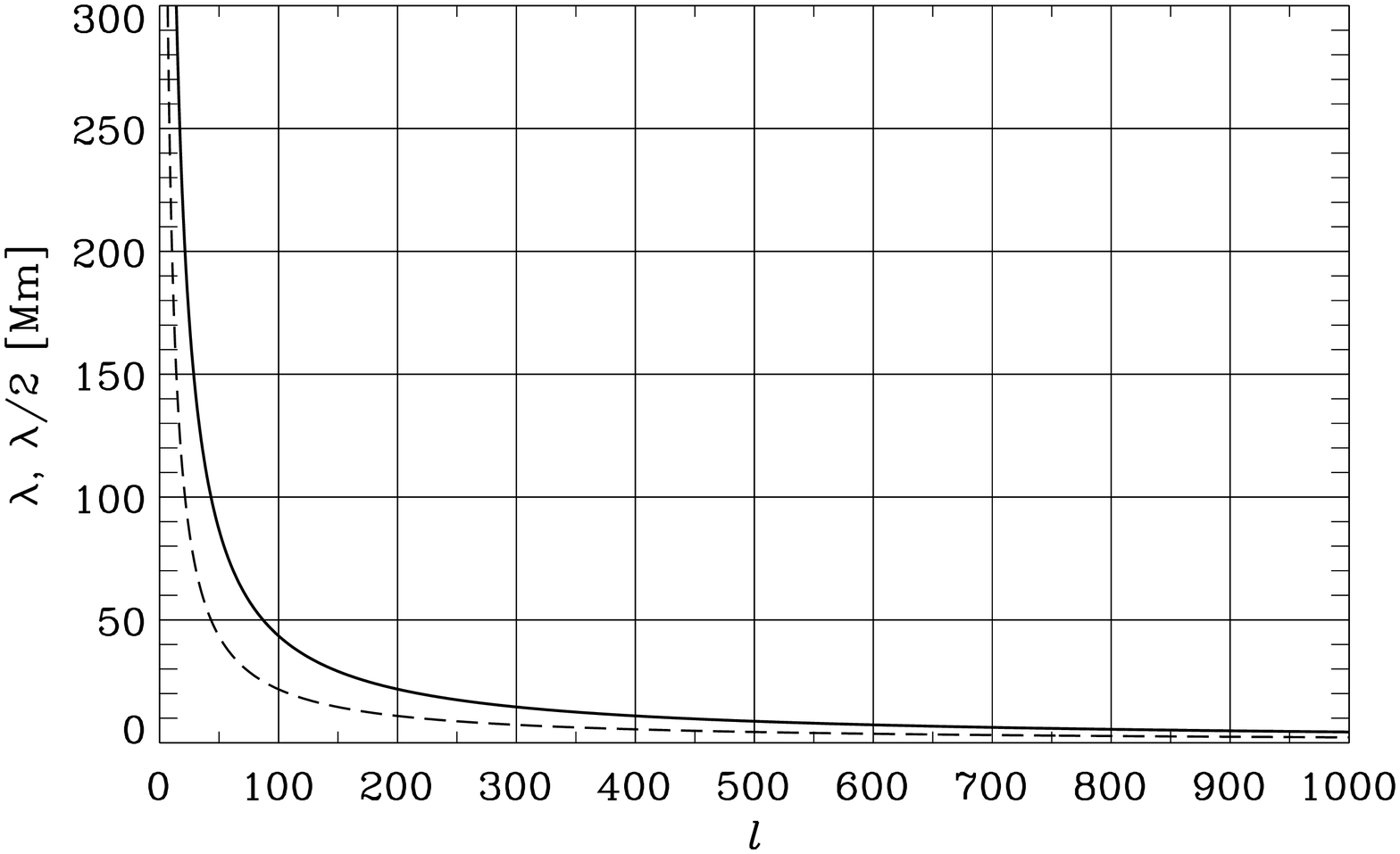}
    \caption{The full wavelength, $\lambda$, of the spherical harmonic  $Y_l^{m}$ and $\lambda/2$ for $r=R_\odot$ as determined by the Jeans formula~(\ref{Jeanseq}).
		\label{Jeans4}}
\end{figure}

\begin{figure*} 
	\centering
    \vspace{-2cm}
    \includegraphics[width=0.2674\textwidth]{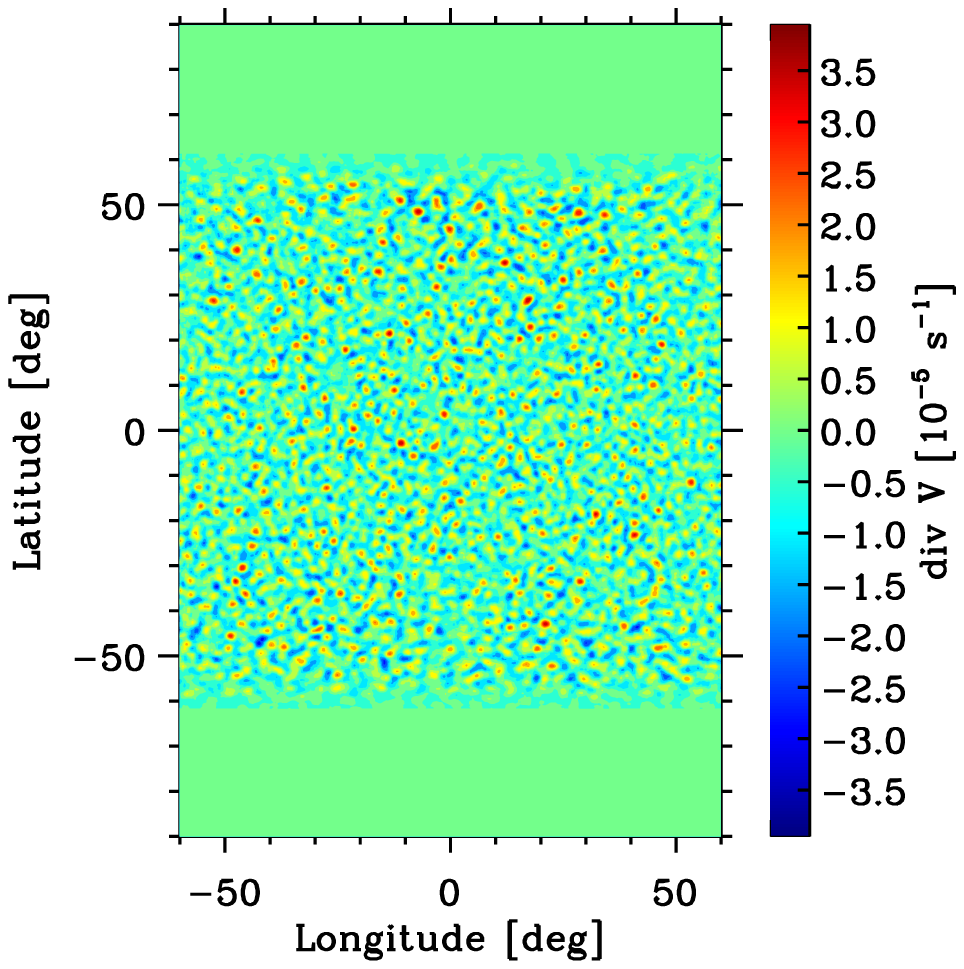}\quad
    \includegraphics[width=0.22\textwidth,bb=50 0 283 425,clip]{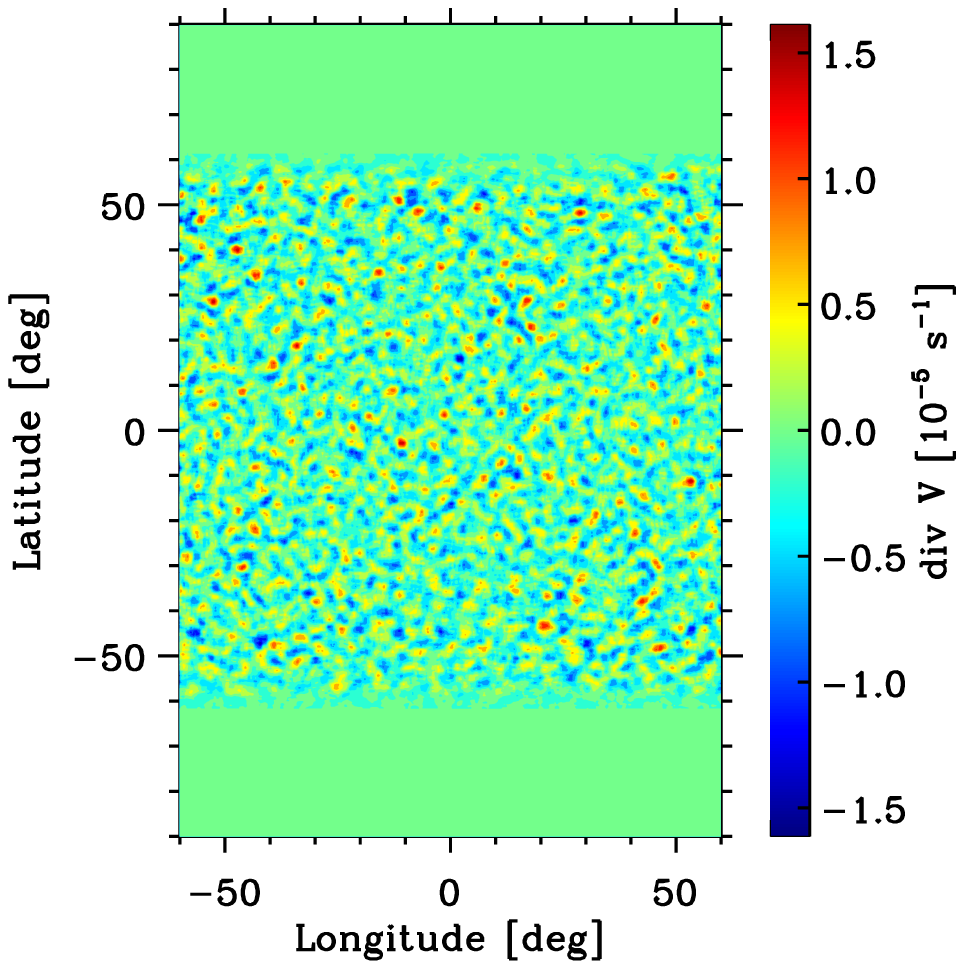}\quad
    \includegraphics[width=0.22\textwidth,bb=50 0 283 425,clip]{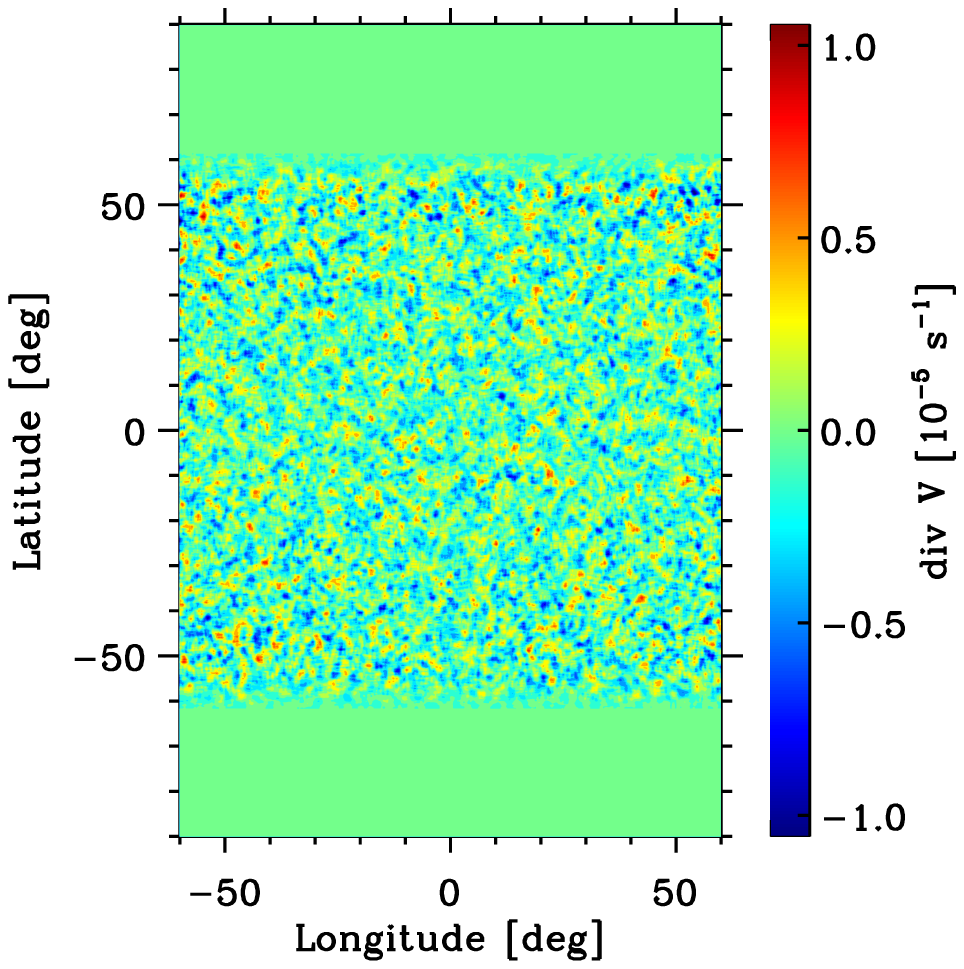}\quad
    \includegraphics[width=0.22\textwidth,bb=50 0 283 425,clip]{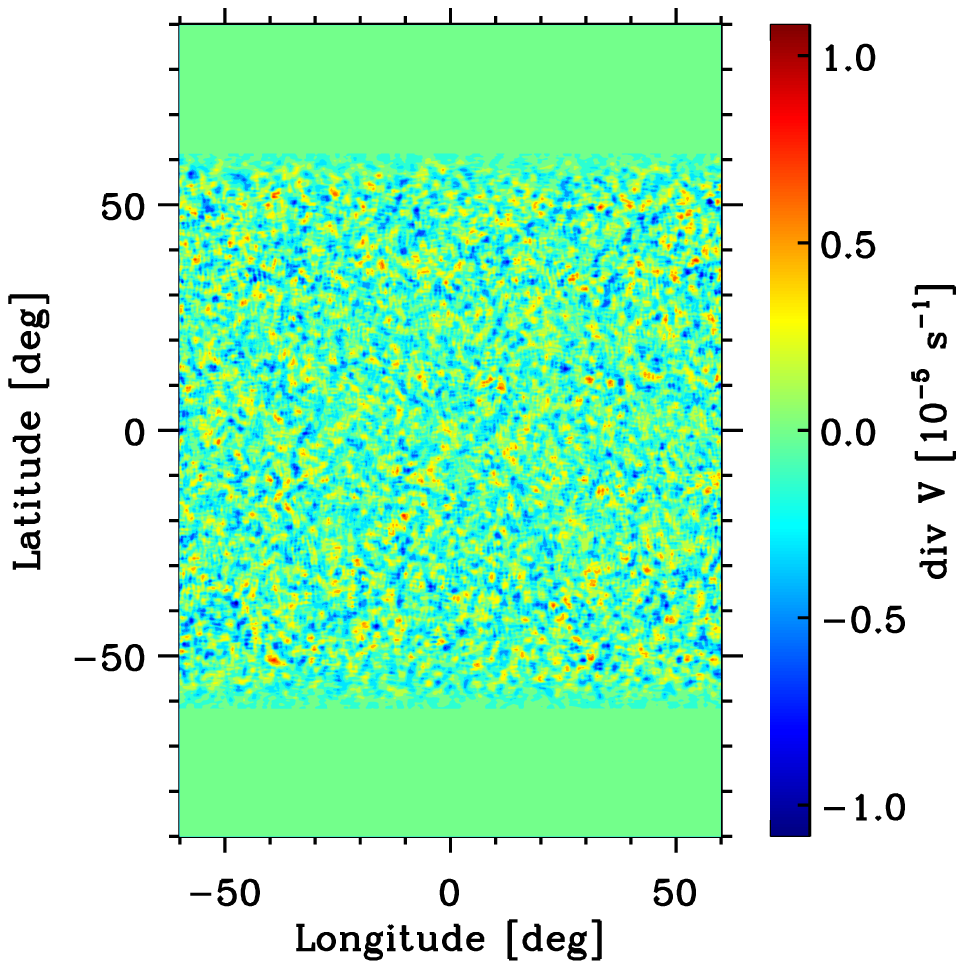}
    \caption{Sample maps of the divergence field at levels $d=0.5, 4.0, 6.0, 11.5$~Mm (from left to right) and the same time.
		\label{images}}
\end{figure*}

\begin{figure} 
	\centering
    \includegraphics[width=0.45\textwidth]{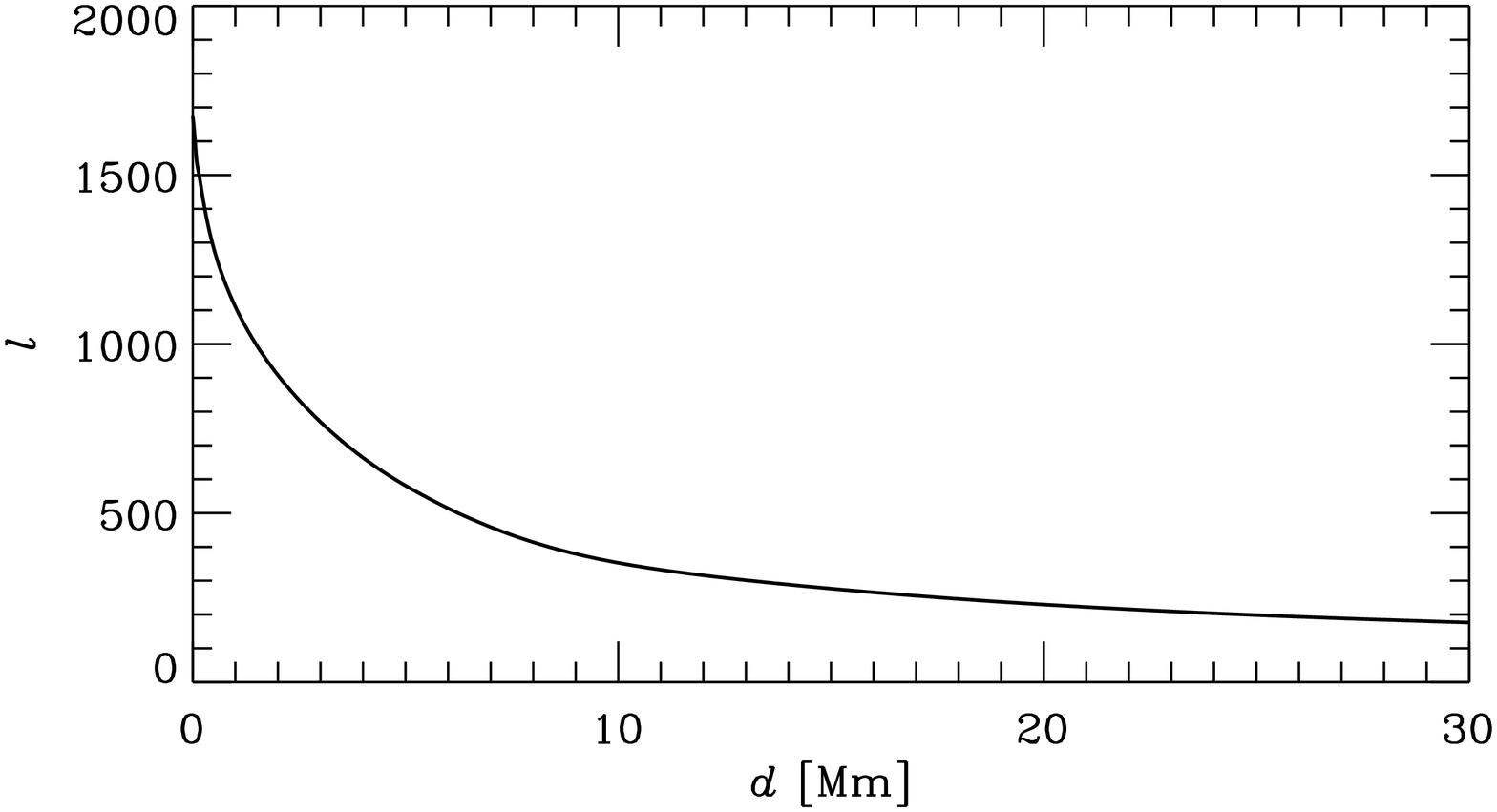}
    \caption{Estimated long-wavelength bound of the realization noise for an acoustic frequency of $3\times 10^{-3}$~Hz as a function of depth.
		\label{lvsdepth}}
\end{figure}

\subsection{Spectral Representations}
	
For convenience, we analyze the scalar field of the horizontal-velocity divergence rather than the velocity-vector field, $\mathbf V$. On a sphere of radius $r$, the divergence of the vector $\mathbf V=\{V_\theta, V_\varphi\}$ is
	\begin{equation}
		f(\theta,\varphi) = \mathop{\mathrm{div}}\mathbf{V}(\theta,\varphi) = \frac {1}{r\sin \theta}\frac {\partial}{\partial \theta}(V_\theta \sin \theta) + \frac {1}{r\sin \theta}\frac {\partial}{\partial \varphi}V_\varphi,
		\label{div}
	\end{equation}
where $\theta$ and $\varphi$ are the polar and azimuthal angles and $r$ plays merely the role of a parameter. We represent the divergence field as a polynomial expansion
	\begin{equation}
		f(\theta,\varphi) = \sum_{l=0}^{l_{\max}}\sum_{m=-l}^{l}A_{lm}Y_{l}^{m}(\theta,\varphi)
		\label{series}
	\end{equation}
in spherical harmonics of angular degree $l$ and azimuthal order $m$,
	\begin{gather}
		Y_{l}^{m} = \sqrt {\frac{(2l+1)}{4\pi}\frac{(l-m)!}{(l+m)!}} P_l^m (\cos \theta) \mathrm {e}^{\mathrm{i}{m\varphi}},
		\label{spherharm}
	\end{gather}
	$$l=0,...,l_{\max},\quad m=0,...,l$$
(where $P_l^m$ are the associated Legendre polynomials and $l_{\max}$  is a properly chosen upper spectral boundary). The spectral coefficients (amplitudes of harmonics) can be determined in a standard way by the equations
	\begin{gather}
		A_{lm}=\frac{1}{4\pi}\int\limits_0^\pi\!\!\int\limits_0^{2\pi} f(\theta,\varphi)\,Y_l^{m} (\theta, \varphi)\sin \theta \ \mathrm d\varphi\, \mathrm d\theta.
		\label{coefts}
	\end{gather}

We are interested here in the power spectra of the flow,
\begin{equation}\label{power}
    p_{lm}=|A_{lm}|^2.
\end{equation}
According to Parseval's theorem, the integrated power of the flow represented by the spectrum (\ref{series}) is
\begin{equation}
p_\mathrm{tot} \equiv \int\limits_\Omega f^2 \mathrm d\Omega =\sum_{l=0}^\infty \sum_{m=-l}^l |A_{lm}|^2.
\end{equation}
The normalized interior sum
\begin{equation}
p_l=\frac{1}{2l+1}\sum_{m=-l}^l |A_{lm}|^2
\label{powerph}
\end{equation}
{is, by definition, the power per degree $l$ and per steradian \citep[see, e.g., Equations~(B.94) and (B.95) on page 858 in][]{Dahlen_Tromp_1998}. Such normalization is chosen because it ensures a ``flat'' spectral representation, $p_l=1/4\pi$, of a Dirac delta function on the unit sphere, $$(\sin\theta)^{-1}\, \delta(\theta-\theta^\prime)\,\delta(\varphi-\varphi^\prime).$$}
In addition, we consider the total power of degree $l$,
\begin{equation}
p^\Sigma_l=\sum_{m=-l}^l |A_{lm}|^2.
\label{powertot}
\end{equation}
To reduce the possible effects of short-term temporal fluctuations of the velocity field, we apply a running-averaging procedure with a 45-day window to the power spectra, $p_{lm}$, and to the power functions, $p_l$ and $p^\Sigma_l$.

Our source data do not cover the whole spherical surface. For this reason, we cut a longitudinal 120\degree-wide sector out of each flow map and complement it with the same data shifted by 120\degree\ and 240\degree\ to fill the complete longitudinal angle. The resultant spectra thus contain nonzero harmonics only with $m$ multiple of 3; we interpolate them to all missing $m$ values and smooth the spectra with a two-point window for better visual perceptibility.
	
Since our source data are restricted to a latitudinal range of $\pm 61\fdg 5$, we have to investigate the effect of the ``empty'' polar caps on the spectrum. To this end, we apply our spectral analysis to a sample model velocity field obtained by G.~Guerrero and A.M.~Stejko using numerical simulations (2021, private communication) and compare the spectra obtained with and without artificially introduced zero velocity in the polar caps $-90\degree < \varphi < -61\fdg 5$ and $61\fdg 5< \varphi < 90\degree$. In addition, we introduce a latitudinal tapering of the flow fields to reduce possible spurious effects due to the Gibbs phenomenon. In other words, we multiply the divergence fields by a window function, which smoothes the sharp drop of velocities at latitudes of $\pm 61\fdg 5$---the boundaries of the ``empty'' polar caps. The analysis of the simulations shows that the ``empty'' polar caps only result in a moderate narrowing of the spectral $l$-band. The latitudinal tapering also results in minimal changes in the flow spectrum.
	
According to the \citet{Jeans_1923} formula, the full wavelength of the harmonic $Y_l^{m}$ on a sphere of radius $r$ is
	\begin{equation}
		\lambda=\frac{2\pi r}{\sqrt{l(l+1)}}
		\label{Jeanseq}
	\end{equation}
(the layer that we consider is much thinner than the convection zone, and $r$ can be put equal to the radius of the Sun, $R_\odot$). This wavelength, determined by the angular degree, $l$, of the spherical harmonic $Y_l^m$, can be used to estimate the characteristic size of the flow structures corresponding to this harmonic (Figure~\ref{Jeans4}).
	
\begin{figure*} 
	\centering
    \includegraphics[width=0.36\textwidth,bb=50 0 800 650,clip] {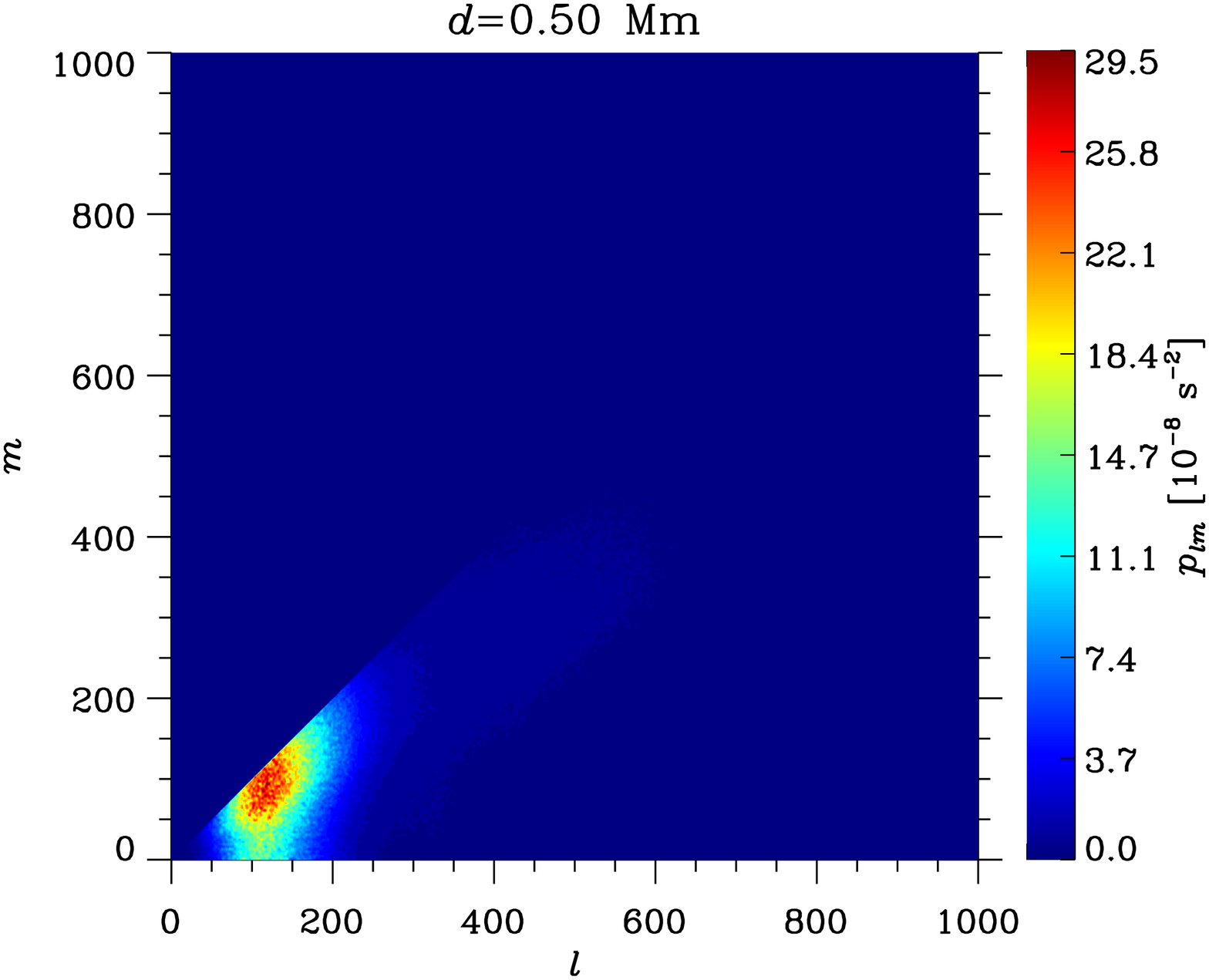}
    \includegraphics[width=0.36\textwidth,bb=50 0 800 650,clip] {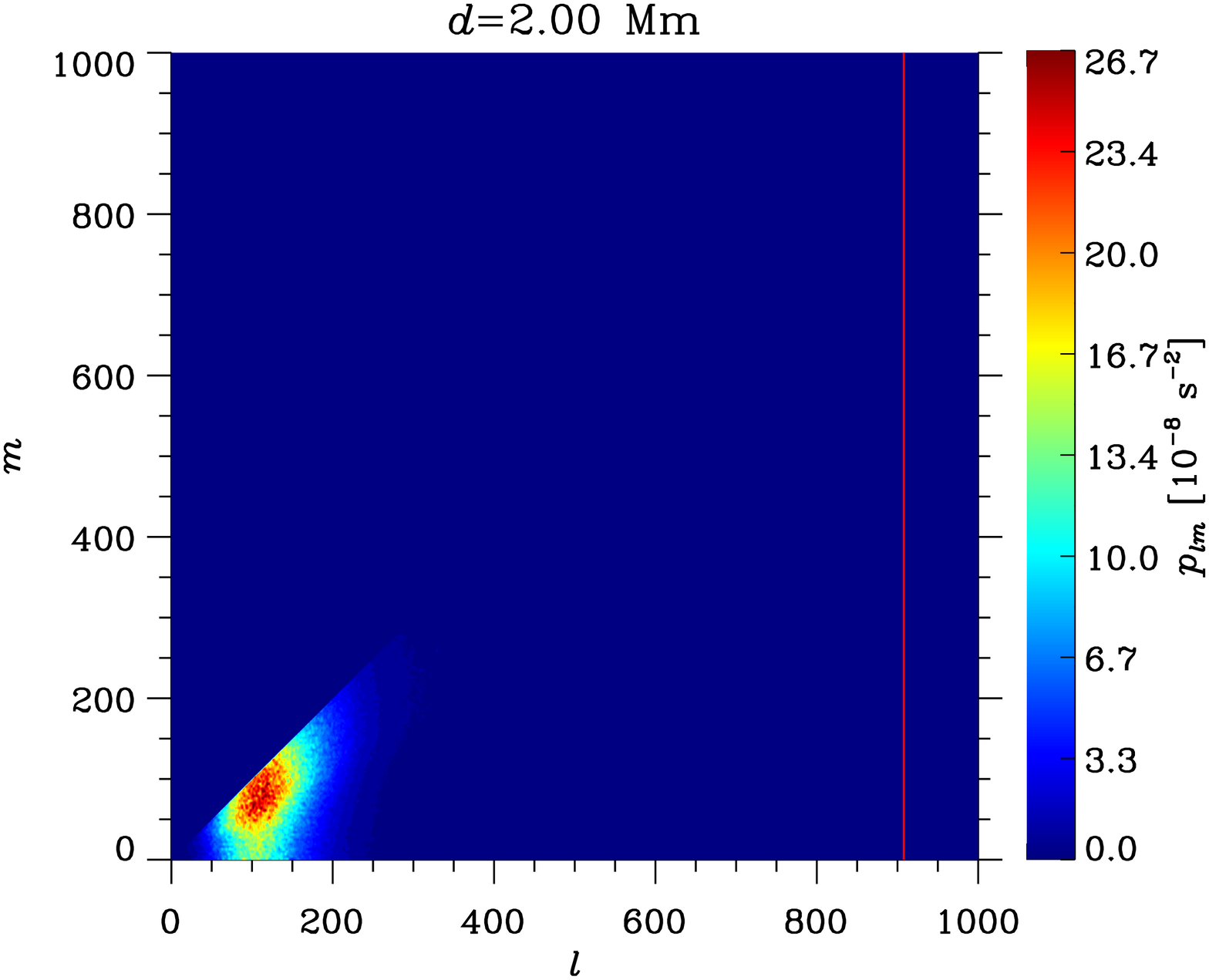}
    \includegraphics[width=0.36\textwidth,bb=50 0 800 650,clip] {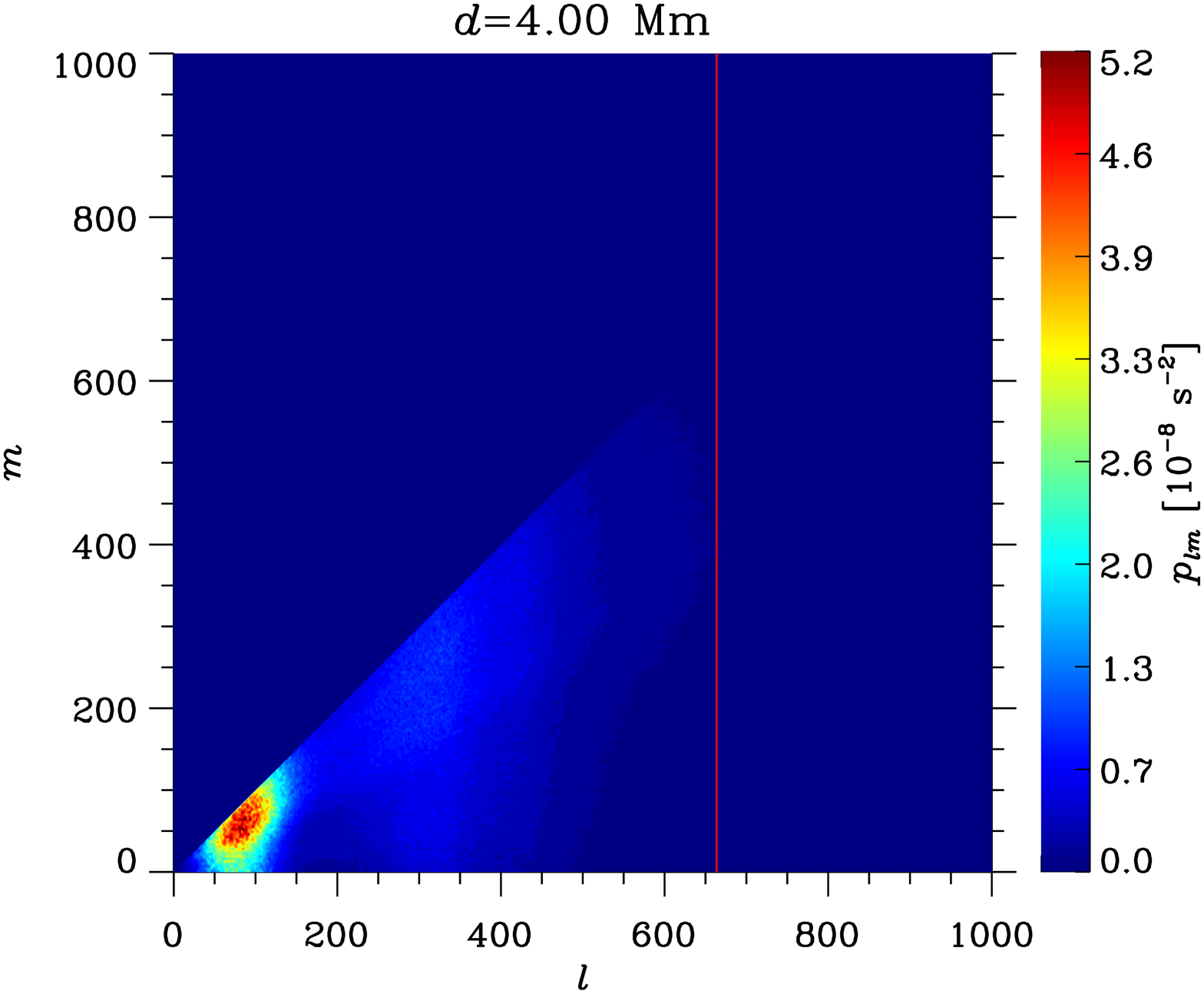}
    \includegraphics[width=0.36\textwidth,bb=50 0 800 650,clip] {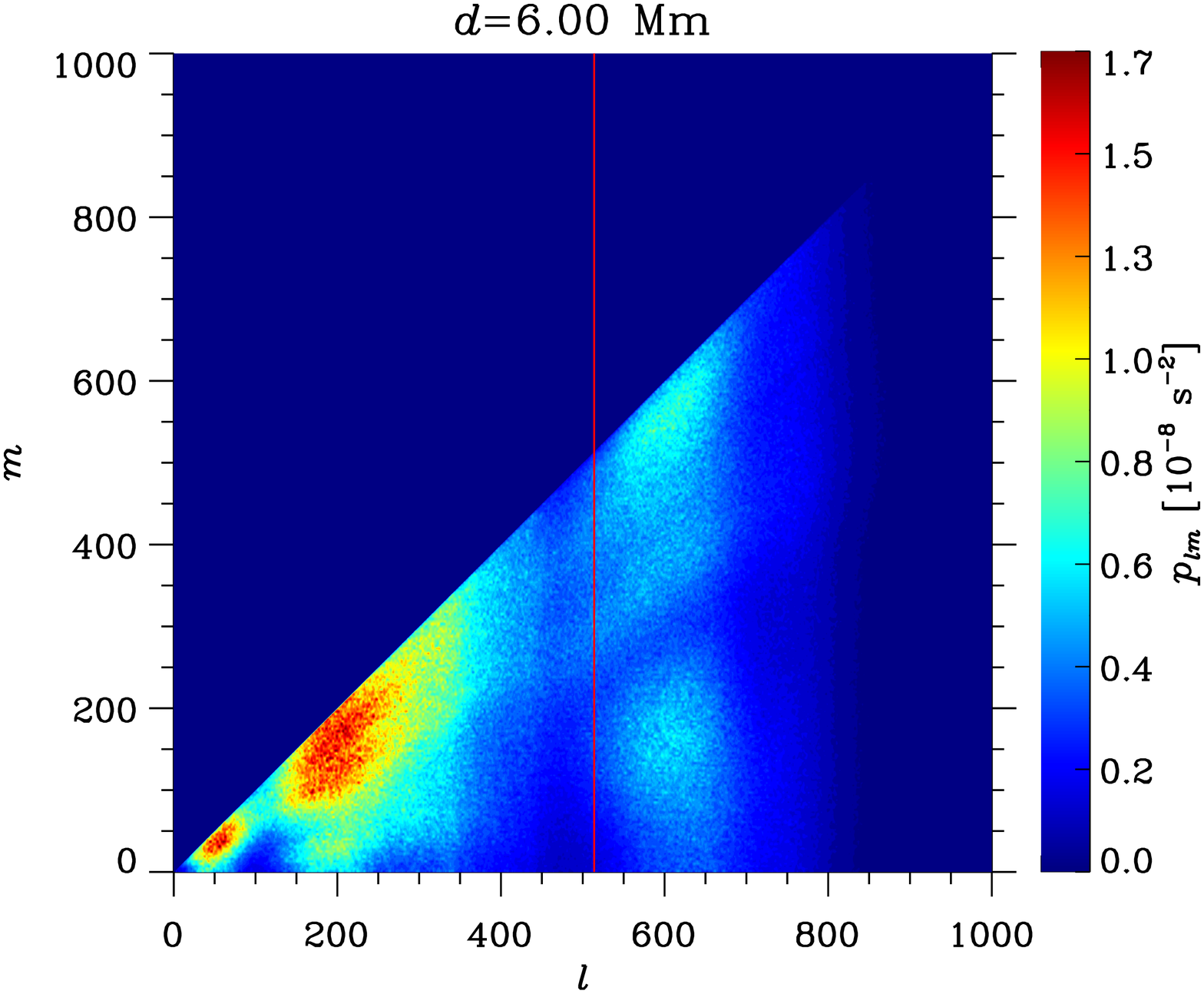}
    \includegraphics[width=0.36\textwidth,bb=50 0 800 650,clip] {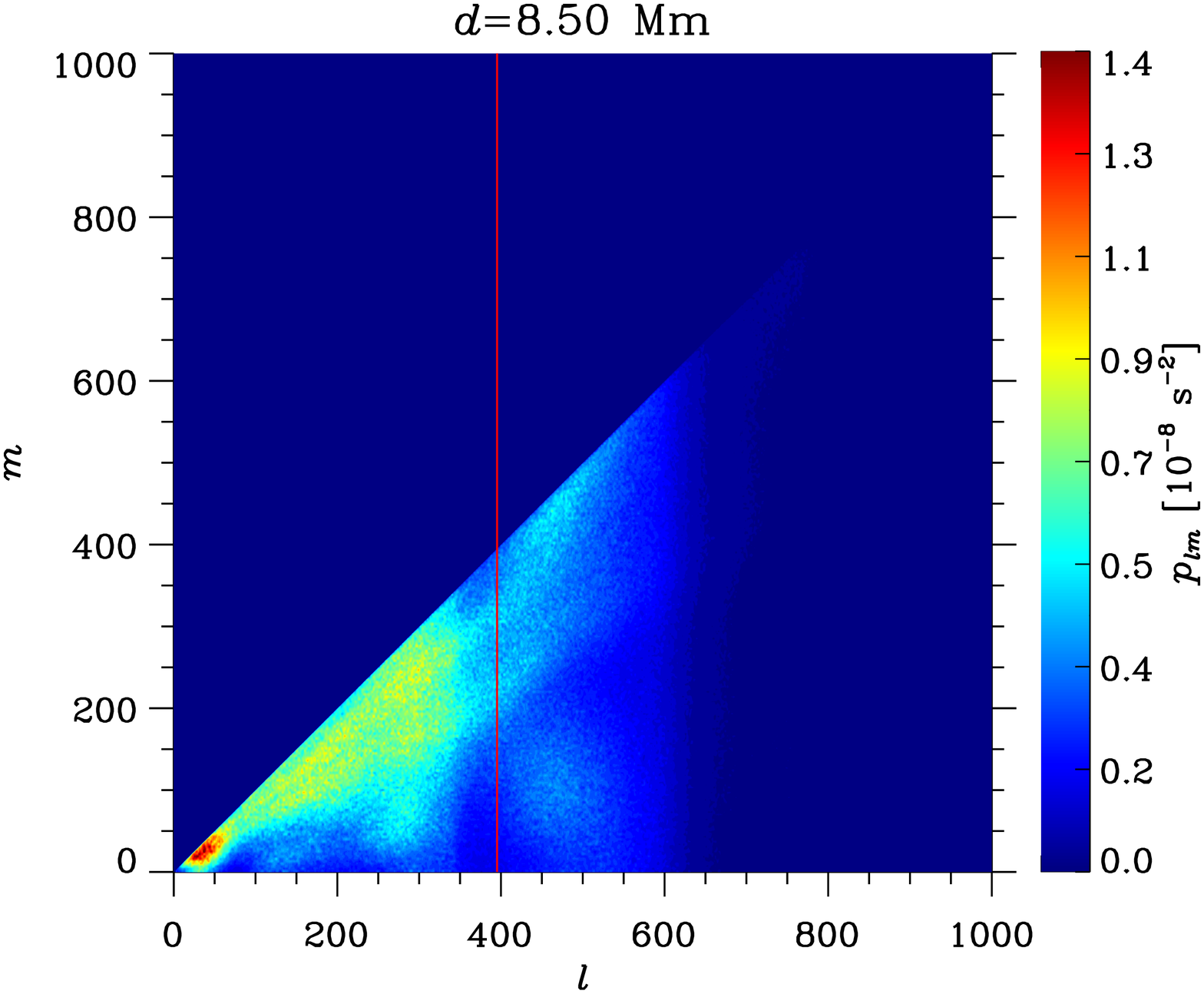}
    \includegraphics[width=0.36\textwidth,bb=50 0 800 650,clip] {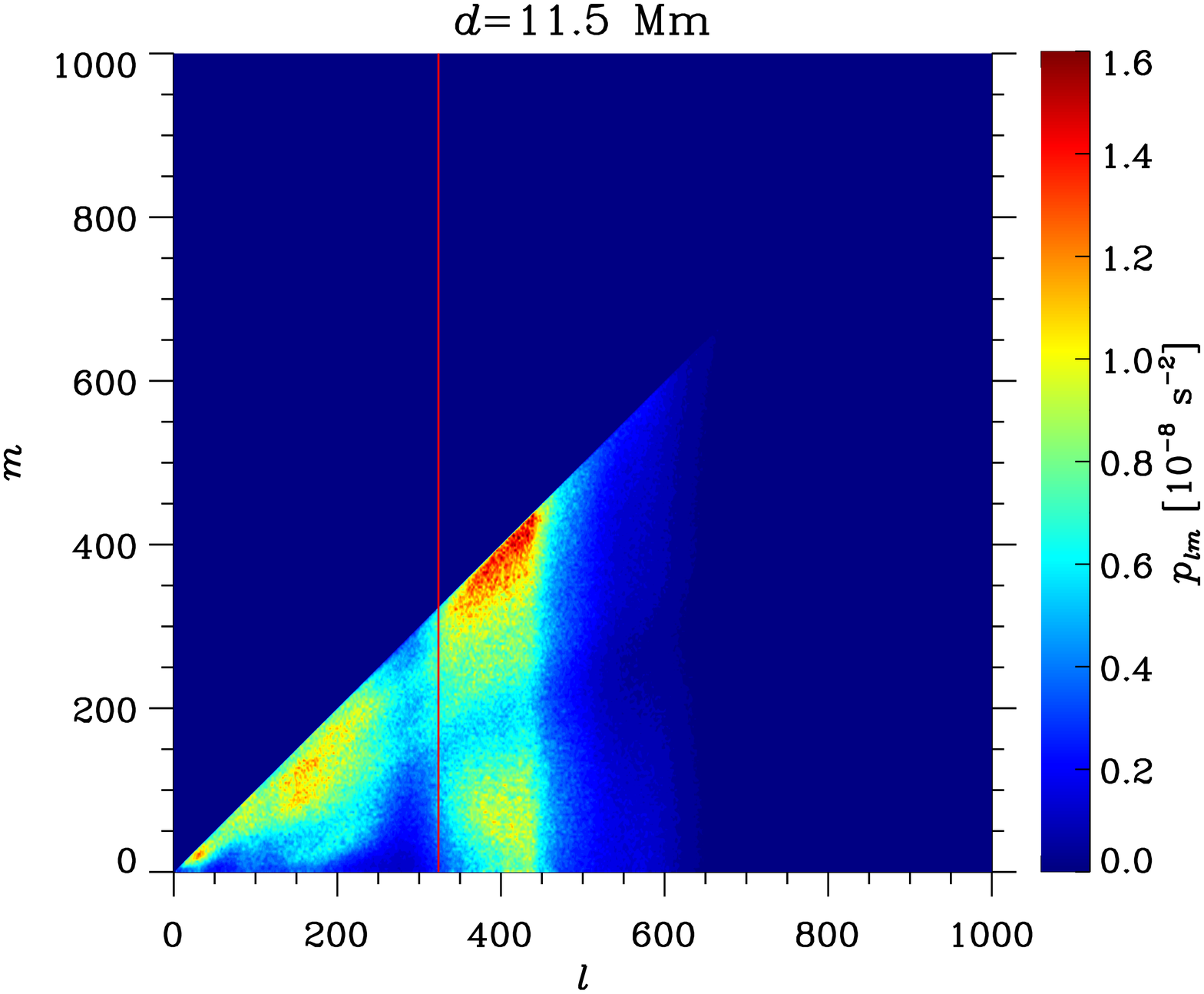}
    \includegraphics[width=0.36\textwidth,bb=50 0 800 650,clip] {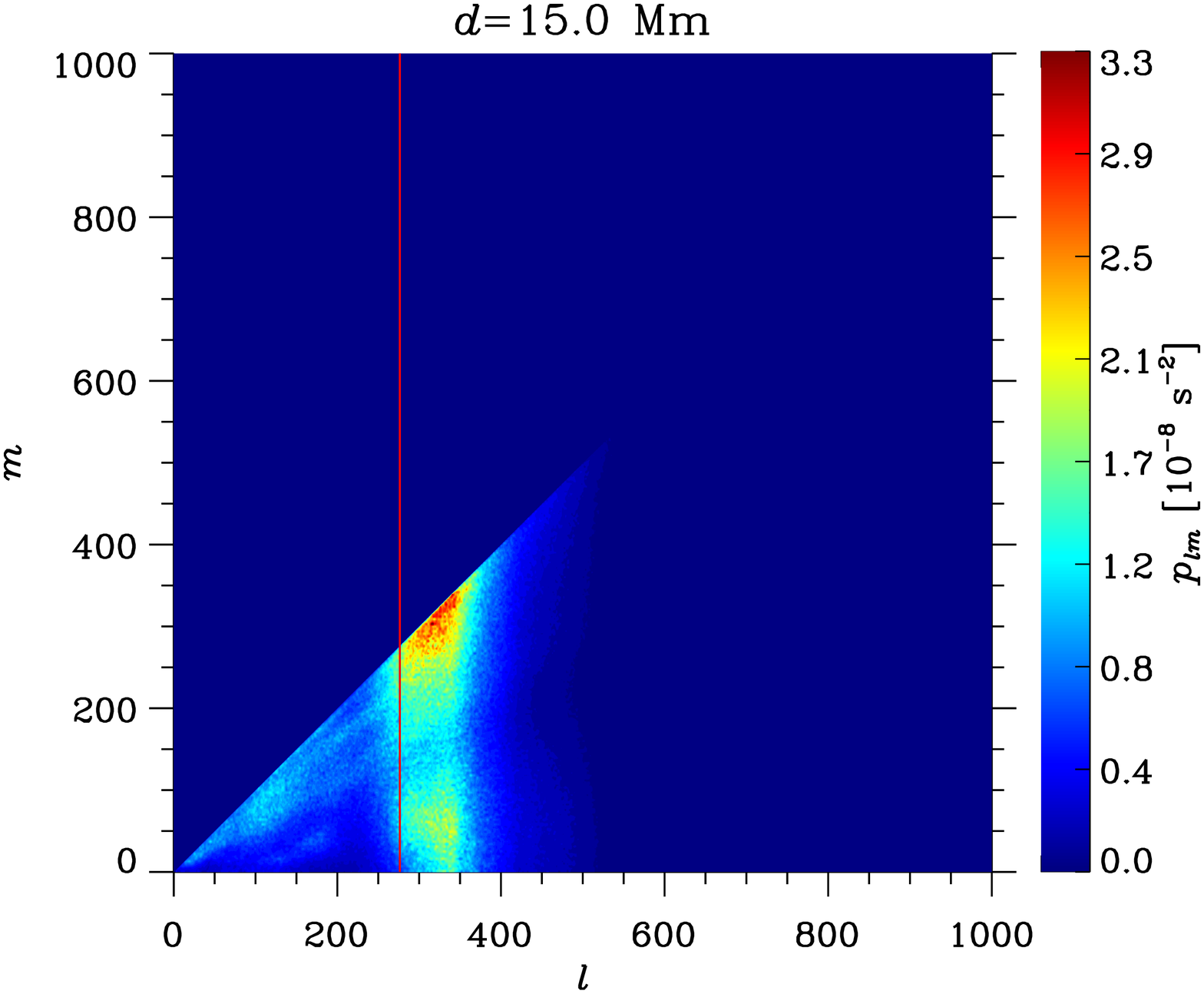}
    \includegraphics[width=0.36\textwidth,bb=50 0 800 650,clip] {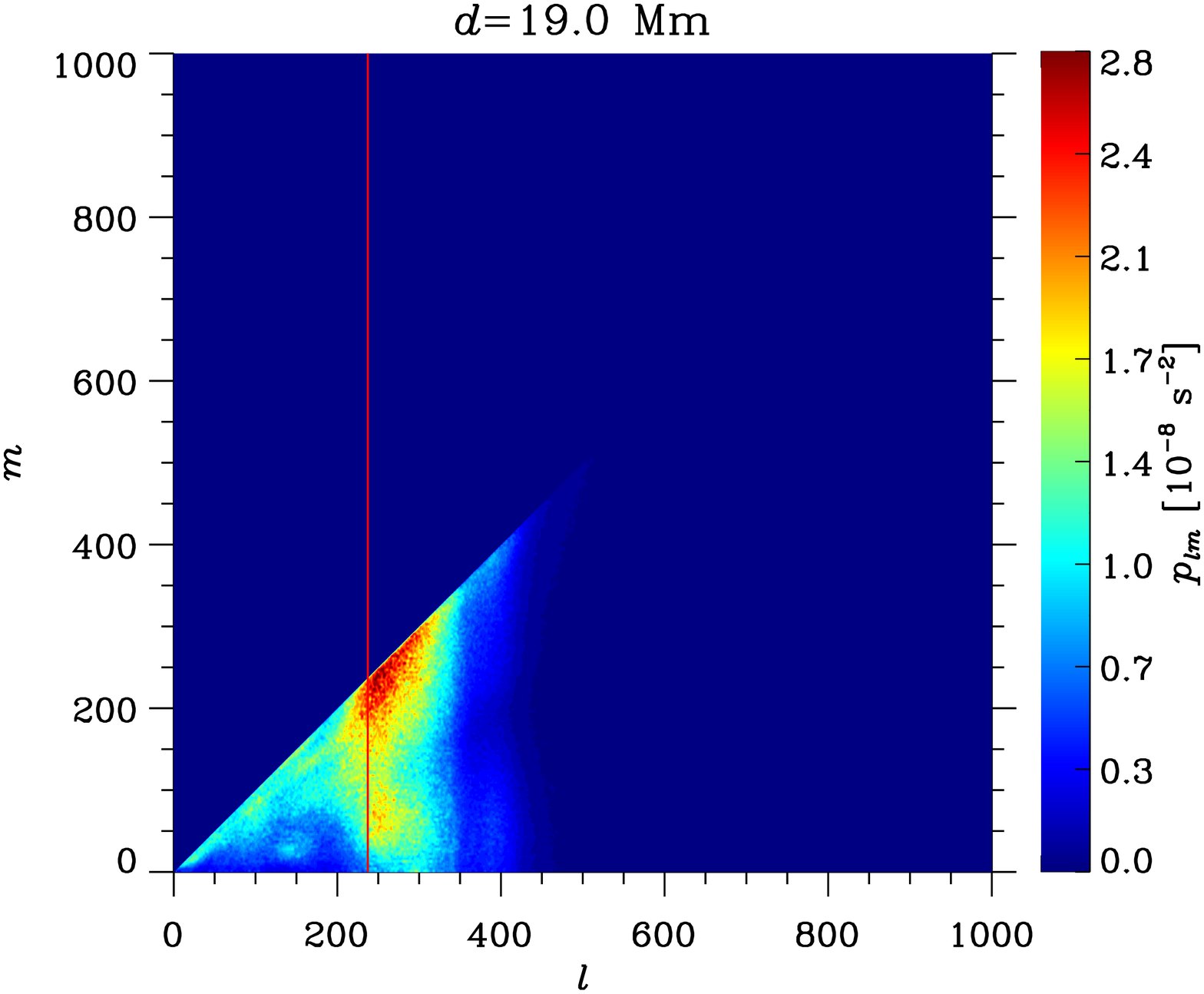}\\
\caption{Depth variation of a sample power spectrum of the divergence of the unsmoothed velocity field, $p_{lm}$, with $l_{\max}=1000$ obtained by 45-day averaging over the low-activity period from 2019 December 21 to 2020 February 4. The red vertical line in each spectrum marks the estimated long-wavelength spectral bound of the realization noise (Figure~\ref{lvsdepth}). The depth values are indicated at the top of each panel.
		\label{spectra1000}}
\end{figure*}

\section{Results}\label{results}
	
\subsection{Unsmoothed fields. Realization noise}

\begin{figure*} 
	\centering
    \includegraphics[width=0.36\textwidth,bb=50 0 800 650,clip] {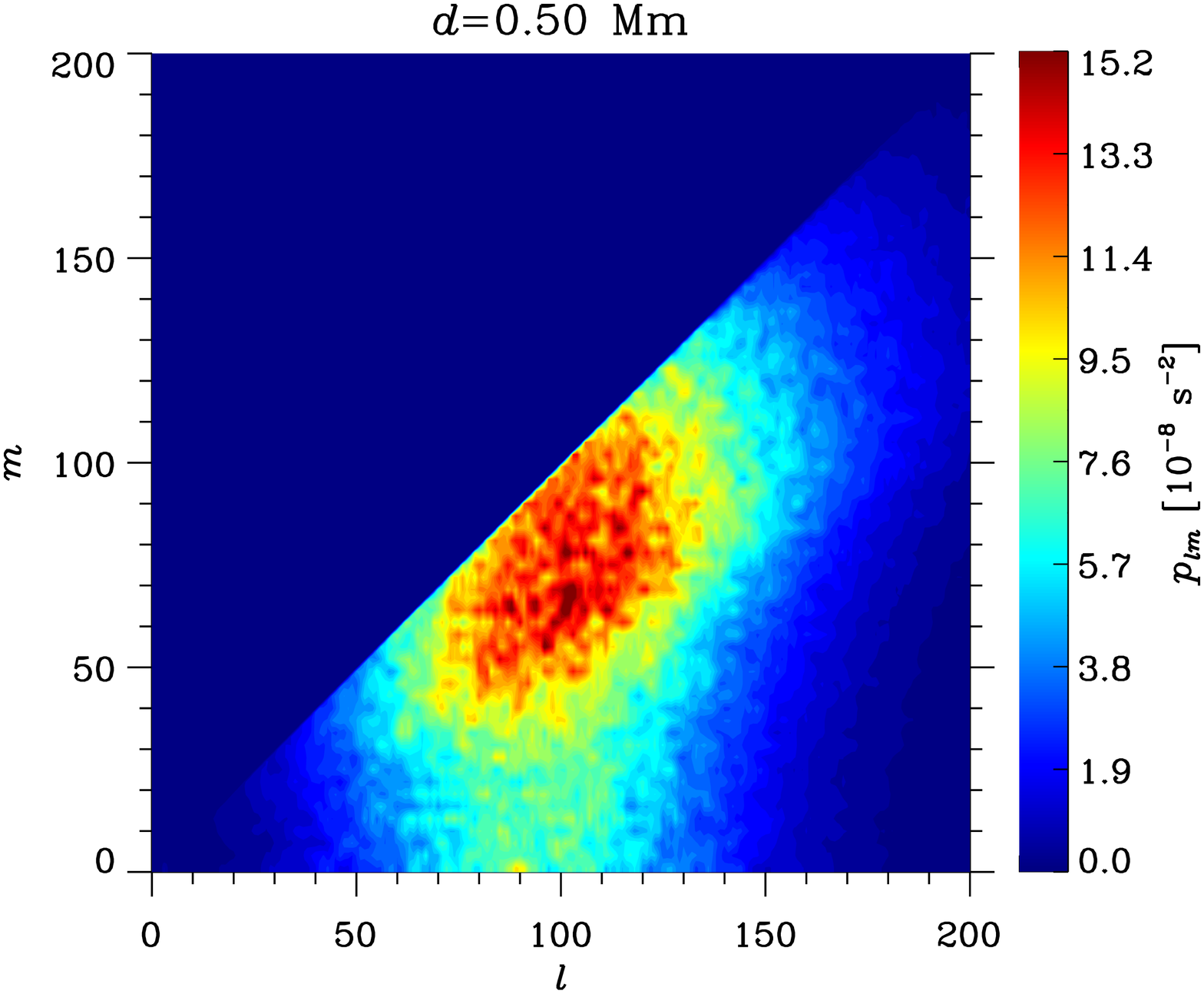}
    \includegraphics[width=0.36\textwidth,bb=50 0 800 650,clip] {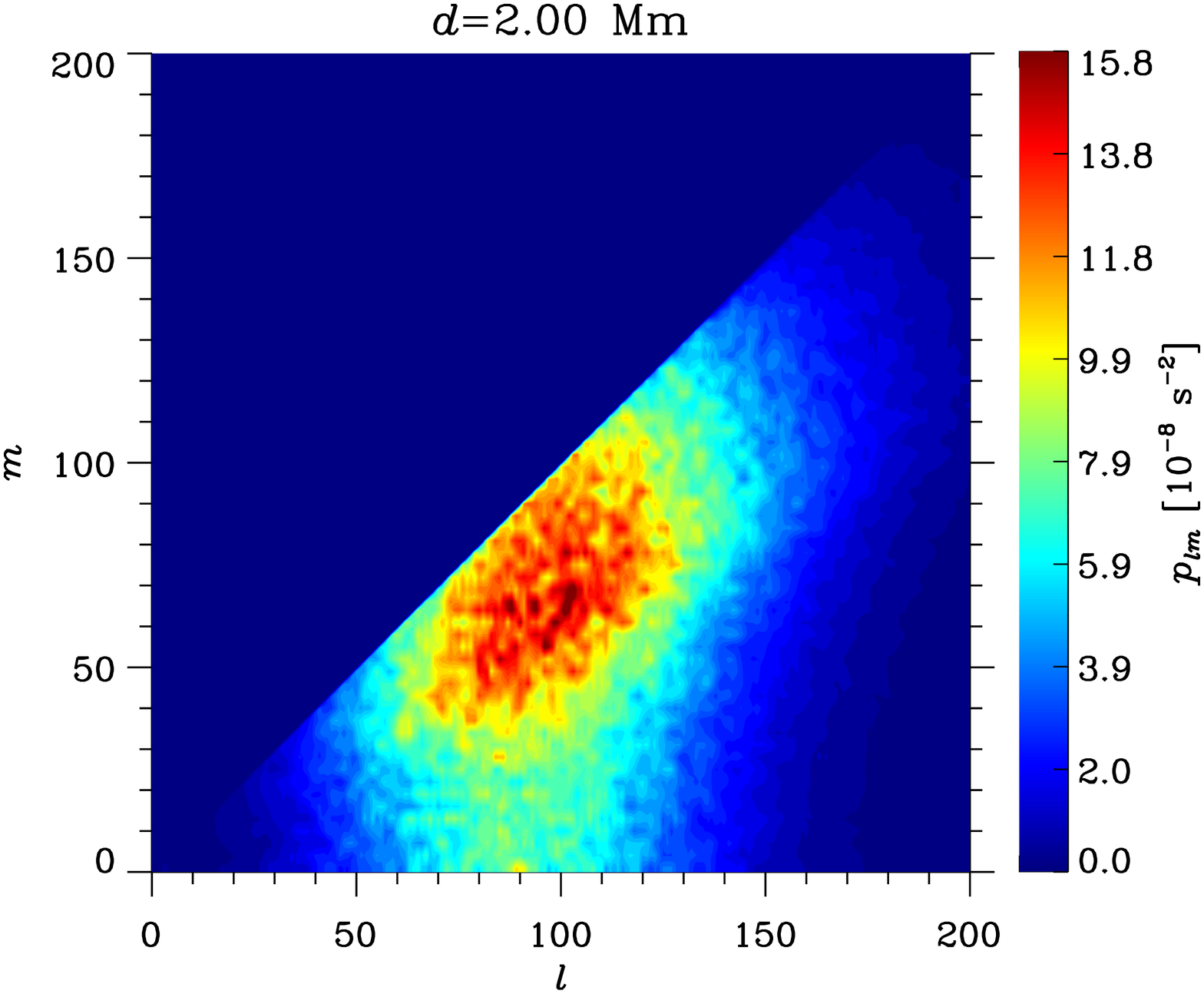}
    \includegraphics[width=0.36\textwidth,bb=50 0 800 650,clip] {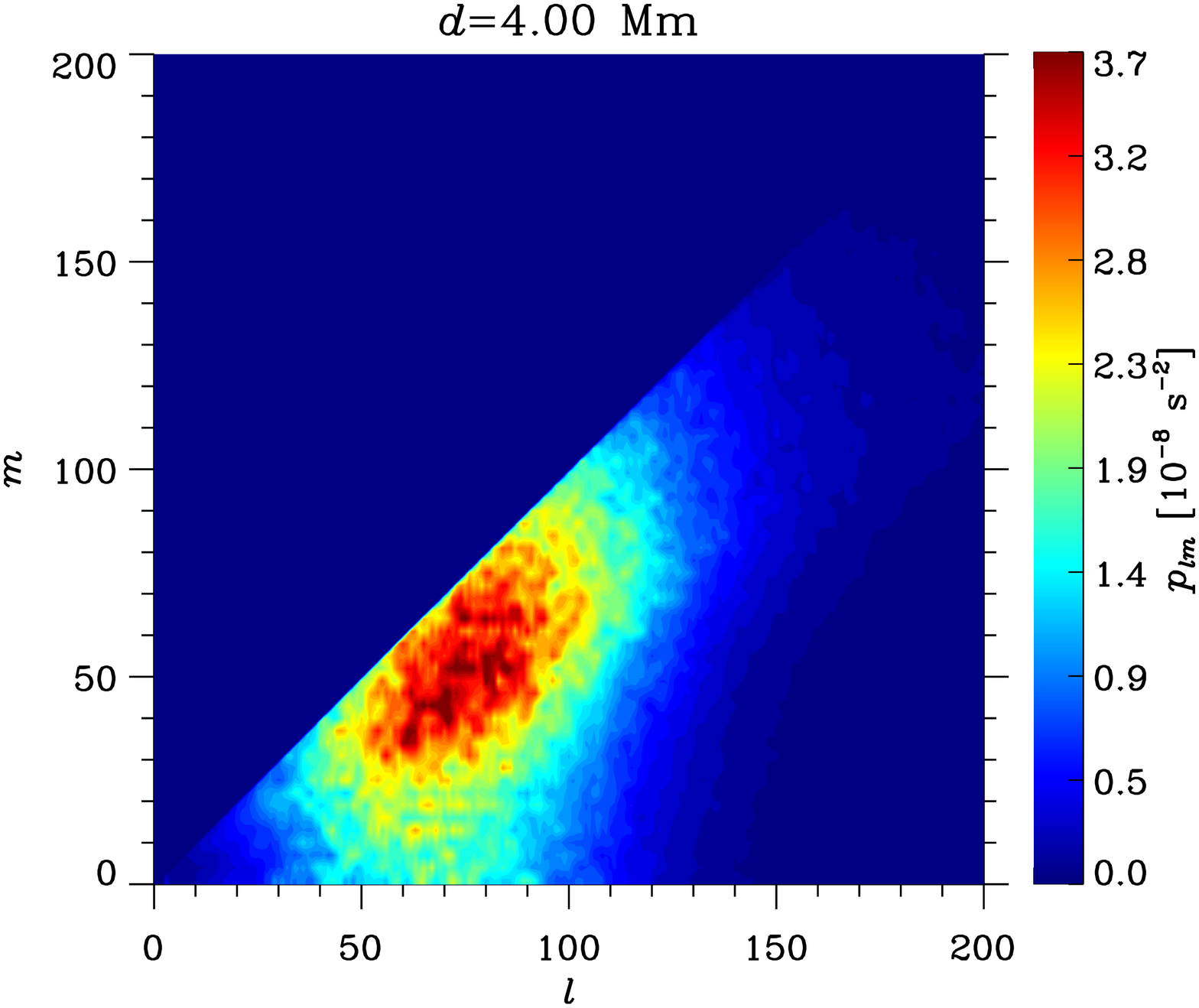}
    \includegraphics[width=0.36\textwidth,bb=50 0 800 650,clip] {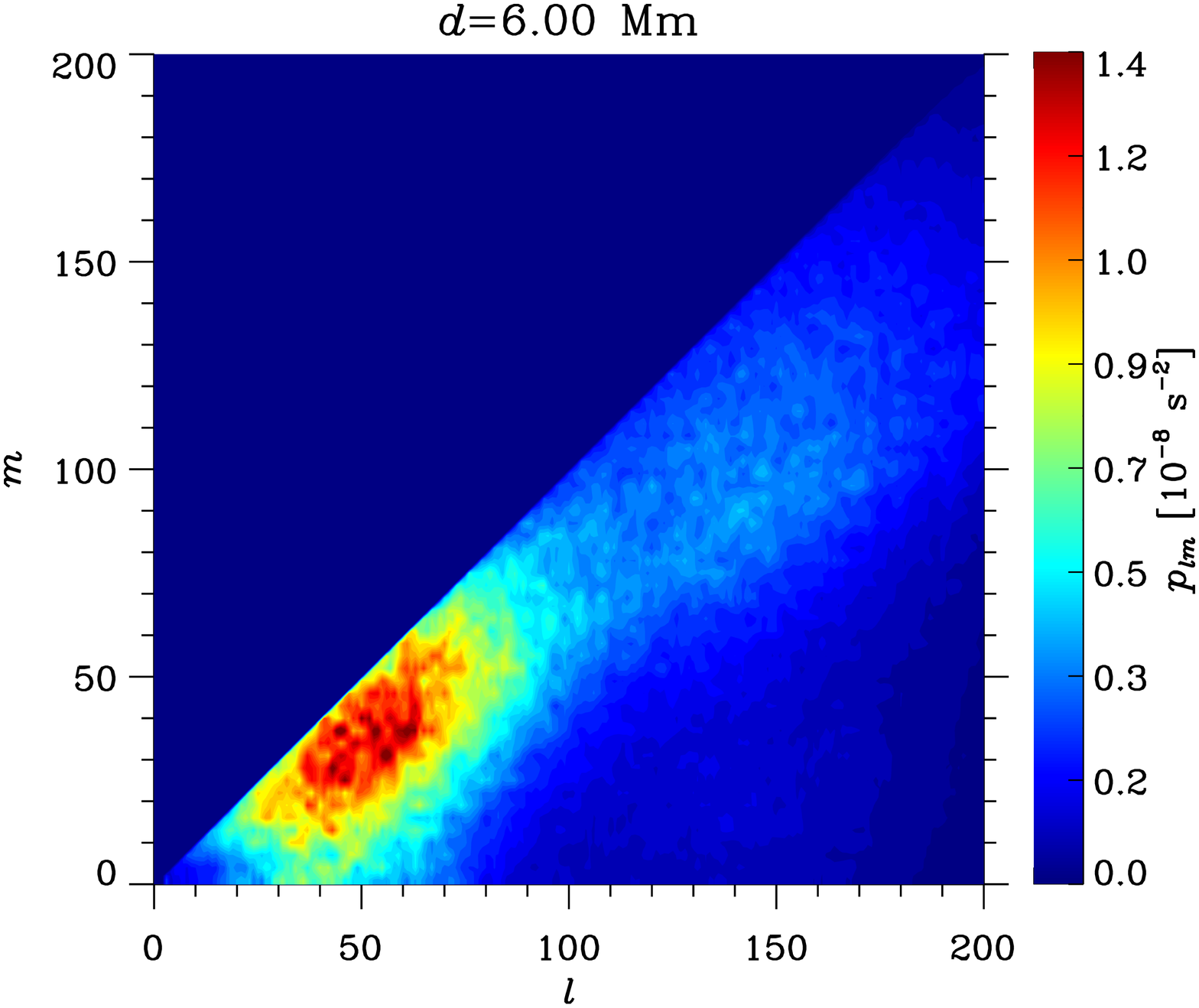}
    \includegraphics[width=0.36\textwidth,bb=50 0 800 650,clip] {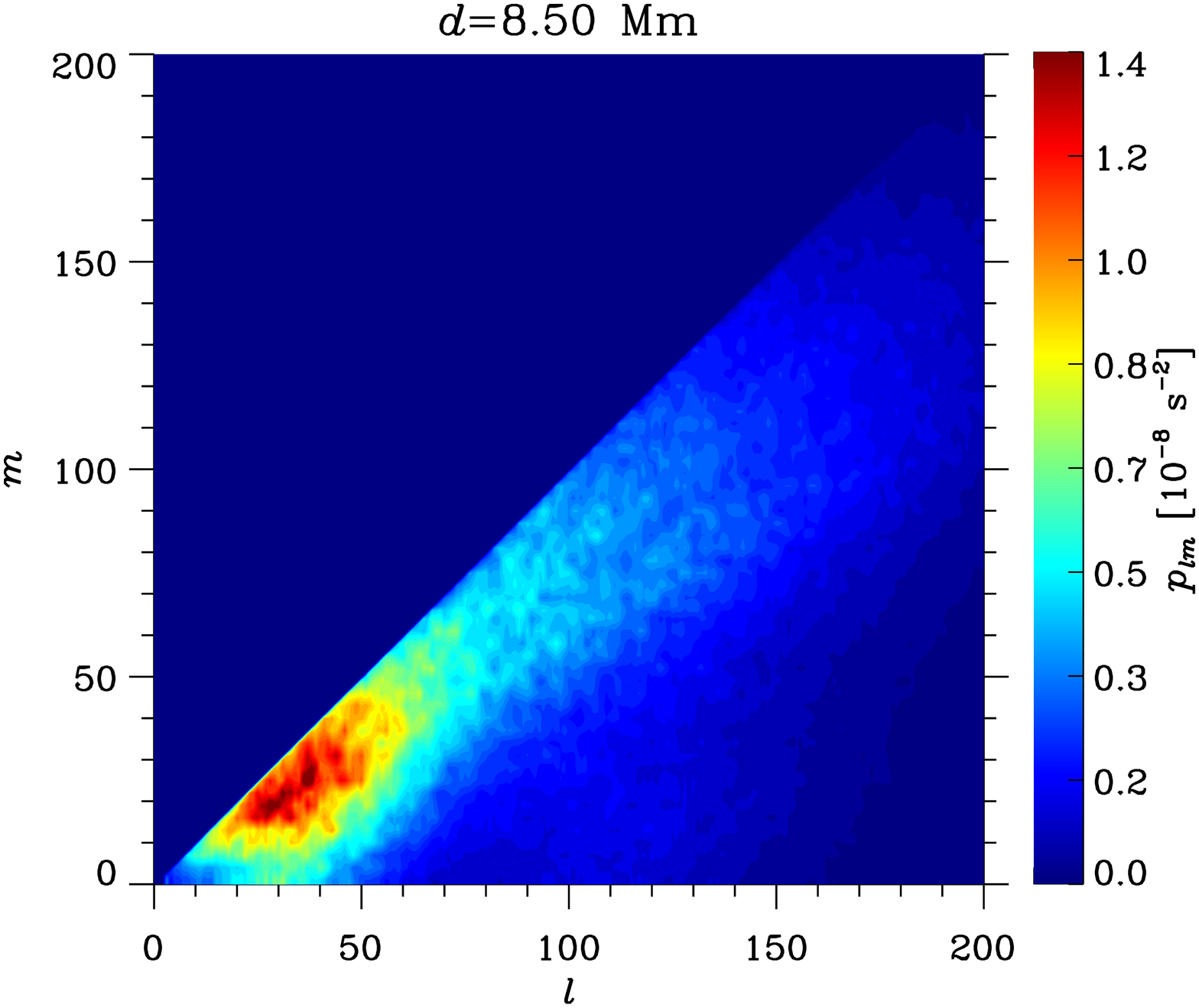}
    \includegraphics[width=0.36\textwidth,bb=50 0 800 650,clip] {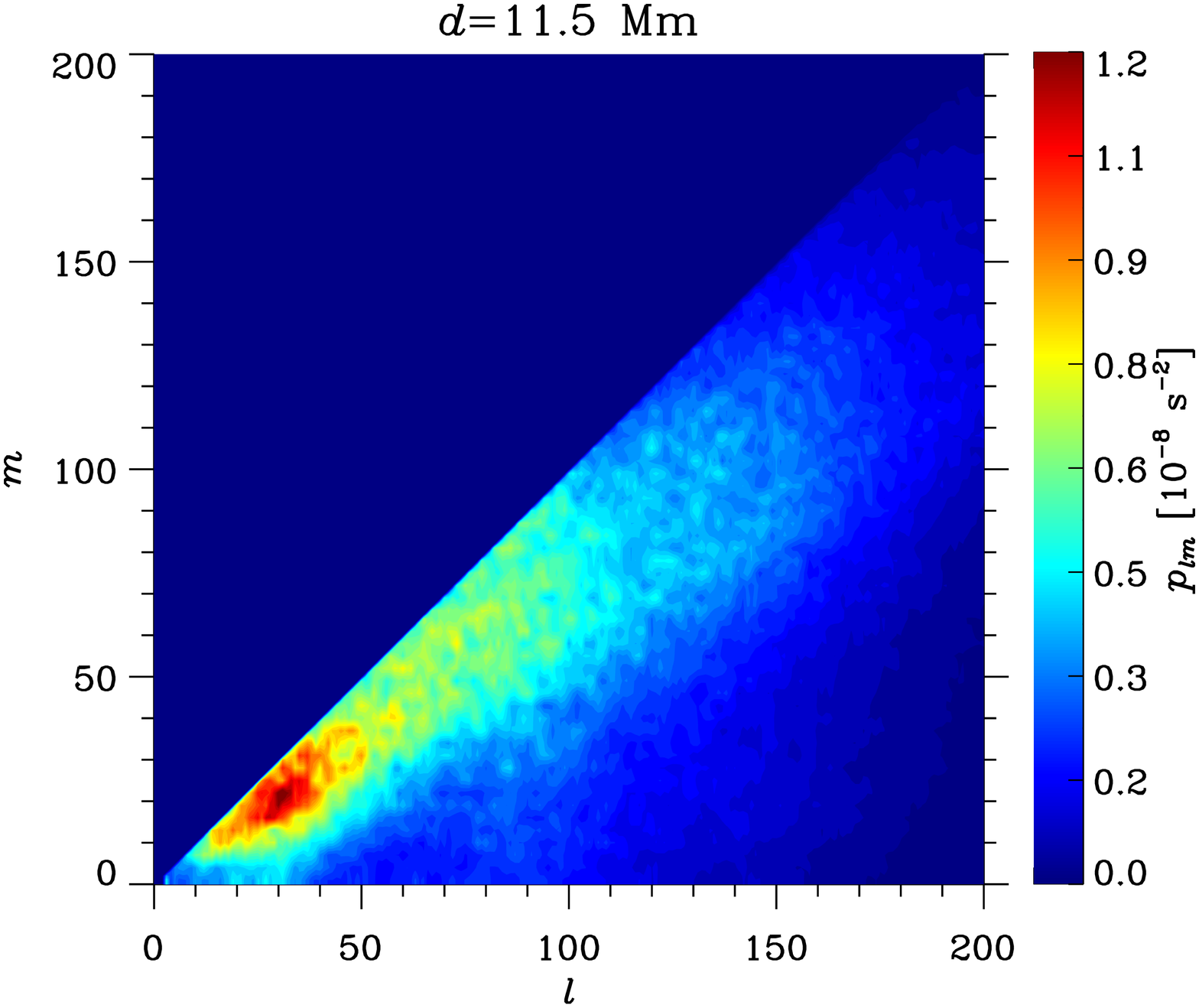}
    \includegraphics[width=0.36\textwidth,bb=50 0 800 650,clip] {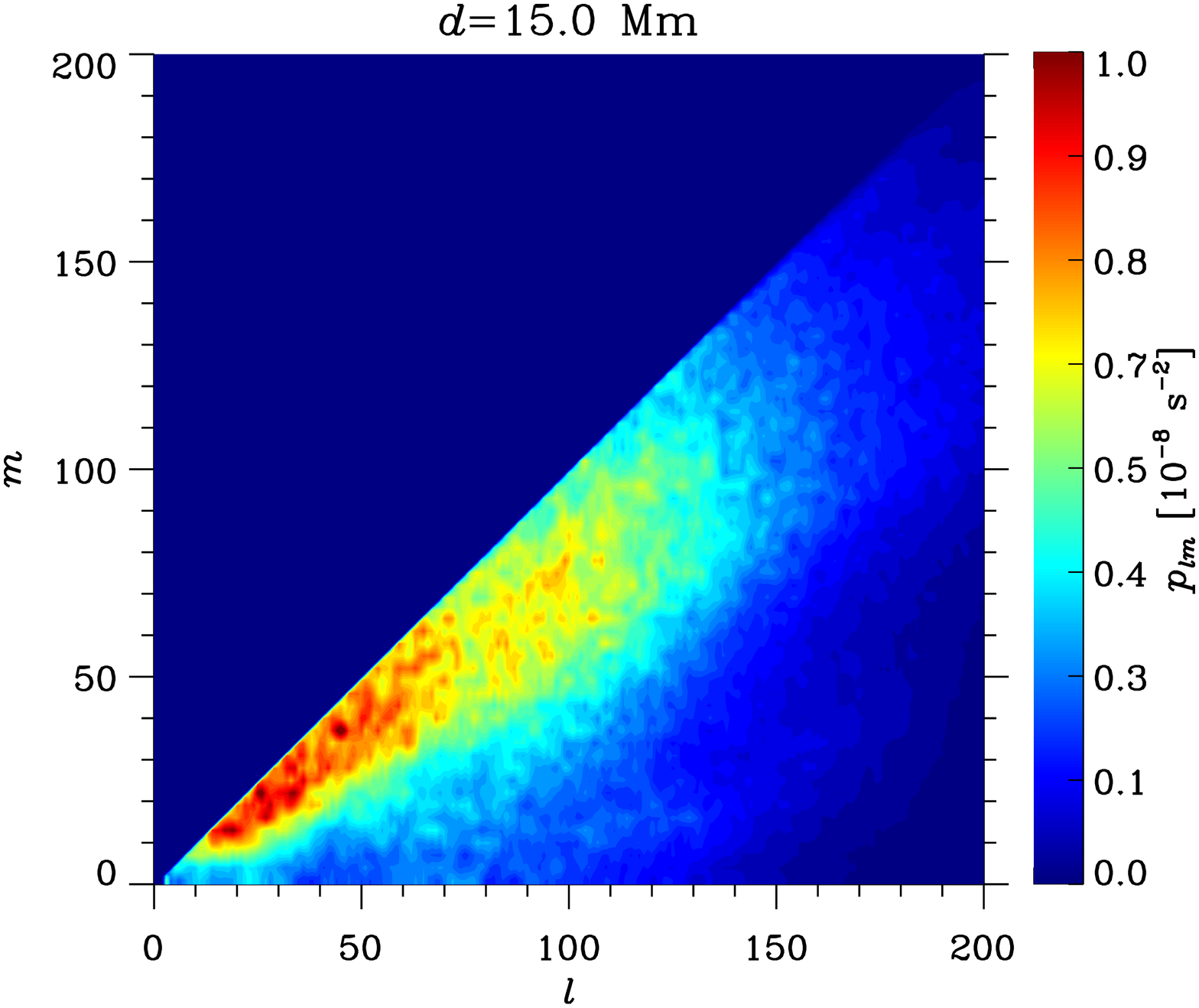}
    \includegraphics[width=0.36\textwidth,bb=50 0 800 650,clip] {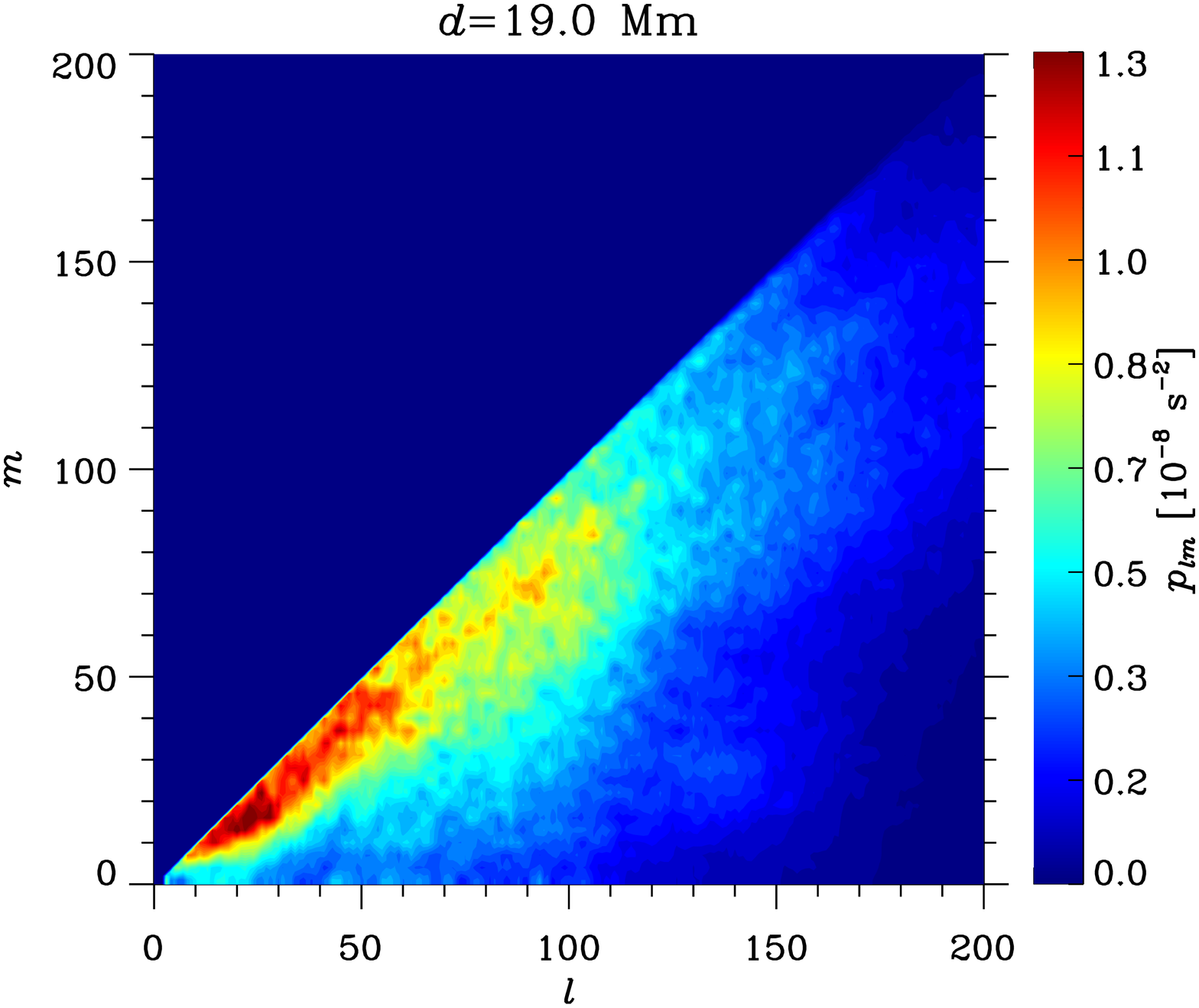}\\
\caption{Depth variation of a sample power spectrum of the divergence of the velocity field smoothed with a 17.5-Mm window, $p_{lm}$, with $l_{\max}=200$ obtained by 45-day averaging over the low-activity period from 2019 December 21 to 2020 February 4. The depth values are indicated at the top of each panel.
		\label{spectra200}}
\end{figure*}

A visual inspection of the velocity-divergence maps suggests that the characteristic scale of convection flows increases with depth (Figure~\ref{images}). The regularities of the depth variation of the flow scales are among the principal aims of our study. We start describing the results with the fields whose spectral representations contain harmonics of orders $l \leqslant l_{\max}=1000$. Therefore, the shortest wavelengths present in the spectra are about 2--4~Mm, which exceeds the spatial sampling interval, or the pixel size of the maps, 1.46~Mm.

Due to the stochastic excitation of solar acoustic oscillations and their finite wavelength, the flow maps contain so-called ``realization noise'' \citep{Gizon_Birch_2004}, the spatial scale of which increases with depth. This scale can be estimated as the horizontal wavelength (or the corresponding angular degree, $l$) of acoustic waves at the inner turning point, where the wave phase speed, $\lambda/T$, is equal to the local sound speed, $c$ (here, $\lambda$ is the horizontal wavelength and $T$ is the period of the wave). If $\lambda$ is determined by Equation~(\ref{Jeanseq}) and $c$ is calculated from the standard solar model \citep[see][]{Crist-Dals_etal_1996}, then, for a frequency of $1/T\sim 3\times 10^{-3}$~Hz, the dependence of the angular degree on the turning-point depth appears as shown in Figure~\ref{lvsdepth}. It indicates that the scale of the realization noise due to five-minute oscillations varies from $l\sim 1000$ in the top layers to $l\sim 250$ at depths of about 20~Mm. Higher-frequency waves have correspondingly shorter wavelengths. The $l$ values plotted in Figure~\ref{lvsdepth} can thus be used as an estimate for the long-wavelength bound of the realization noise.

Figure~\ref{spectra1000} shows the power spectra of the divergence field computed for all the eight depths and averaged over a 45-day time interval from 2019 December 21 to 2020 February 4, when the solar activity was low. The long-wavelength bound of realization noise is marked with red vertical lines in these spectra. Noise patterns can clearly be seen to the right of the red lines in the panels for $d\geqslant 6$~Mm. The left-hand parts of the diagrams represent the spectra of convective flows.

The flow maps used here have a pixel size of 1.46~Mm and do not resolve granulation. Supergranulation in the upper layers, $d=$0.5--4~Mm, is mainly manifest in a degree range of $l\sim$~70--150, which corresponds to $\lambda\sim$~30--60~Mm. In the depth range of 6--11.5~Mm, a second spectral peak at $l\sim$150--270 corresponding to a characteristic scale of $\lambda\sim$20--30~Mm emerges. It may be related to changes in the properties of supergranulation in these layers.

In the deep layers, harmonics with $l\sim$20--30, of giant-cell scales $\lambda\sim$150--200~Mm, are prominent. In the context of our particular interest in the steady, large-scale component of the spectrum, we can regard the rapidly changing convection as a noise component, along with the realization noise in the helioseismic measurements due to stochastic excitation of solar oscillations. For this reason, from here on, we will consider spectra with this noise filtered out by a smoothing procedure.

\begin{figure*} 
	\centering
    \includegraphics[width=0.36\textwidth,bb=50 0 800 650,clip] {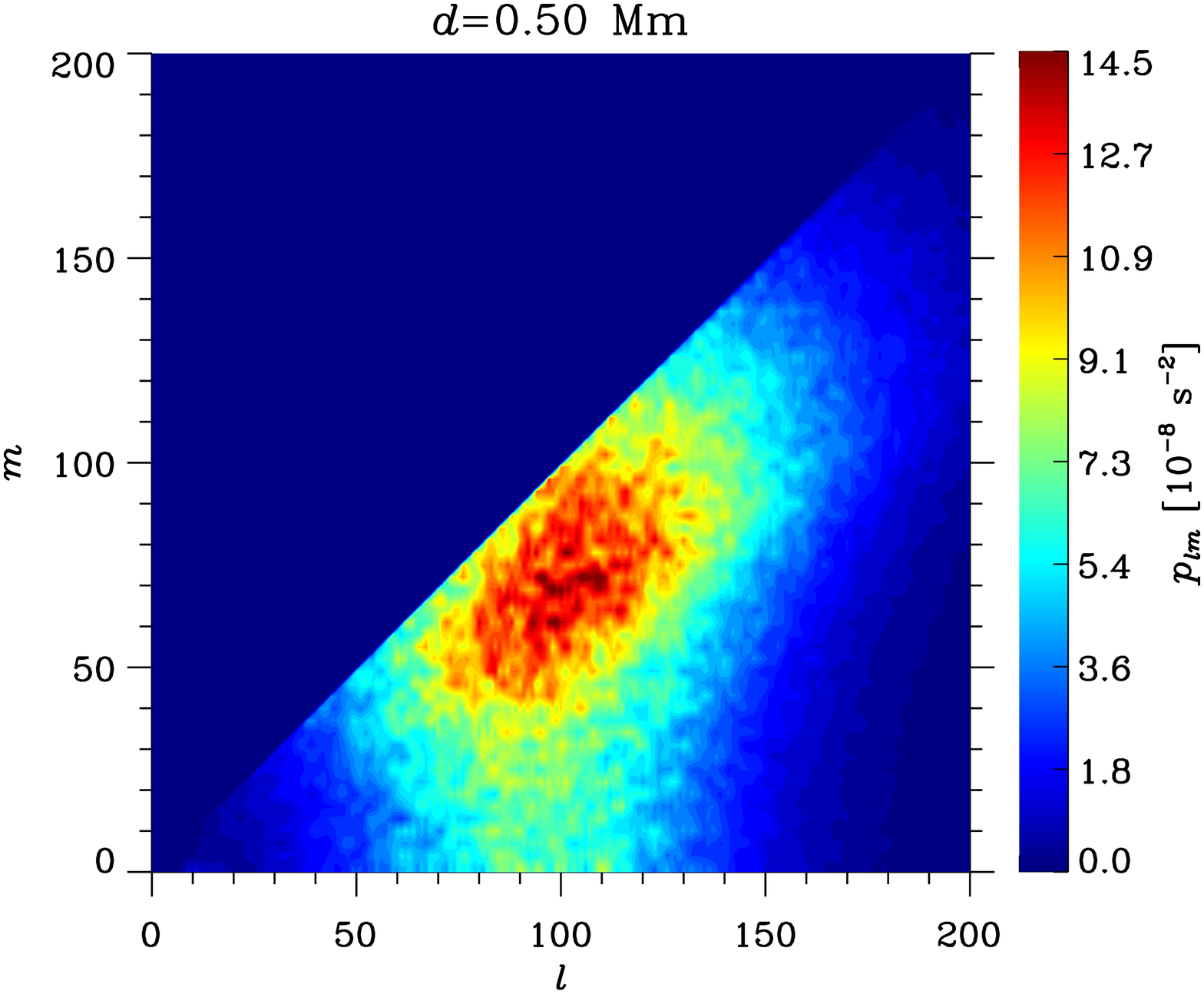}
    \includegraphics[width=0.36\textwidth,bb=50 0 800 650,clip] {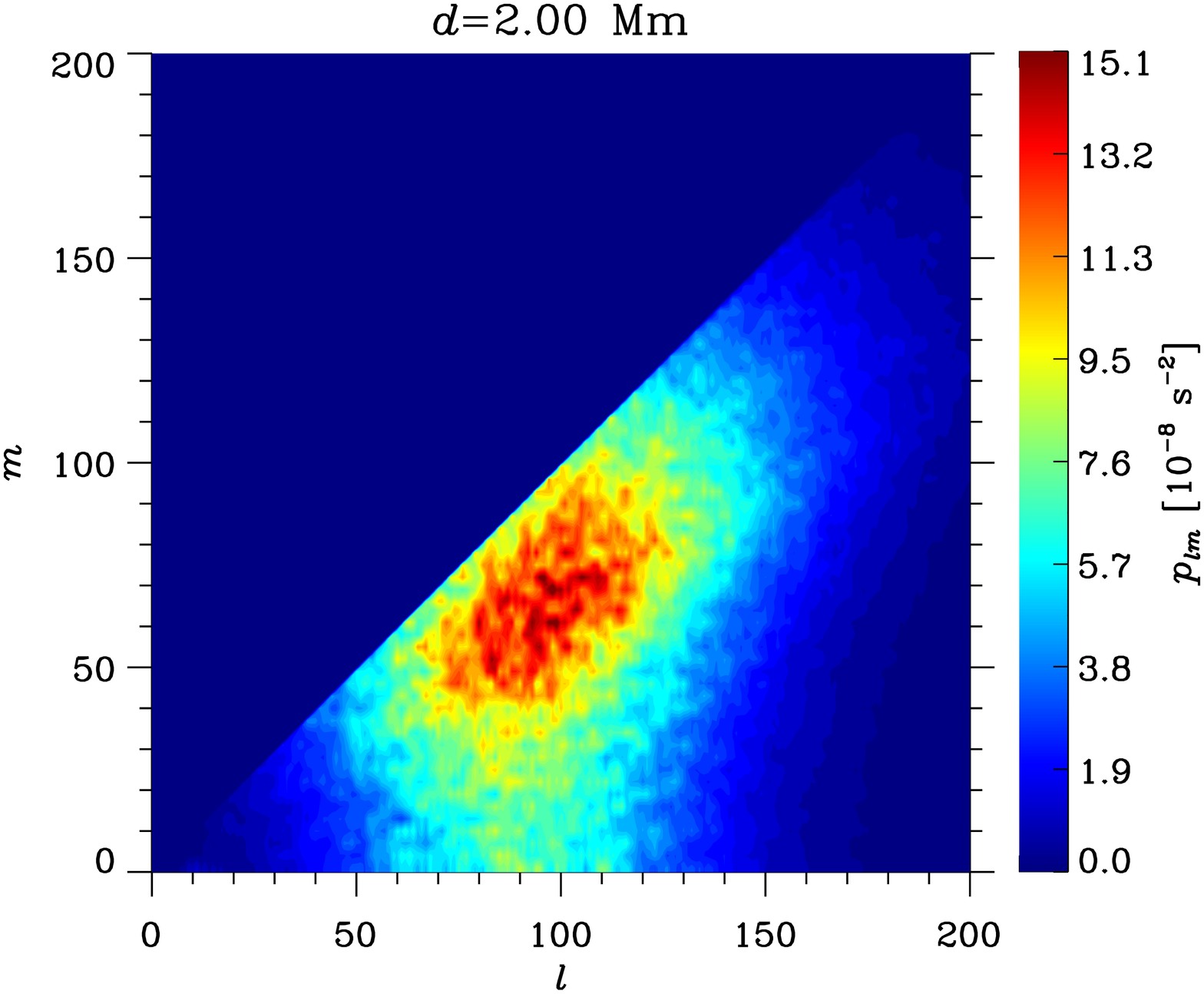}
    \includegraphics[width=0.36\textwidth,bb=50 0 800 650,clip] {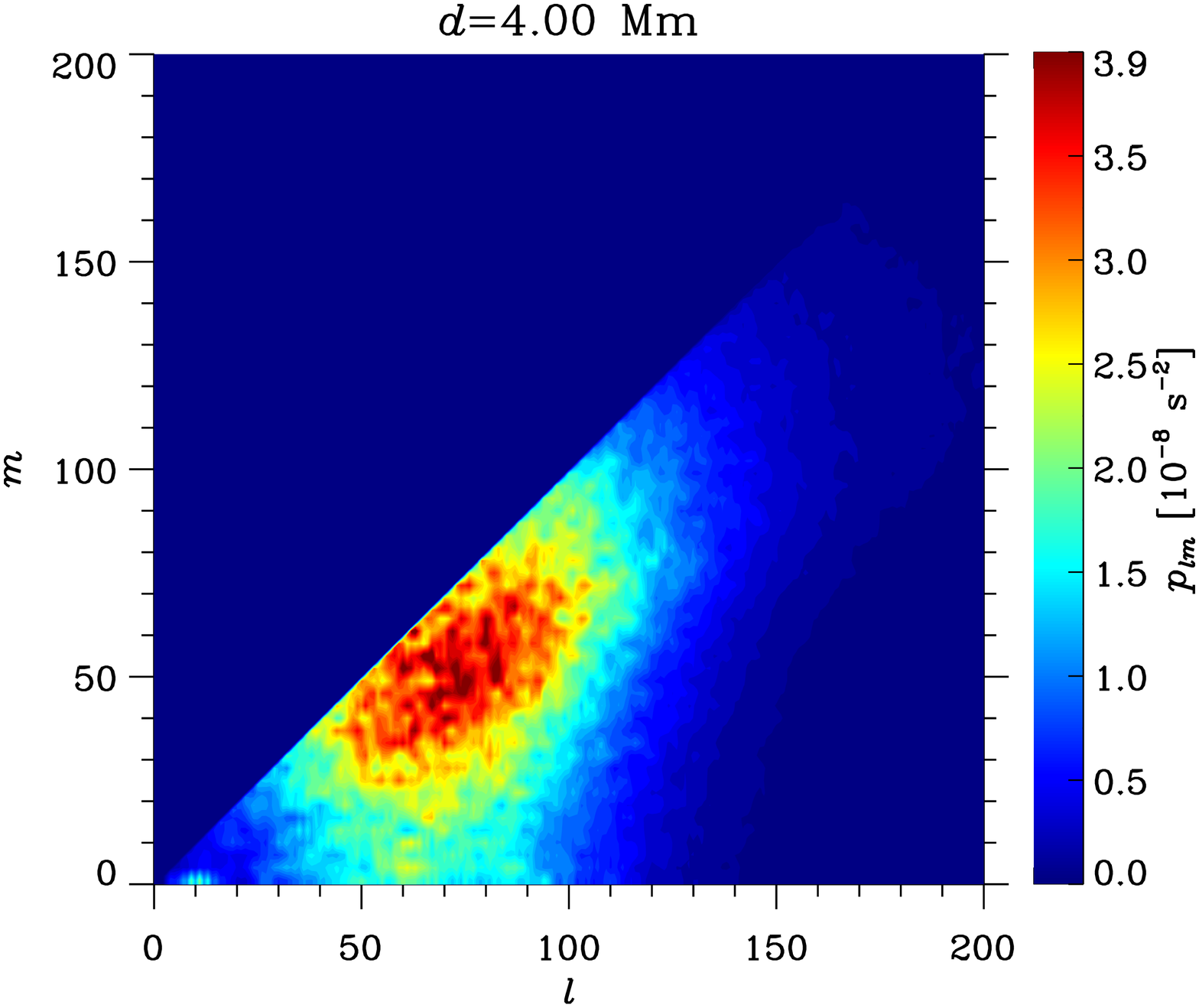}
    \includegraphics[width=0.36\textwidth,bb=50 0 800 650,clip] {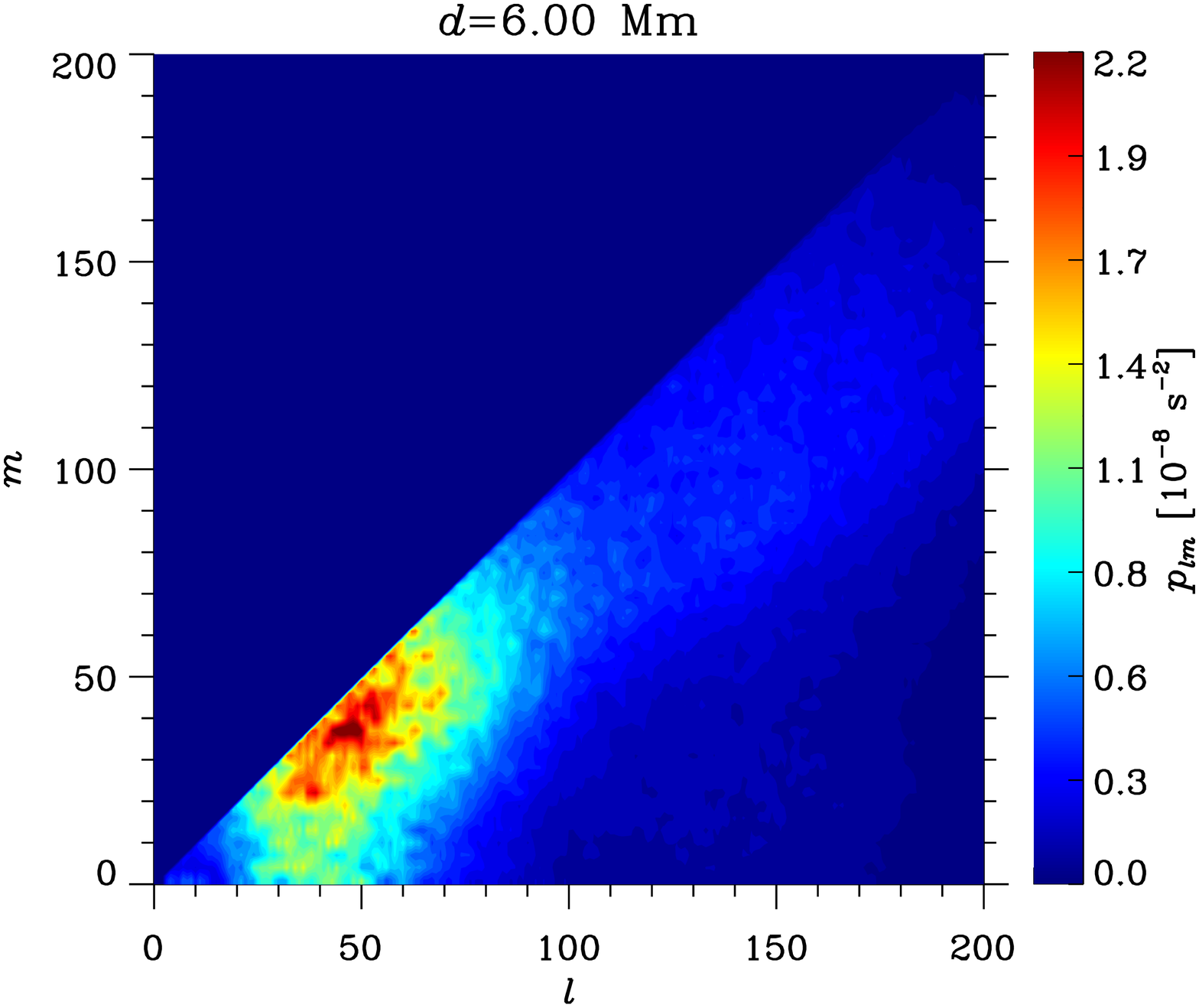}
    \includegraphics[width=0.36\textwidth,bb=50 0 800 650,clip] {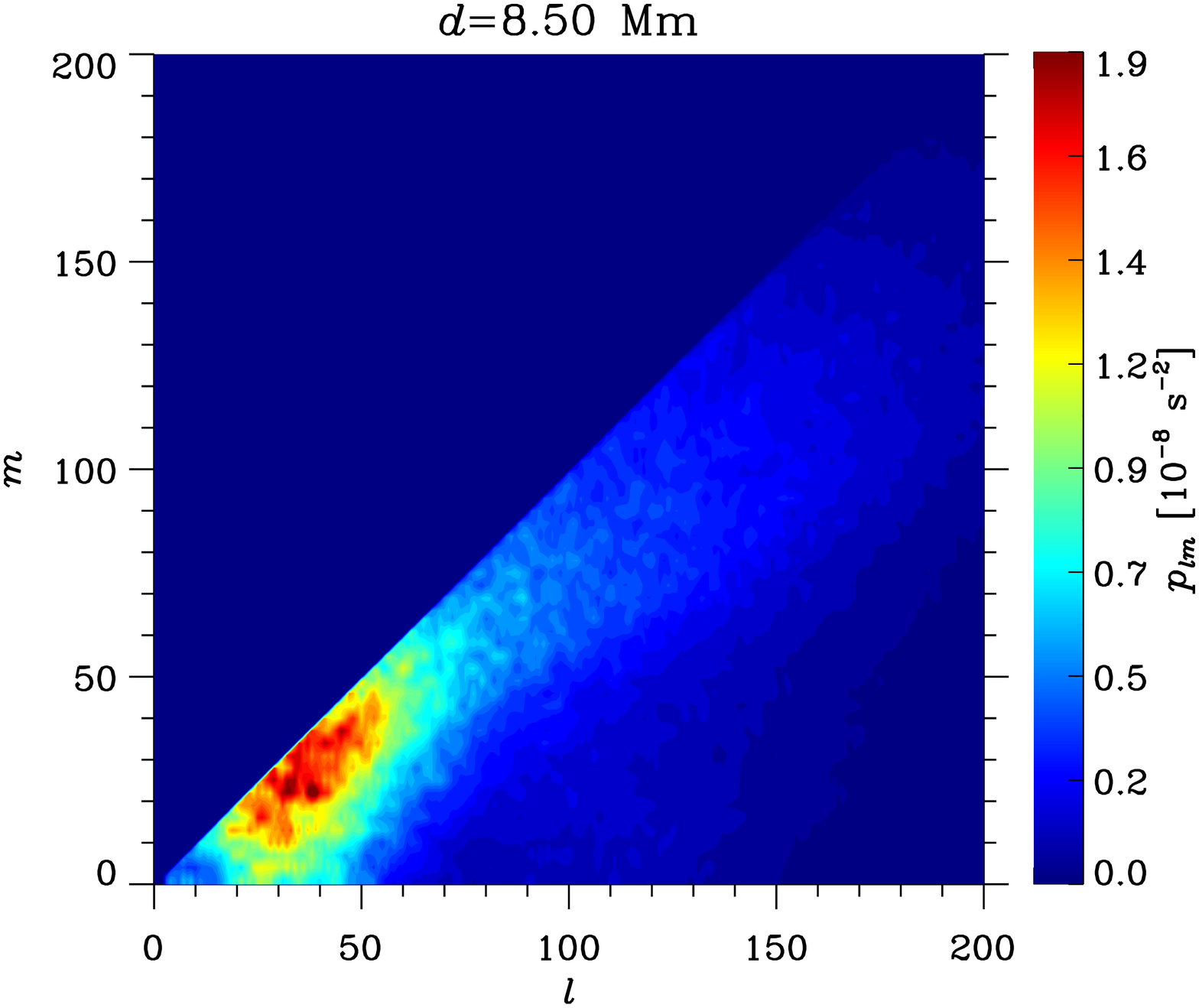}
    \includegraphics[width=0.36\textwidth,bb=50 0 800 650,clip] {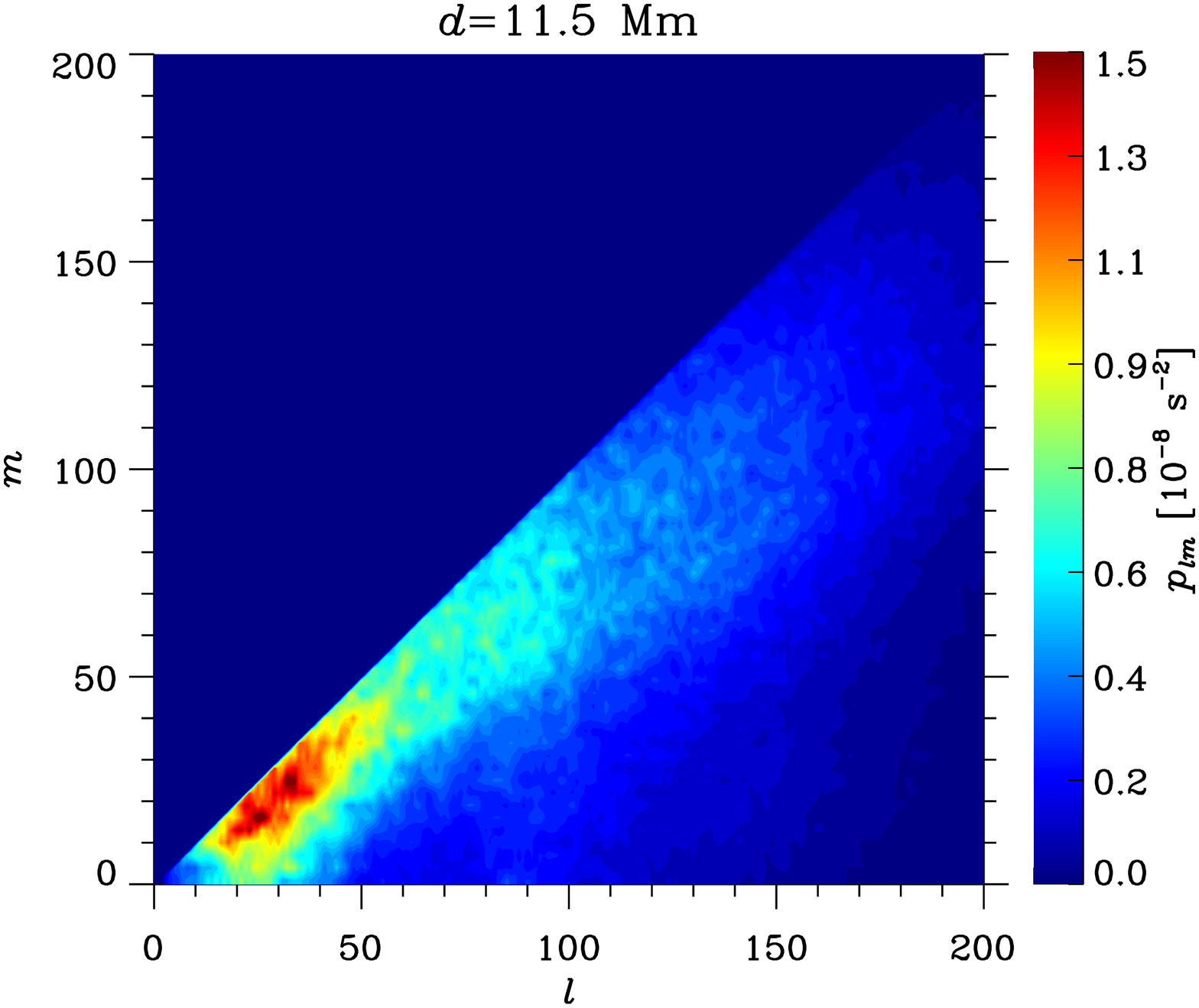}
    \includegraphics[width=0.36\textwidth,bb=50 0 800 650,clip] {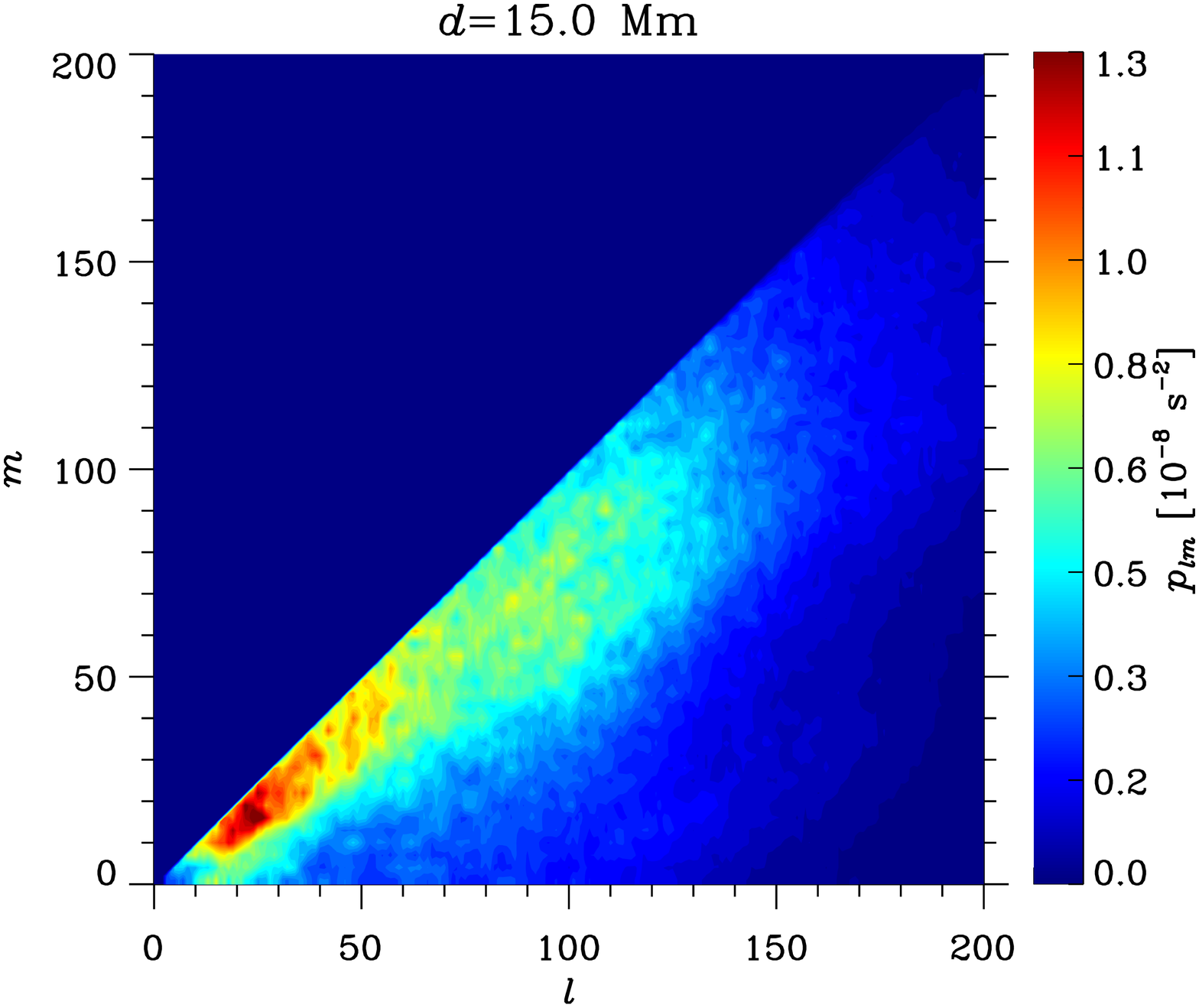}
    \includegraphics[width=0.36\textwidth,bb=50 0 800 650,clip] {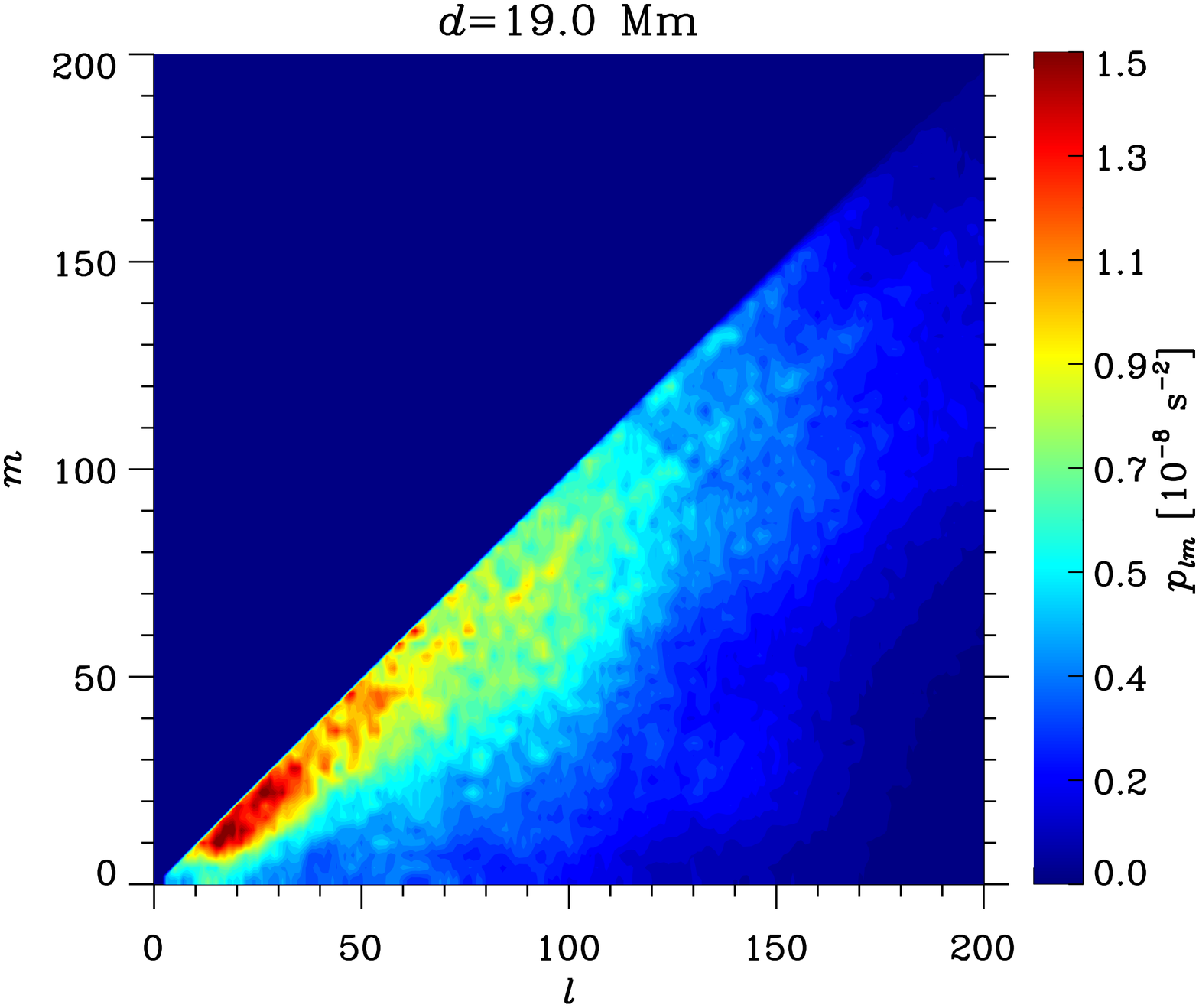}\\
\caption{Same as in Figure~\ref{spectra200} but for the high-activity period from 2013 December 18 to 2014 February 1.
		\label{spectra200HA}}
\end{figure*}

\begin{figure*} 
\centering
\includegraphics[width=0.33 \textwidth,bb=20 0 435 226,clip] {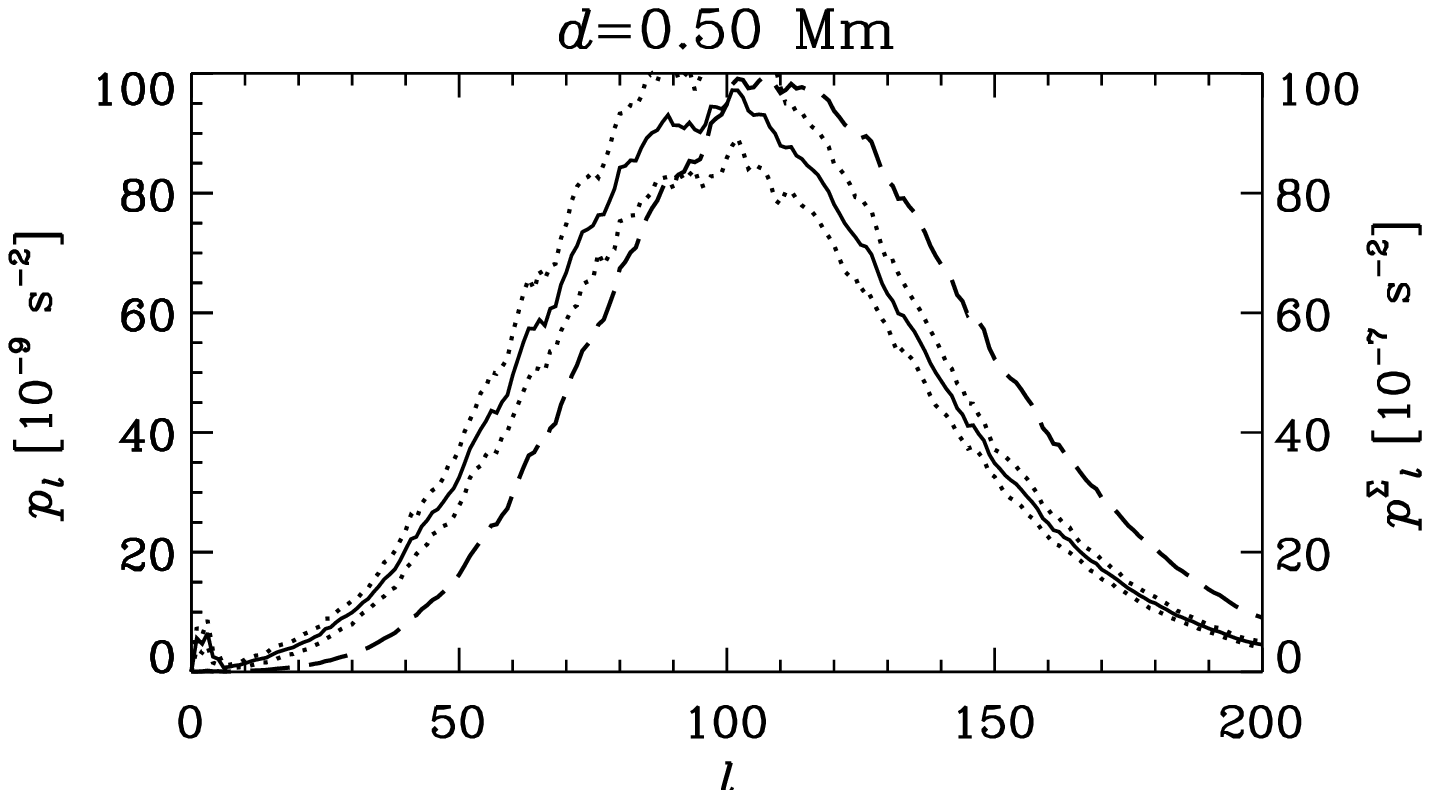}
\includegraphics[width=0.33\textwidth,bb=20 0 435 226,clip] {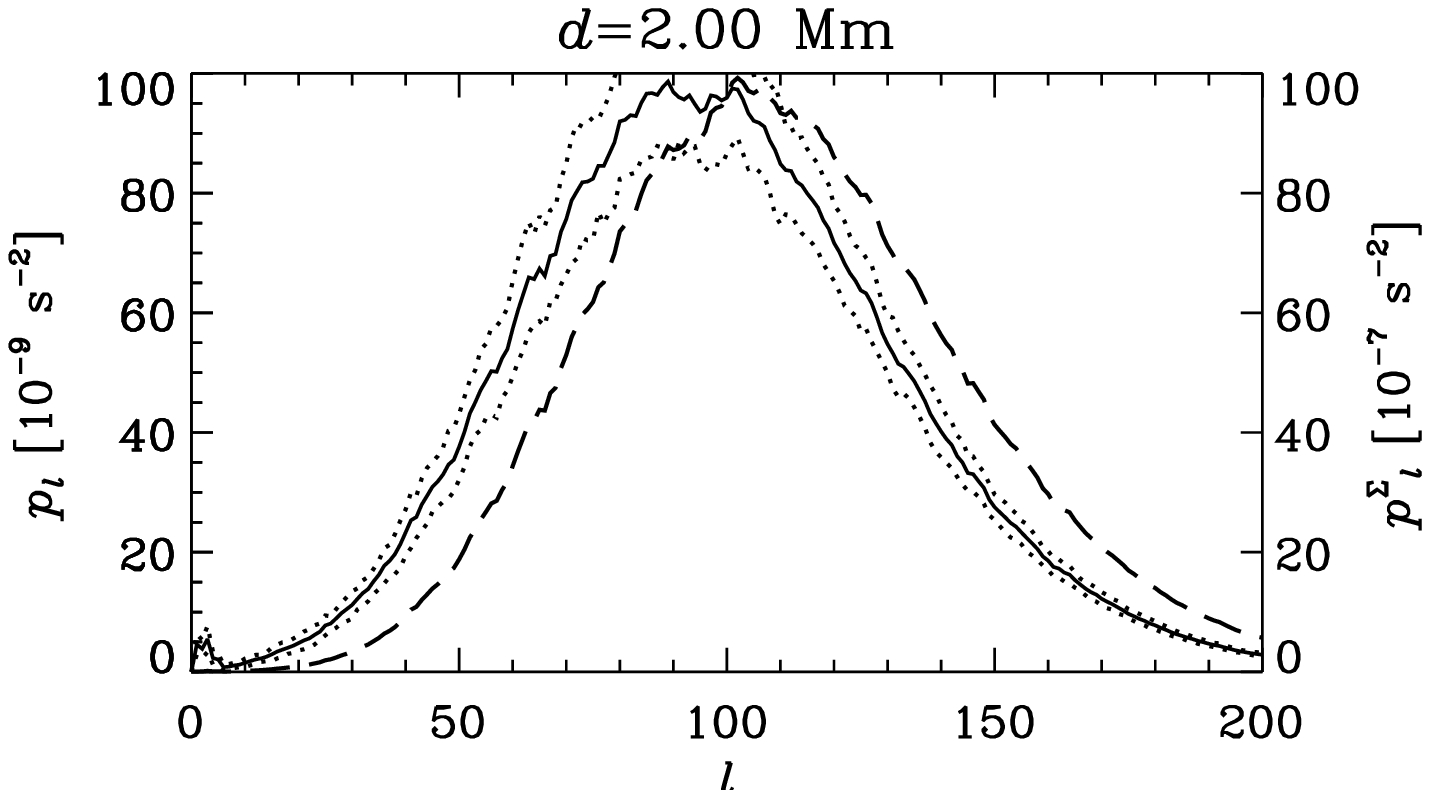}
\includegraphics[width=0.33\textwidth,bb=20 0 435 226,clip] {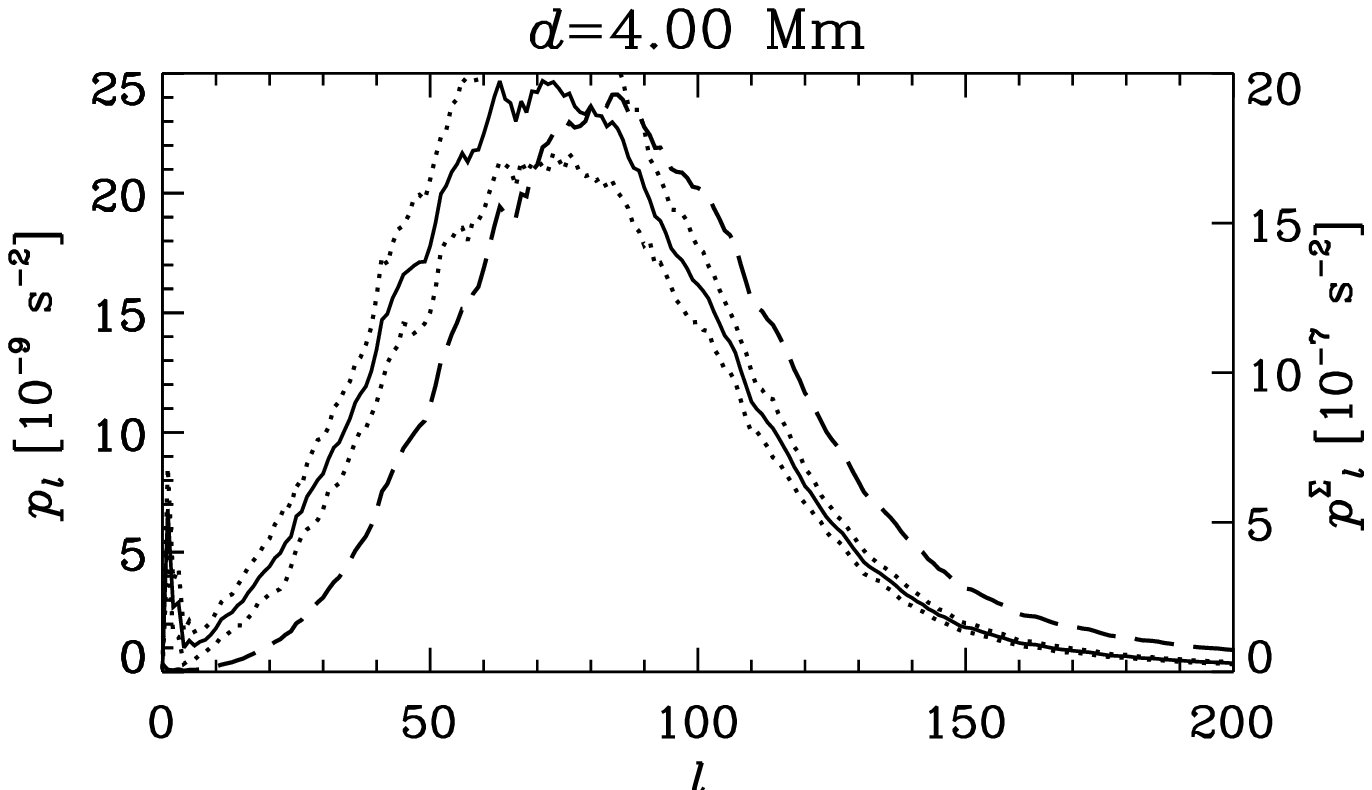}\\[6pt]
\includegraphics[width=0.33\textwidth,bb=20 0 435 226,clip] {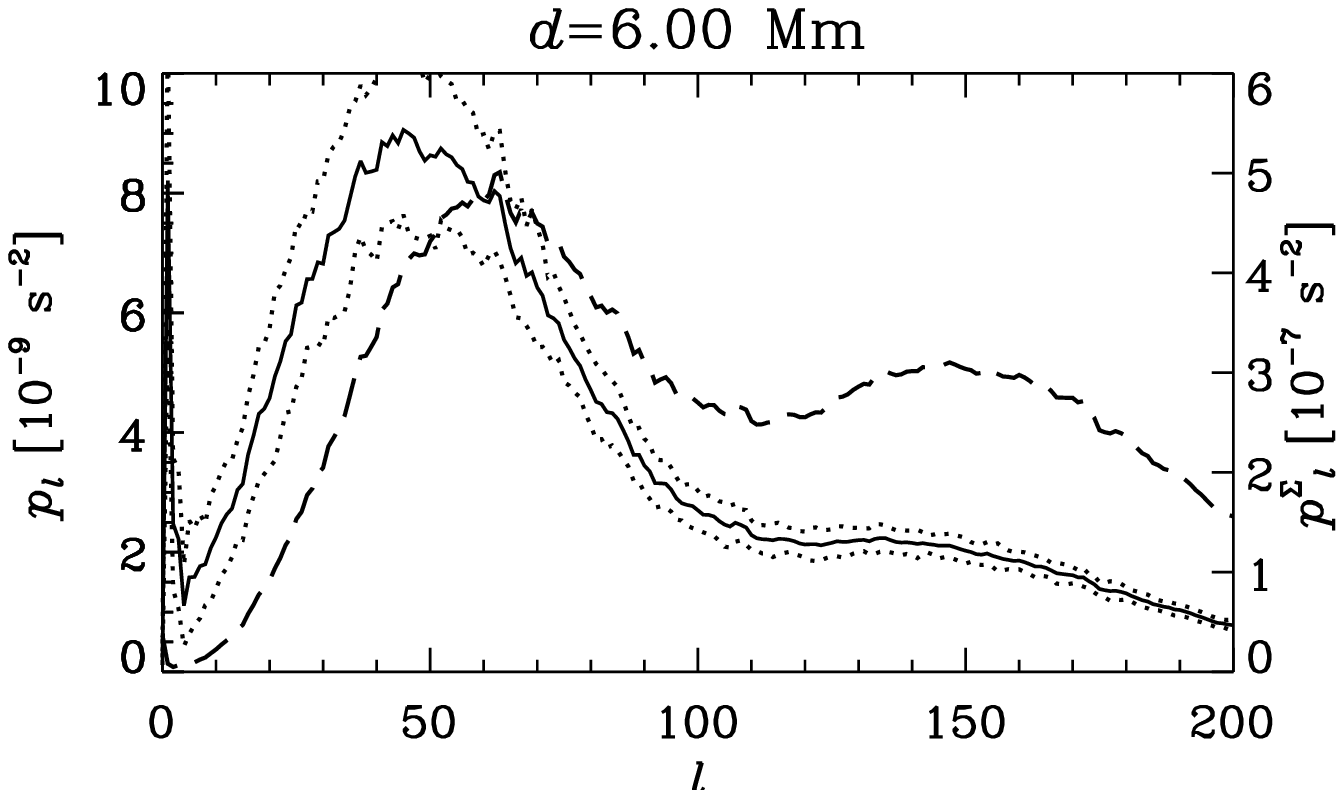}
\includegraphics[width=0.33\textwidth,bb=20 0 435 226,clip] {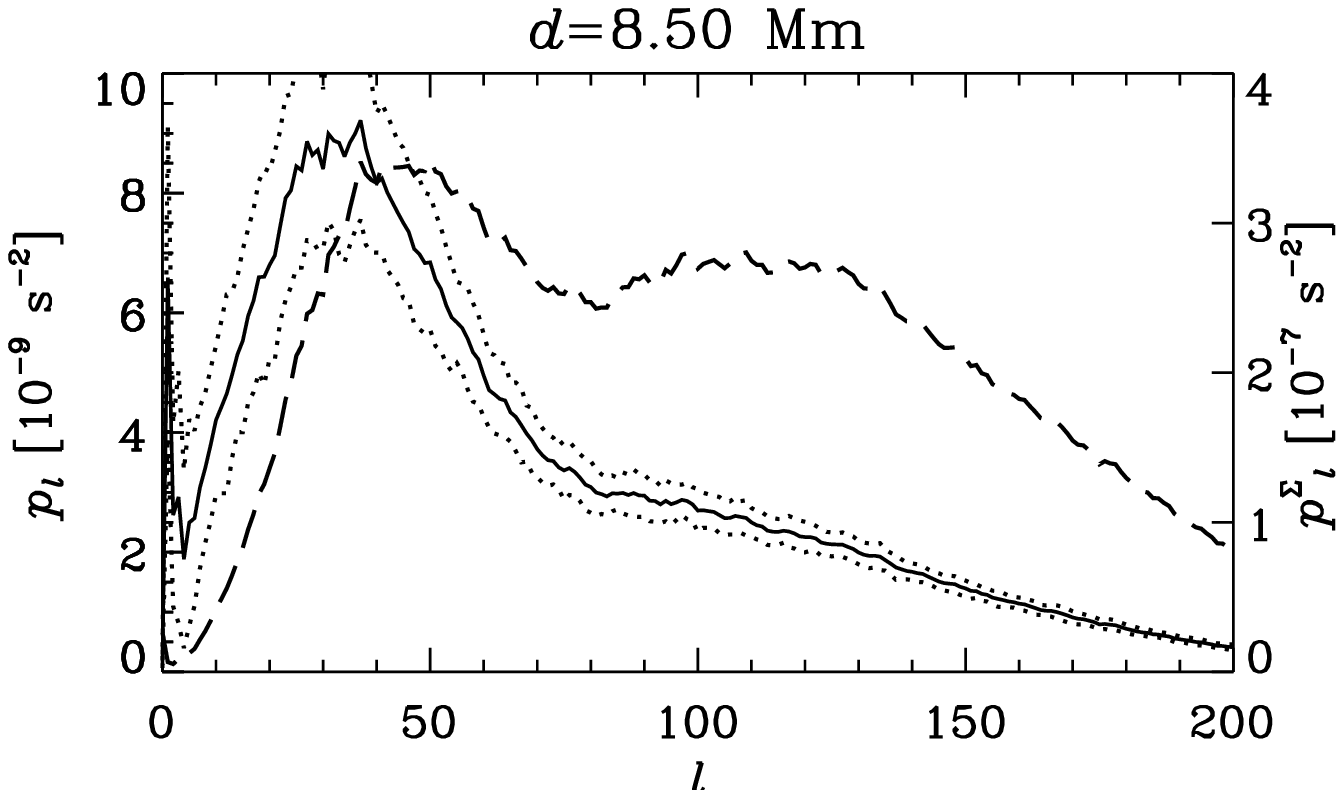}
\includegraphics[width=0.33\textwidth,bb=20 0 435 226,clip] {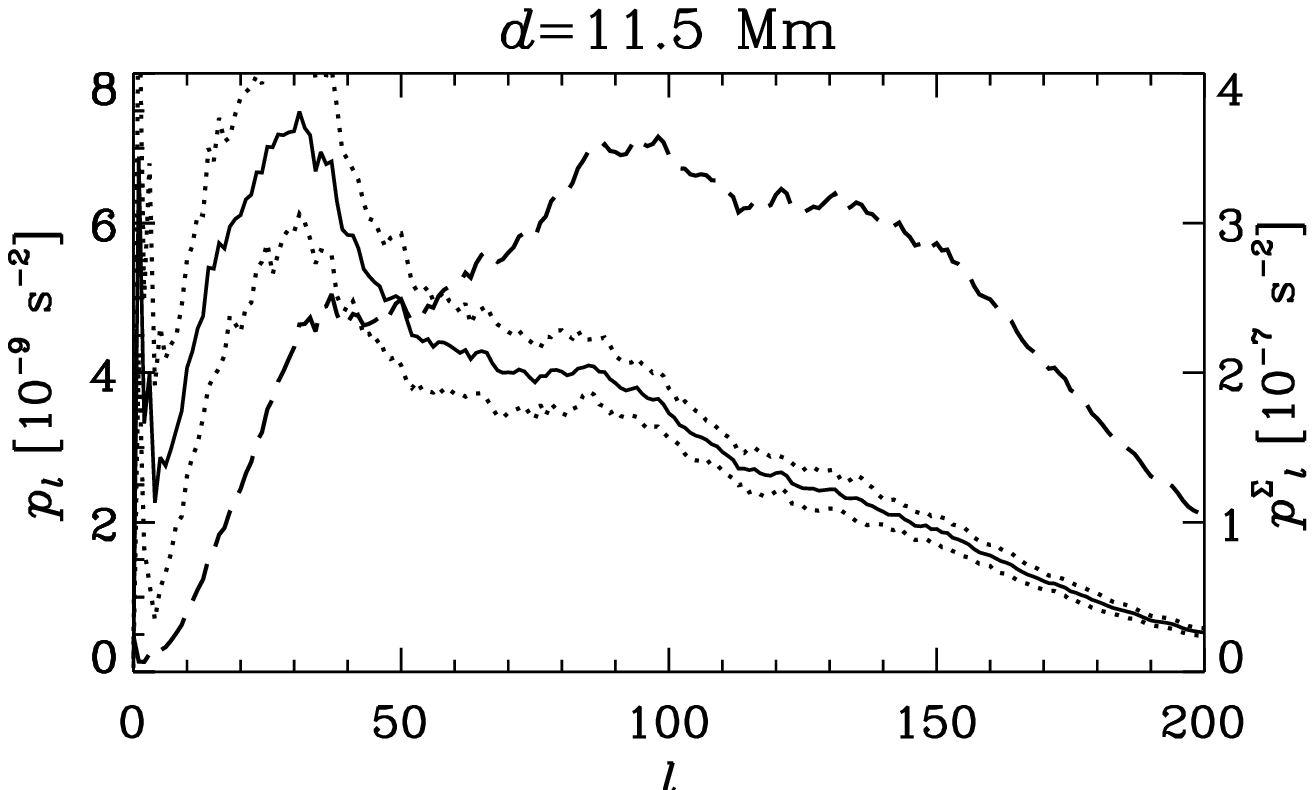}\\[6pt]
\includegraphics[width=0.33\textwidth,bb=20 0 435 226,clip] {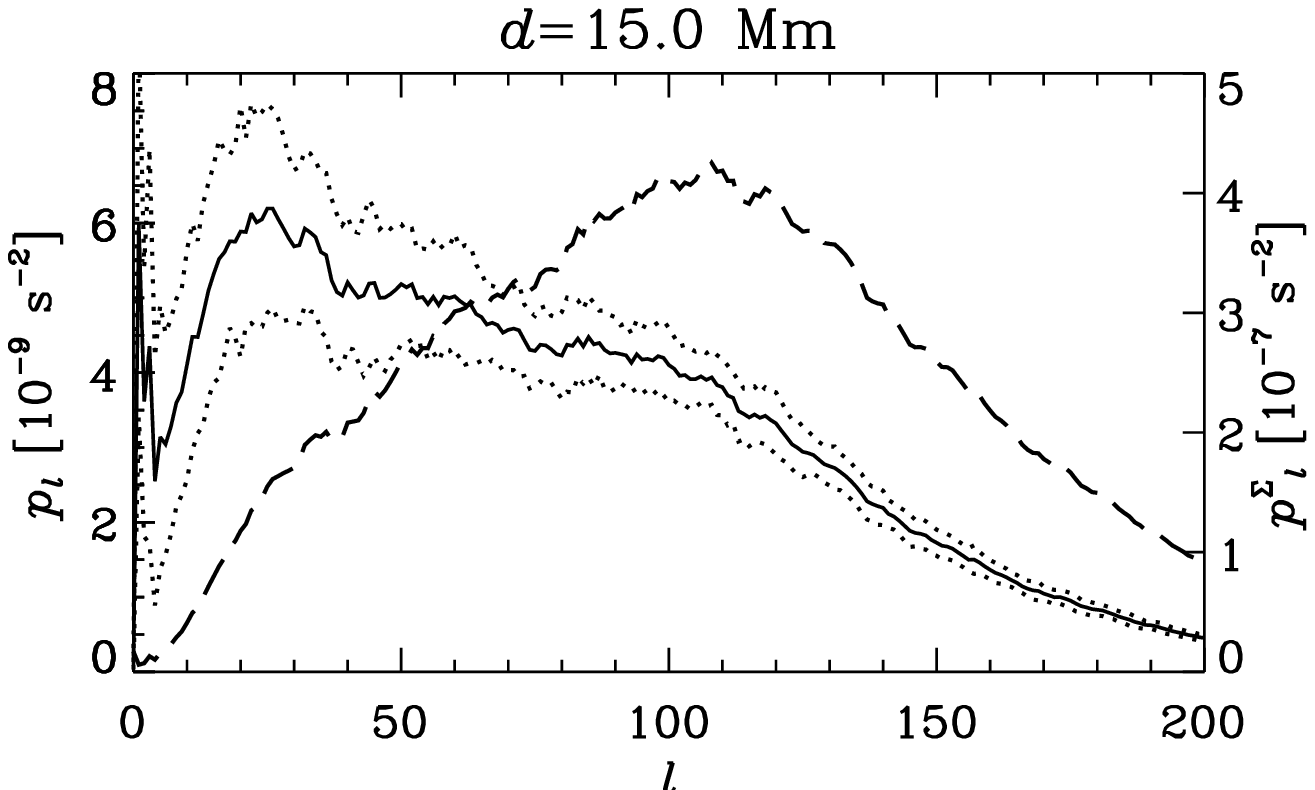}
\includegraphics[width=0.33\textwidth,bb=20 0 435 226,clip] {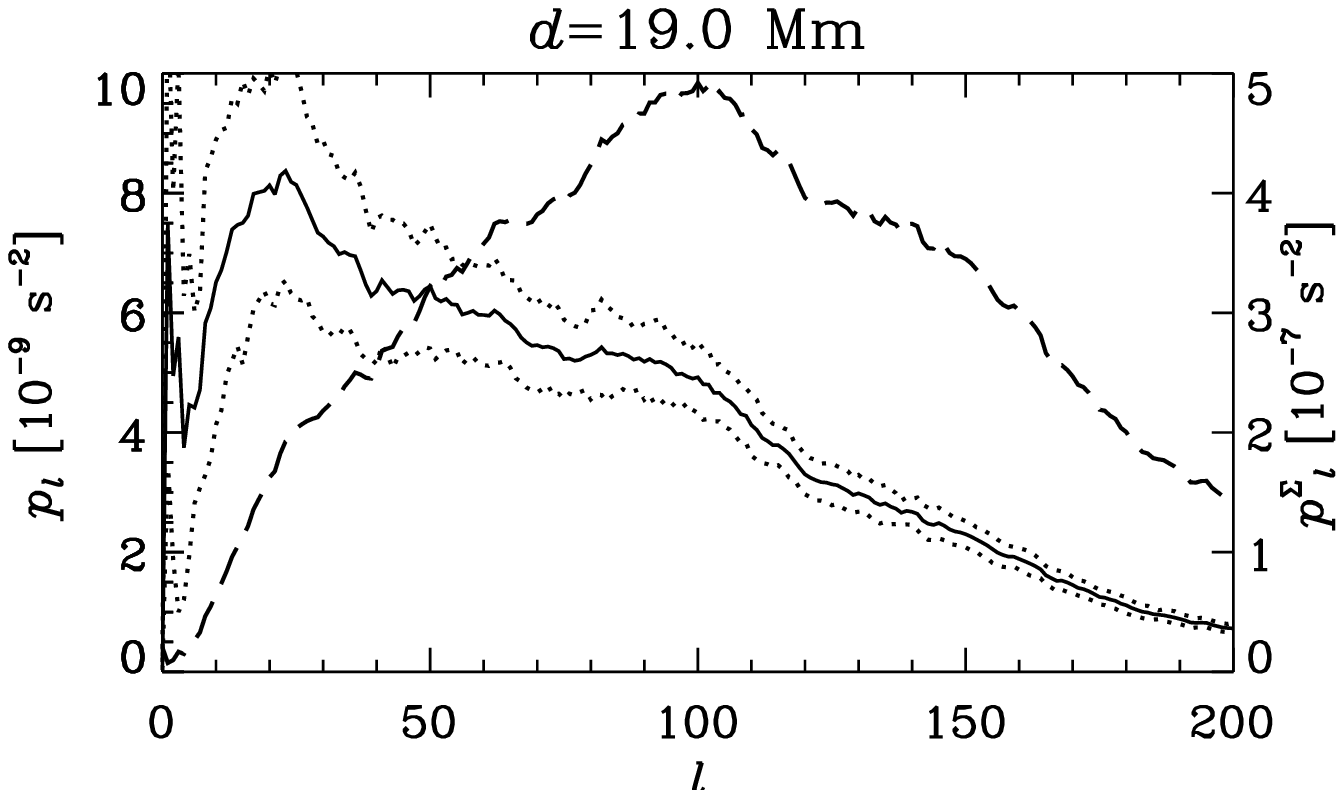}
\includegraphics[width=0.33\textwidth,bb=15 0 430 226,clip]{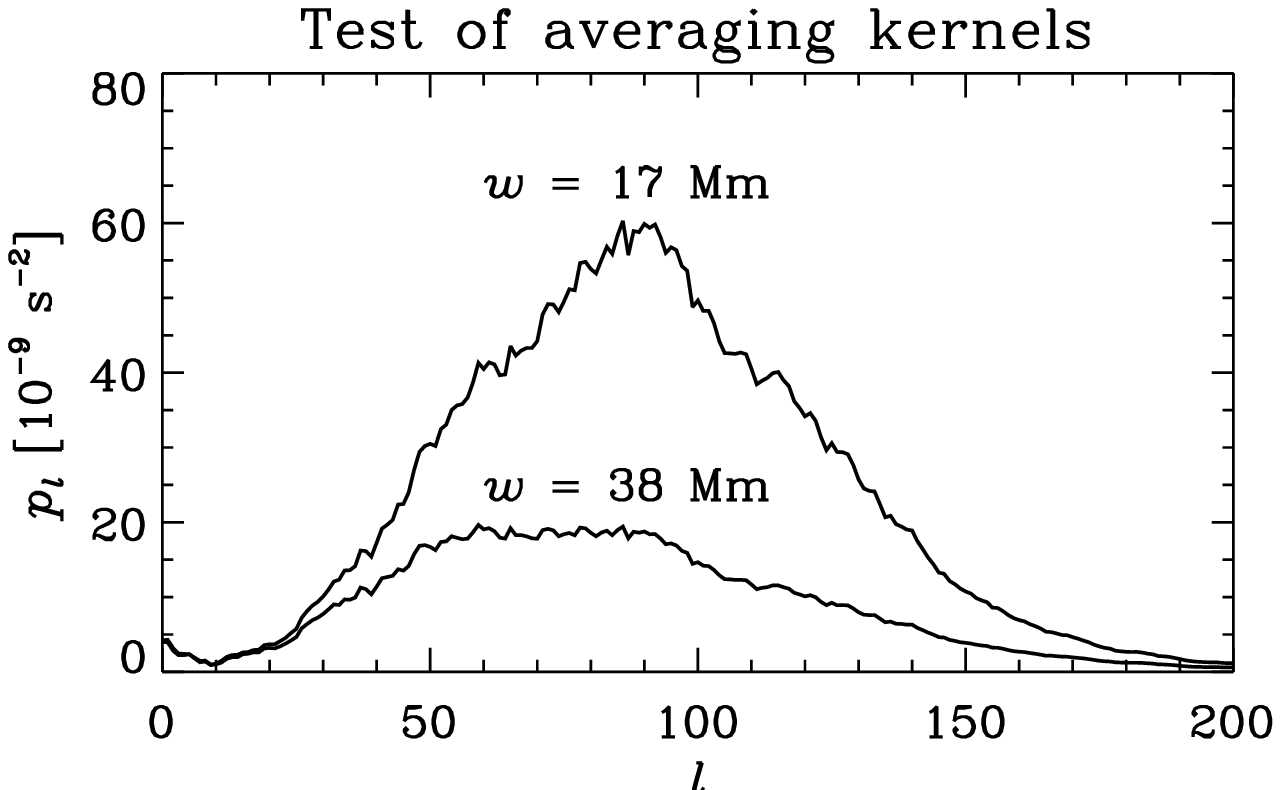}\\
\caption{{Solid curves (in all panels but the last one): $l$-variation of the power $p_l$ based on the above-displayed spectra for different depths (low-activity period, Figure~\ref{spectra200}); dotted curves indicate the standard deviation of $p_l$ from its running average; dashed curves: the total power, $p^\Sigma_l$. The depth values are indicated at the top of each panel. The bottom right panel shows the $p_l$ curves obtained for a velocity-divergence field at $d=0.5$~Mm averaged with the Gaussians corresponding to the helioseismic averaging kernels of two different widths,~$w$.}
	\label{m-averages}}
\end{figure*}

\begin{figure*} 
\centering
\includegraphics[width=0.33 \textwidth,bb=20 0 435 226,clip] {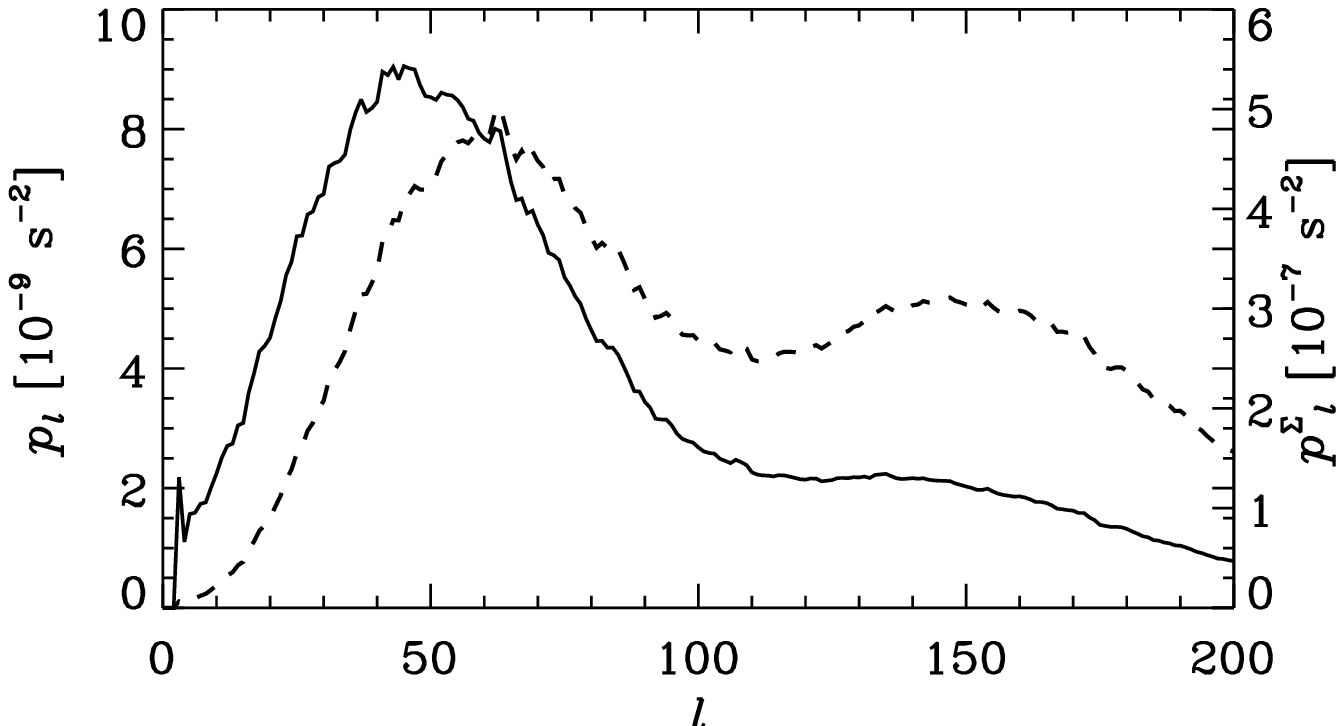}
\includegraphics[width=0.33\textwidth,bb=20 0 435 226,clip] {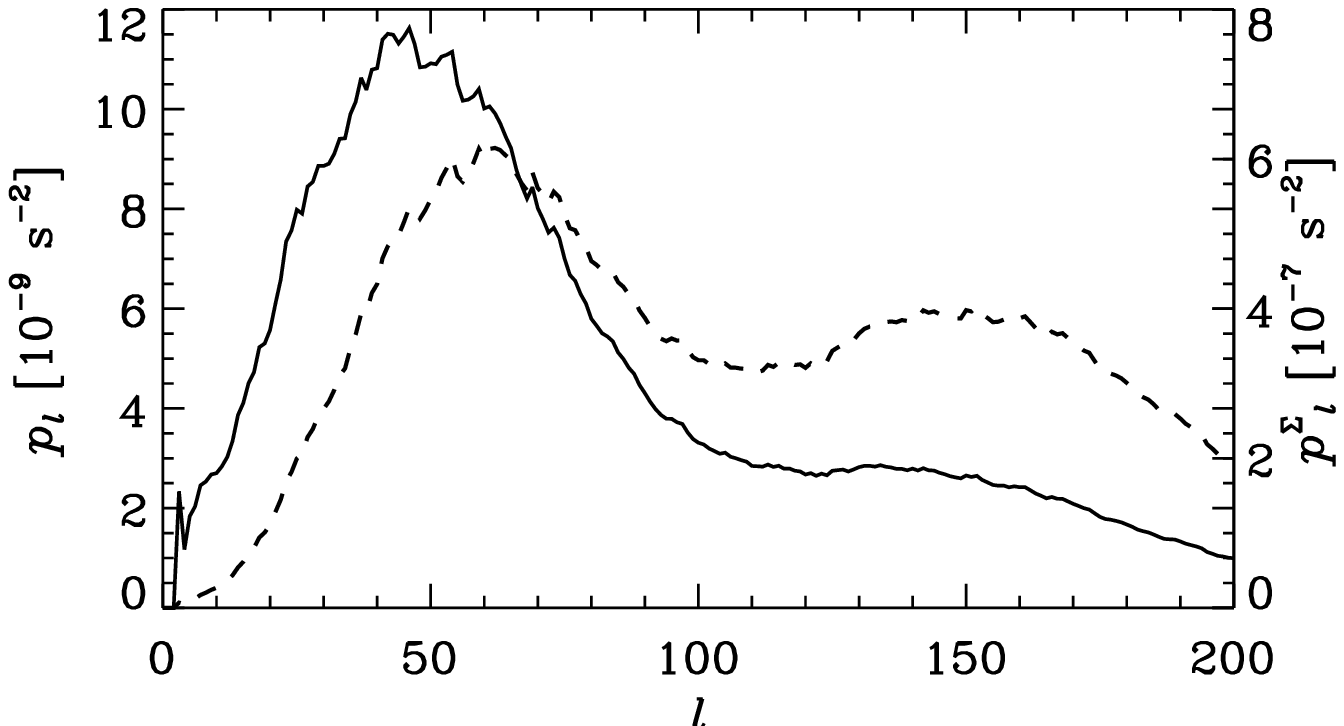}
\includegraphics[width=0.33\textwidth,bb=20 0 435 226,clip] {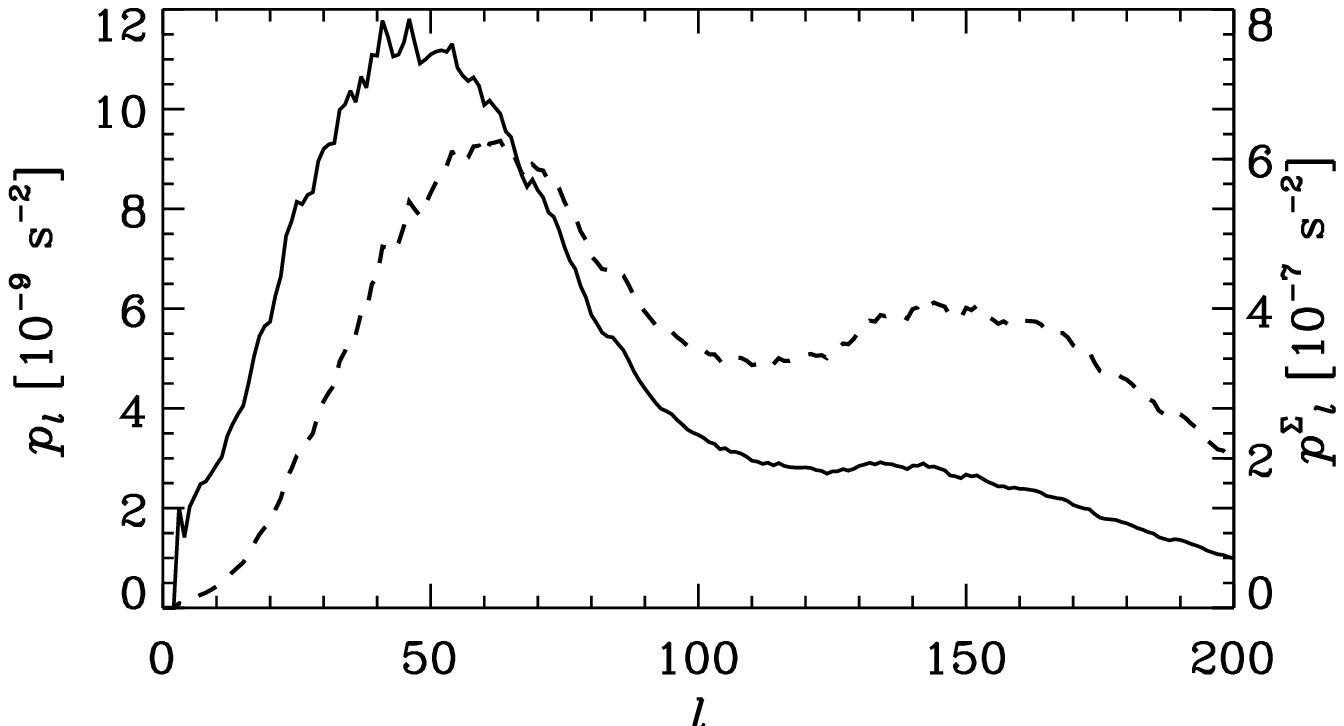}
\includegraphics[width=0.33 \textwidth,bb=20 0 435 226,clip] {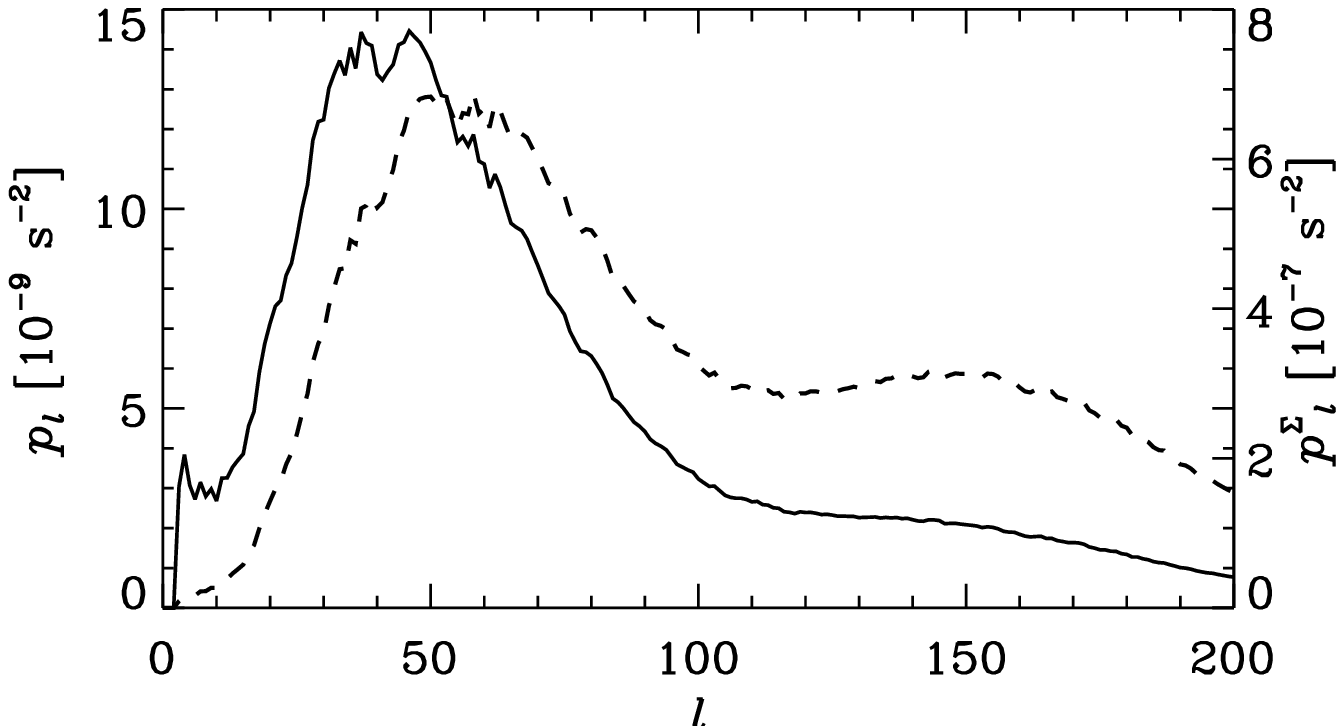}
\includegraphics[width=0.33\textwidth,bb=20 0 435 226,clip] {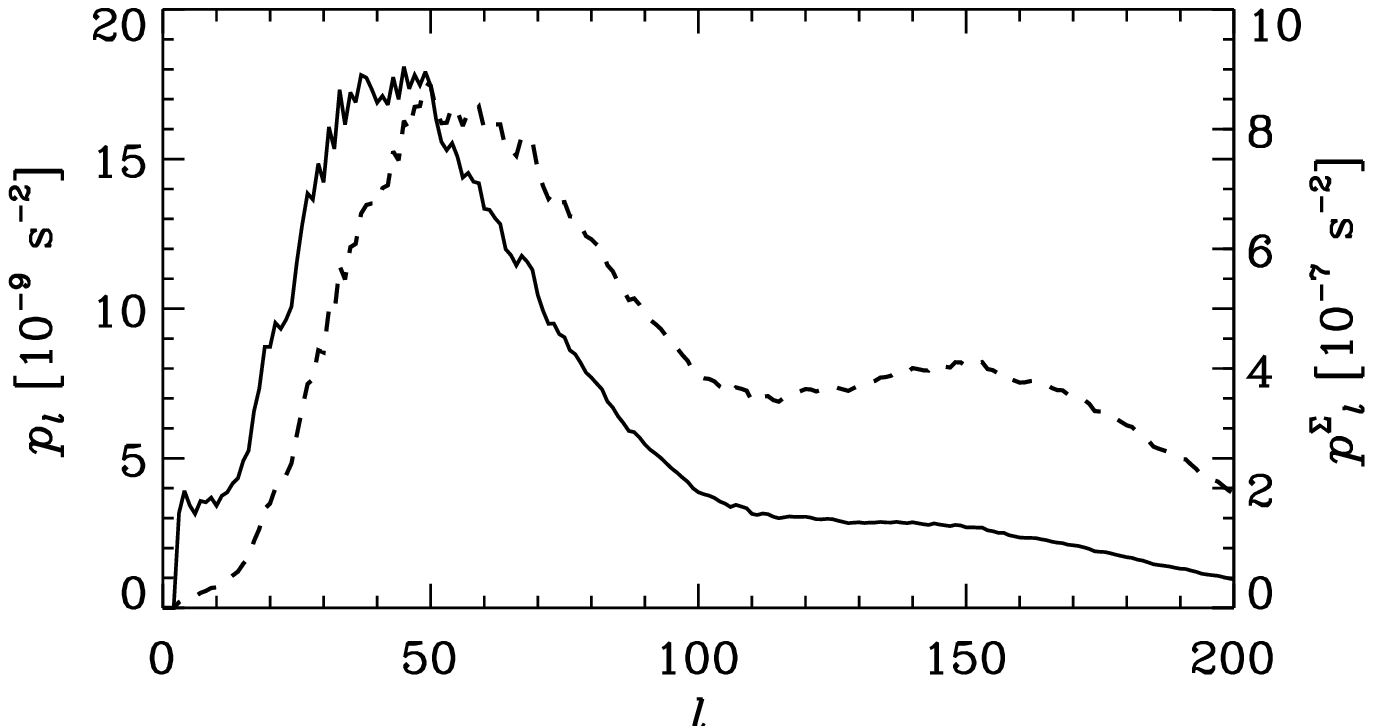}
\includegraphics[width=0.33\textwidth,bb=20 0 435 226,clip] {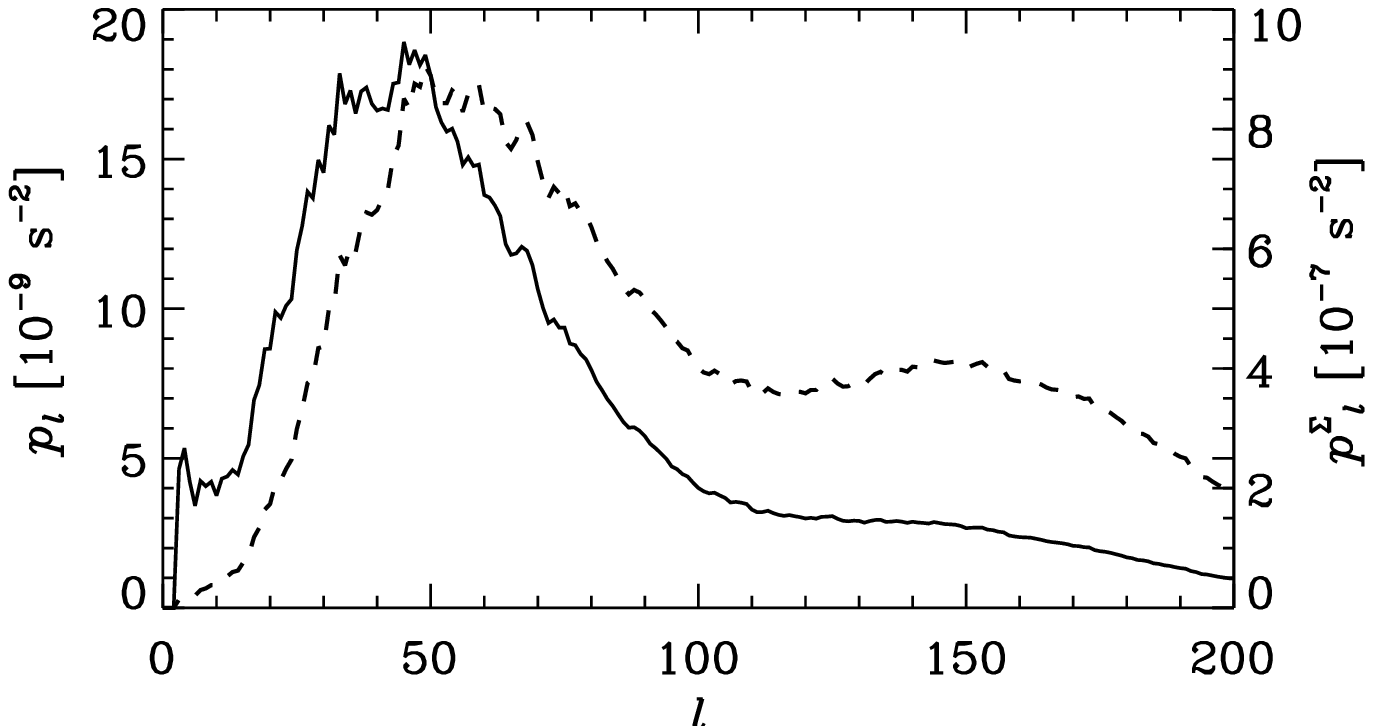}
\caption{The effect of narrowing the source-field sector. The $l$-averaged (solid curves) and $l$-summed (dashed curves) spectra are shown for divergence fields at $d=6$~Mm differently chosen for spherical-harmonic transform: triplicate 120\degree-wide fields (left); quadruplicate 90\degree-wide fields of the $(-45\degree, +45\degree)$ sector (middle); quadruplicate 90\degree-wide field of the $(-60\degree, +30\degree)$ sector (right). Top: low-activity period, from 2019 December 21 to 2020 February 4; bottom: high-activity period, from 2013 December 21 to 2014 February 4.
	\label{narrowing}}
\end{figure*}

\subsection{Smoothed fields. Flow scales}\label{scales}
	
We smooth the velocity field with a 17.5-Mm window and assume the upper spectral boundary in our analysis to be $l_{\max}=200$. The depth variation of the power spectra of smoothed convective flows is illustrated in Figure~\ref{spectra200} for the same 45-day averaging interval as in Figure~\ref{spectra1000}. The range of degrees $l$ and the corresponding range of scales $\lambda$ are fairly wide in the shallow layers. As $d$ increases, the main spectral peak narrows and shifts into the low-$l$ (long-wavelength) region. Although the long-wavelength part of the smoothed spectra (Figure~\ref{spectra200}) for deep layers visually appears to be more extended and more powerful than that of the unsmoothed ones (Figure~\ref{spectra1000}), the power values and the peak widths are similar in both cases.

Such behavior of the spectrum can naturally be explained in terms of a superposition of differently scaled flows: we will see that the supergranular-scale flows in the upper layers ($l\sim$70--130, $\lambda\sim$30--60~Mm) coexist with the upper parts of giant convection cells; they are much less manifest in the deep levels, where the largest-scale energetic harmonics have $l\sim 10$ and the corresponding wavelength range is broad and centered at 300~Mm---a giant-cell scale. These scales, being pronounced in the bottom half of the considered layer, are not so noticeable near the surface because the color scale used for the graphic representation of the spectra is determined by the more energetic, smaller-scale flows. However, the power values for the largest scales in the upper and deep levels are comparable---we discuss this fact below by considering the depth ($d$-) variation of the $m$-averaged power spectrum defined by Equations~(\ref{powerph}), $p_l$ , as a function of $l$.

Spectra obtained for a period of high solar activity, 2013 December 18--2014 February 1, are presented in Figure~\ref{spectra200HA}. Visually, they are very similar to those in Figure~\ref{spectra200}. Differences between the two cases are mainly in the extremum power values rather than in the shape of the spectrum. We discuss the effects of the solar-activity level in Subsection~\ref{timevar}.

Figure~\ref{m-averages} shows the $m$-averaged spectra, $p_l$, obtained from the original spectra displayed in Figure~\ref{spectra200}. The uncertainty of the results is calculated as the standard deviation of the power $p_l$ from its running average and indicated by dotted curves. The total power of the harmonics with a given $l$ calculated according to Equation~(\ref{powertot}), $p^\Sigma_l$, is also shown in Figure~\ref{m-averages}, by dashed curves.

Let us compare the $p_l$ values for the largest-scale harmonics at different depths. The $p_l$ spectra demonstrate the displacement of the main peak to longer wavelengths with depth. The peak of the spectrum for $d=19$~Mm is near $l\approx 13$, its height is about $8.5\times 10^{-9}$~s$^{-2}$.  At $d=0.5$~Mm, the power values for such wavenumbers are about $5\times 10^{-9}$~s$^{-2}$. Similar $p_l$ values in this long-wavelength range are also typical of the intermediate depths. Therefore, the flows characterized by the horizontal scales in the range $\lambda\sim$200--300~Mm have comparable power values over the whole depth range. However, near the surface, the supergranulation-scale flows are much more powerful than the large-scale flows. The spectrum thus represents a superposition of flow components with widely differing scales: the supergranular-scale components are dominant in the upper layers, while the weaker larger-scale components extend from deep to shallow layers.

The helioseismic inferences yield estimates of the flow velocities averaged with kernels of characteristic widths increasing with depth. To evaluate the possible effect of the averaging-kernel broadening with depth, we computed the $p_l$ spectra of a convective-velocity-divergence field for $d=0.5$~Mm smoothed with Gaussian kernels whose widths vary over the same range as do the kernels used in helioseismological inversions \citep{Couvidat_etal_2005}. The $p_l$ spectra corresponding to the extreme kernel-width values, $w=17$ and $38$~Mm, are displayed in the bottom right panel of Figure~\ref{m-averages} and demonstrate only a moderate shift of the spectrum to longer wavelengths with depth, which does not affect our conclusions. Small near-surface sidelobes of the averaging kernels might, in principle, contribute to the power enhancement at small scales in the deep layers. However, this is unlikely since the sidelobes in both the vertical and horizontal directions are fairly weak compared to the primary averaging-kernel peak \citep[][Figure 11]{Couvidat_etal_2005}.

It is worth noting one more particular feature of the spectra shown in Figure~\ref{spectra200}. Since the harmonics for $l=m$ are sectorial (latitude-independent), the fact that the main spectral peak approaches the $l=m$ line with the  increase of $d$ can be interpreted as an indirect indication of the presence of meridionally elongated, banana-shaped convection structures.

We have to make an important remark in the context of interpreting our results. \citet{Zhao_etal_CLS_2012} noted a systematic center-to-limb variation in the measured helioseismic travel times, which must be taken into account in determinations of the meridional velocities in the convection zone. They suggested a procedure removing this systematic variation, which was subsequently applied by \citet{Gizon_etal_2020} and others in their analyses.To assess the possible effects of the systematic center-to-limb variation in the measured quantities on the results of our analysis, we have additionally calculated the $m$-averaged power spectra choosing sectors 90\degree\ wide in longitude and repeated by 90\degree, 180\degree, and 270\degree\ rotations so as to obtain fields covering the complete longitudinal angle. The sectors were specified to occupy either a ($-45\degree, +45\degree)$ or a ($-60$\degree, +30\degree) Stonyhurst-longitude range. The computations were done for both high-activity (from 2013 December 21 to 2014 February 4) and low-activity (from 2019 December 21 to 2020 February 4) times. We found that, although such source-data modifications slightly increase the amplitudes of the spectra (enhancing the ``coherence'' of the data: the same field is, in this case, repeated four rather than three times in the compound 360\degree\ field, thus contributing to the spectrum with a greater weight), they do not significantly modify the shape of the $l$-averaged and $l$-summed spectra (Figure~\ref{m-averages}). In particular, they leave the spectral-peak positions unchanged. Thus, the center-to-limb variations do not affect our conclusions.

\begin{figure*} 
	\centering
    \includegraphics[width=0.31\textwidth]{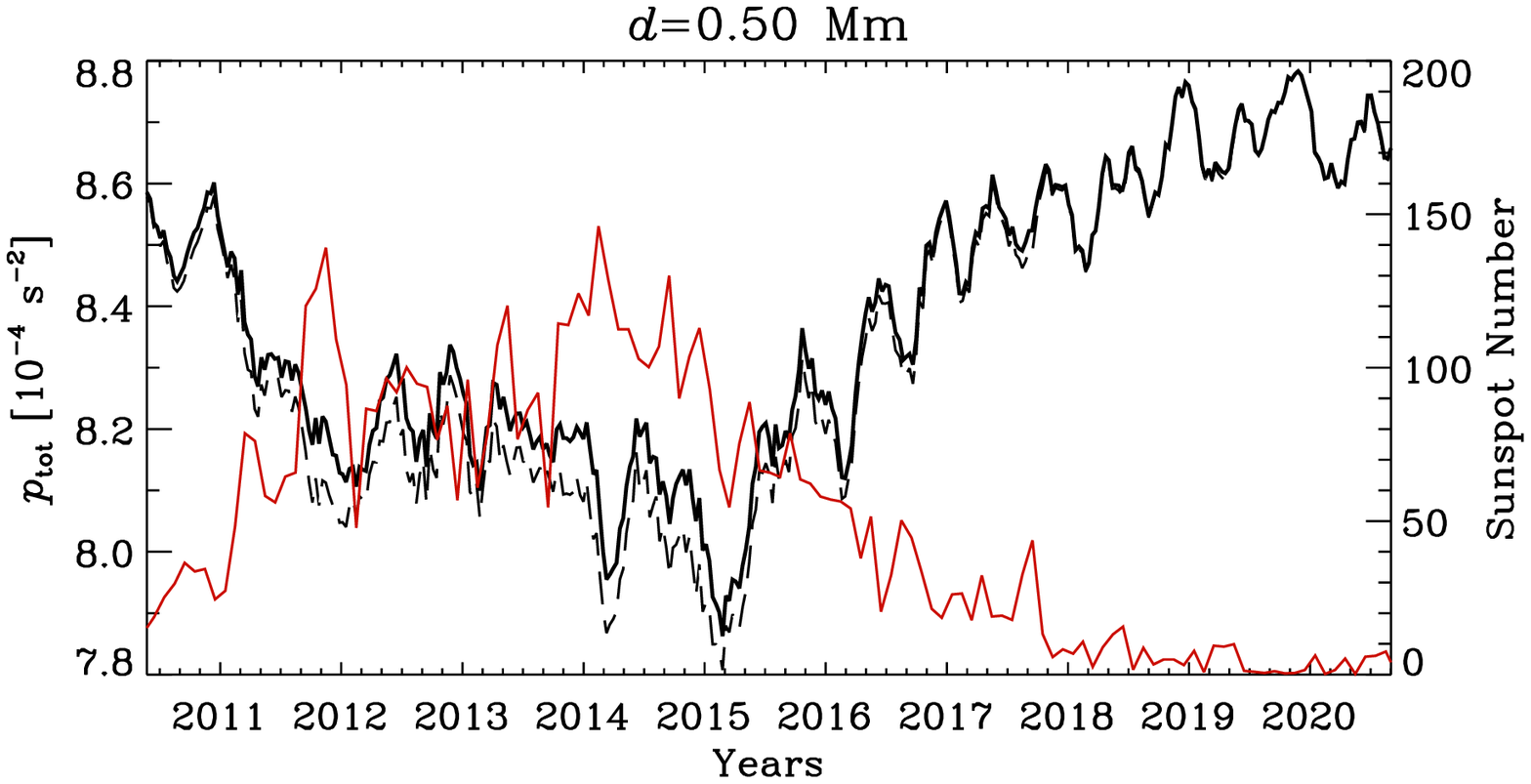}
    \includegraphics[width=0.31\textwidth]{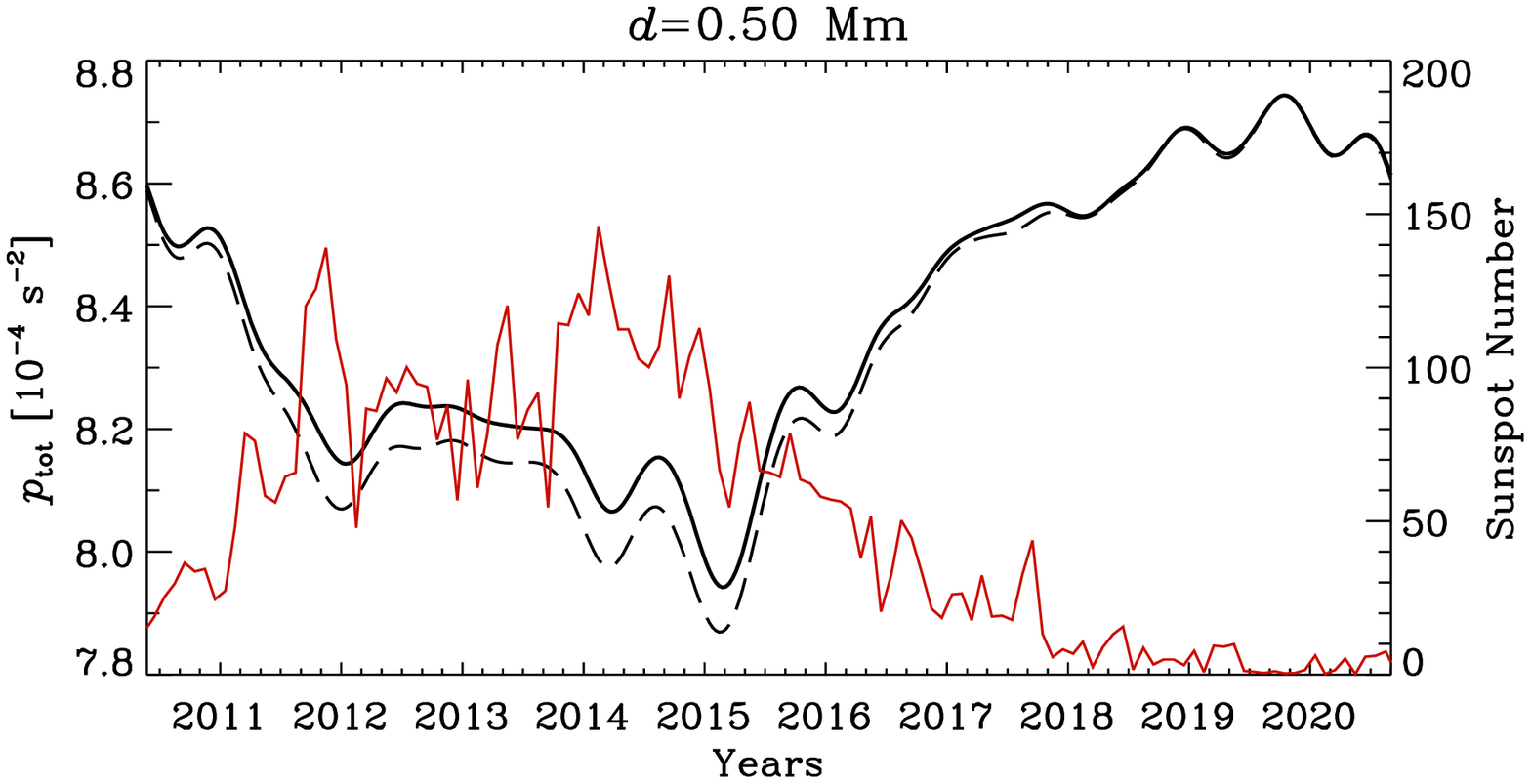}\\
    \includegraphics[width=0.31\textwidth]{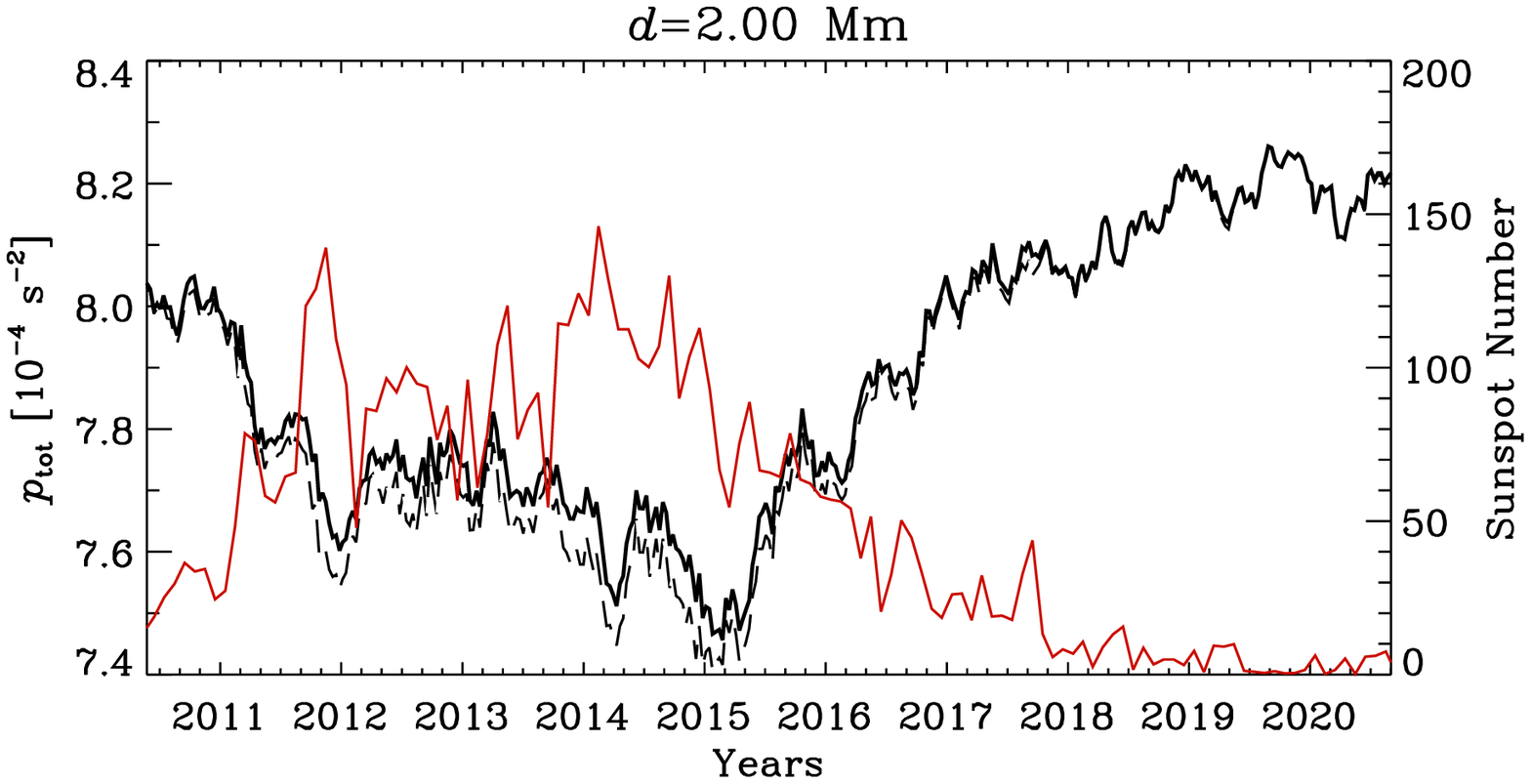}
    \includegraphics[width=0.31\textwidth]{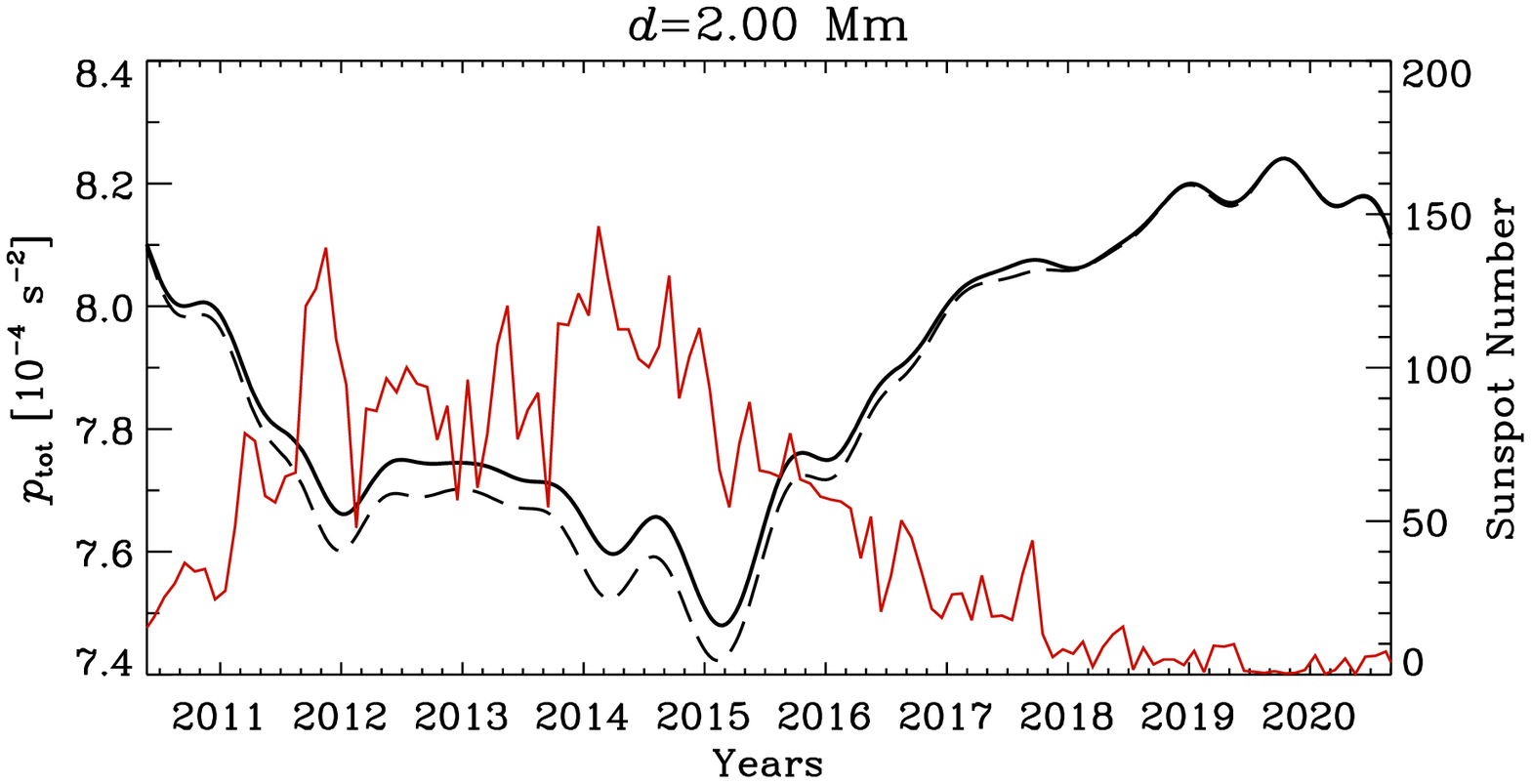}\\
    \includegraphics[width=0.31\textwidth]{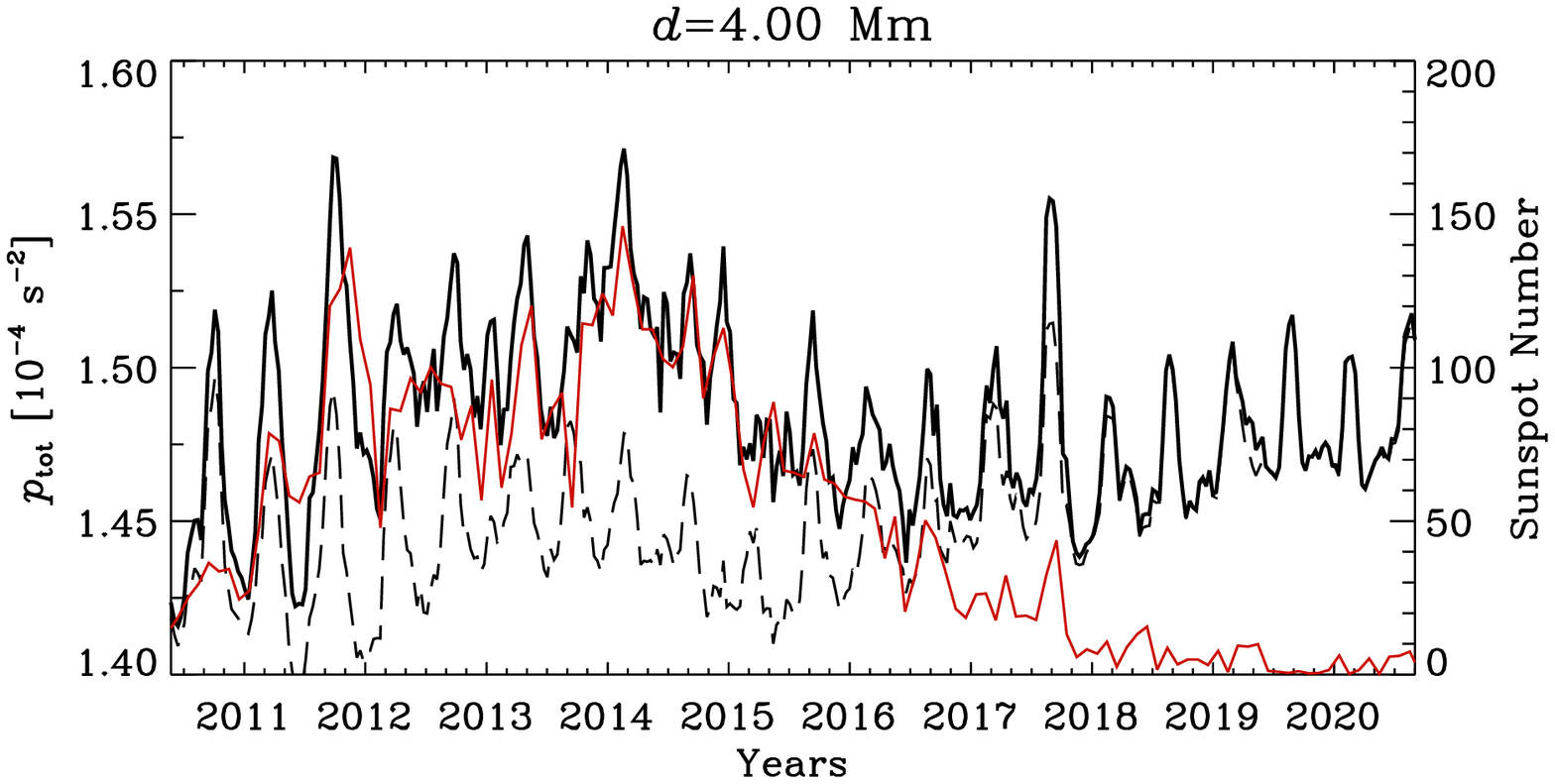}
    \includegraphics[width=0.31\textwidth]{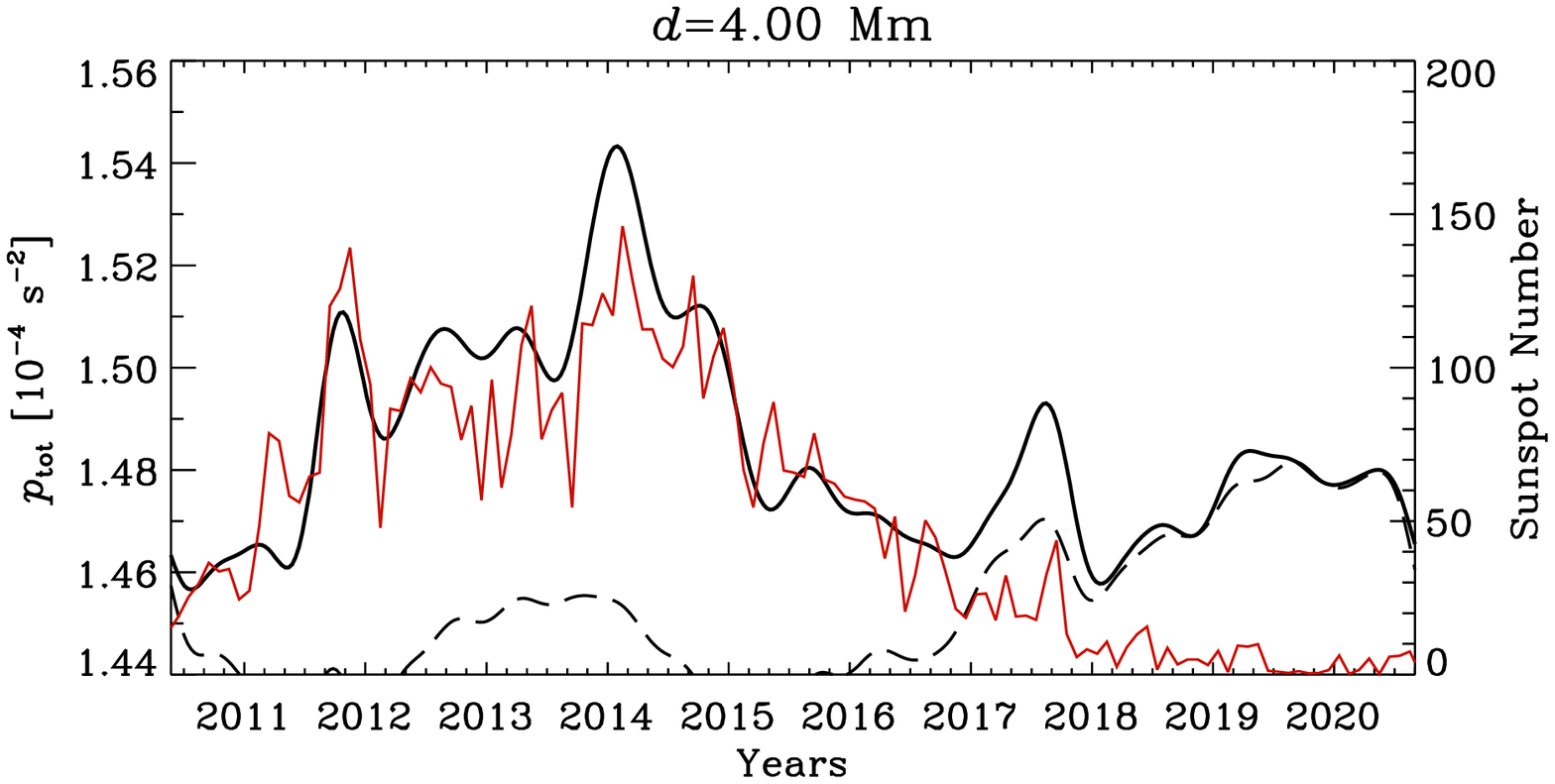}\\
    \includegraphics[width=0.31\textwidth]{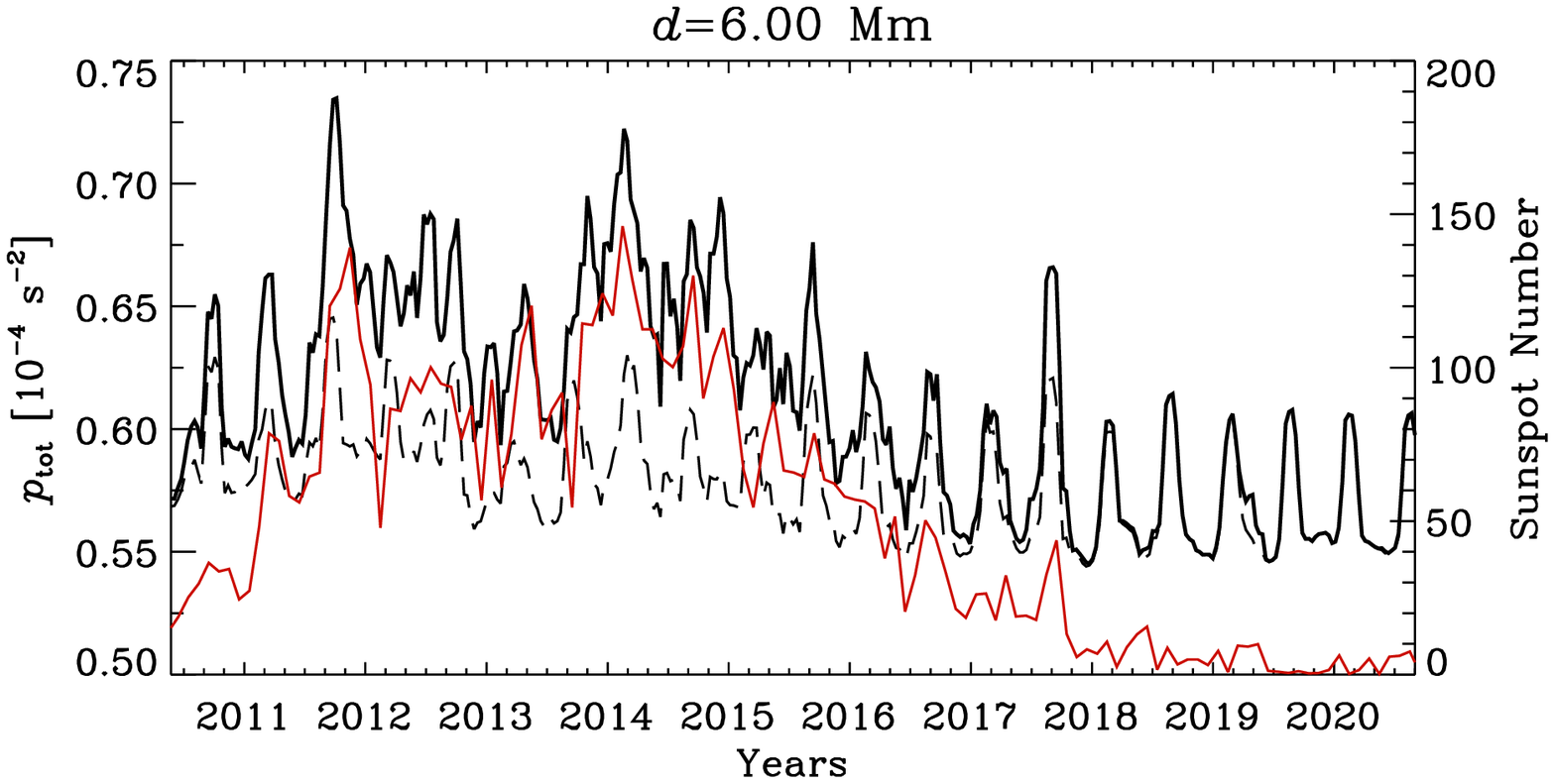}
    \includegraphics[width=0.31\textwidth]{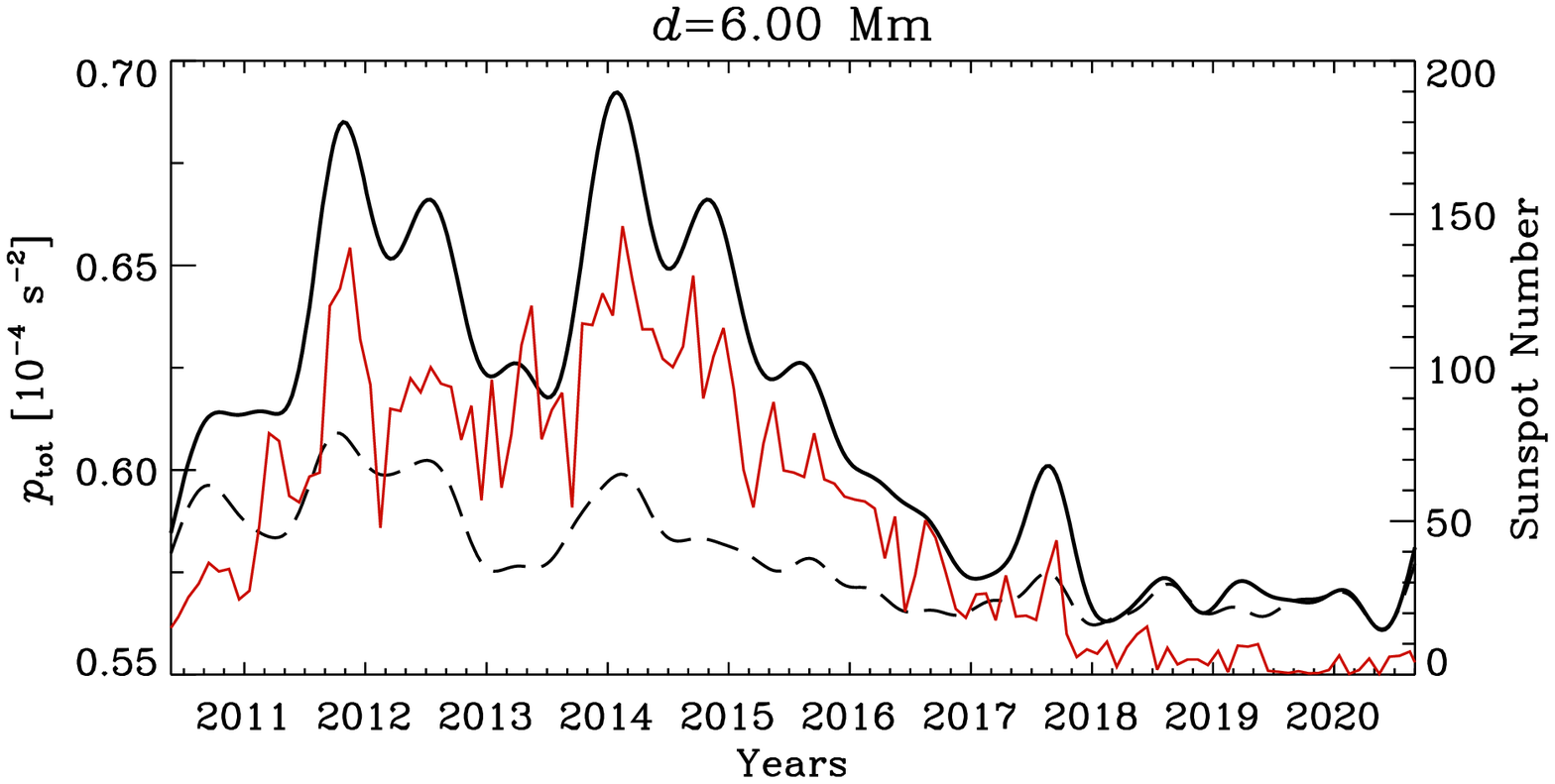}\\
    \includegraphics[width=0.31\textwidth]{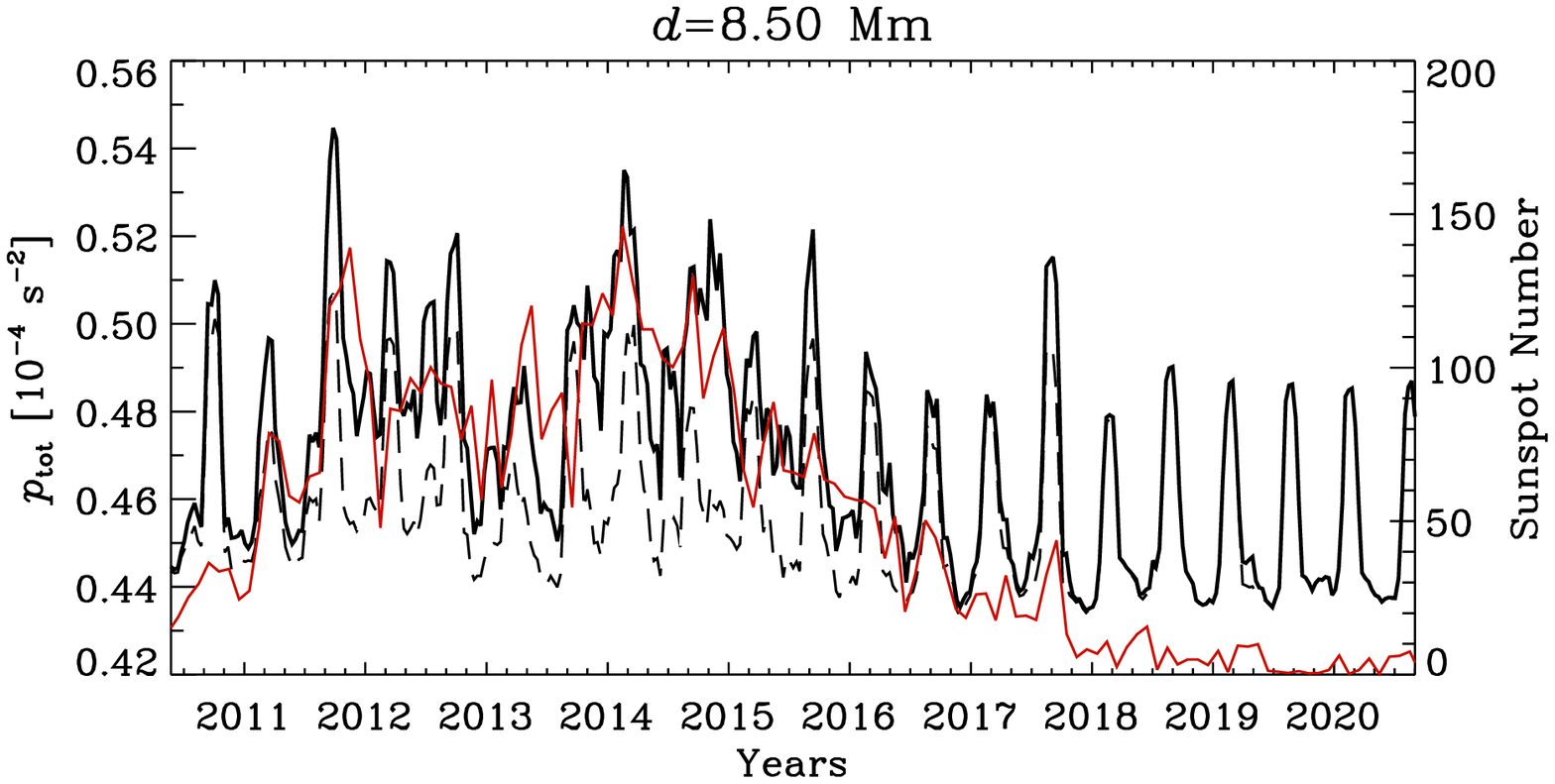}
    \includegraphics[width=0.31\textwidth]{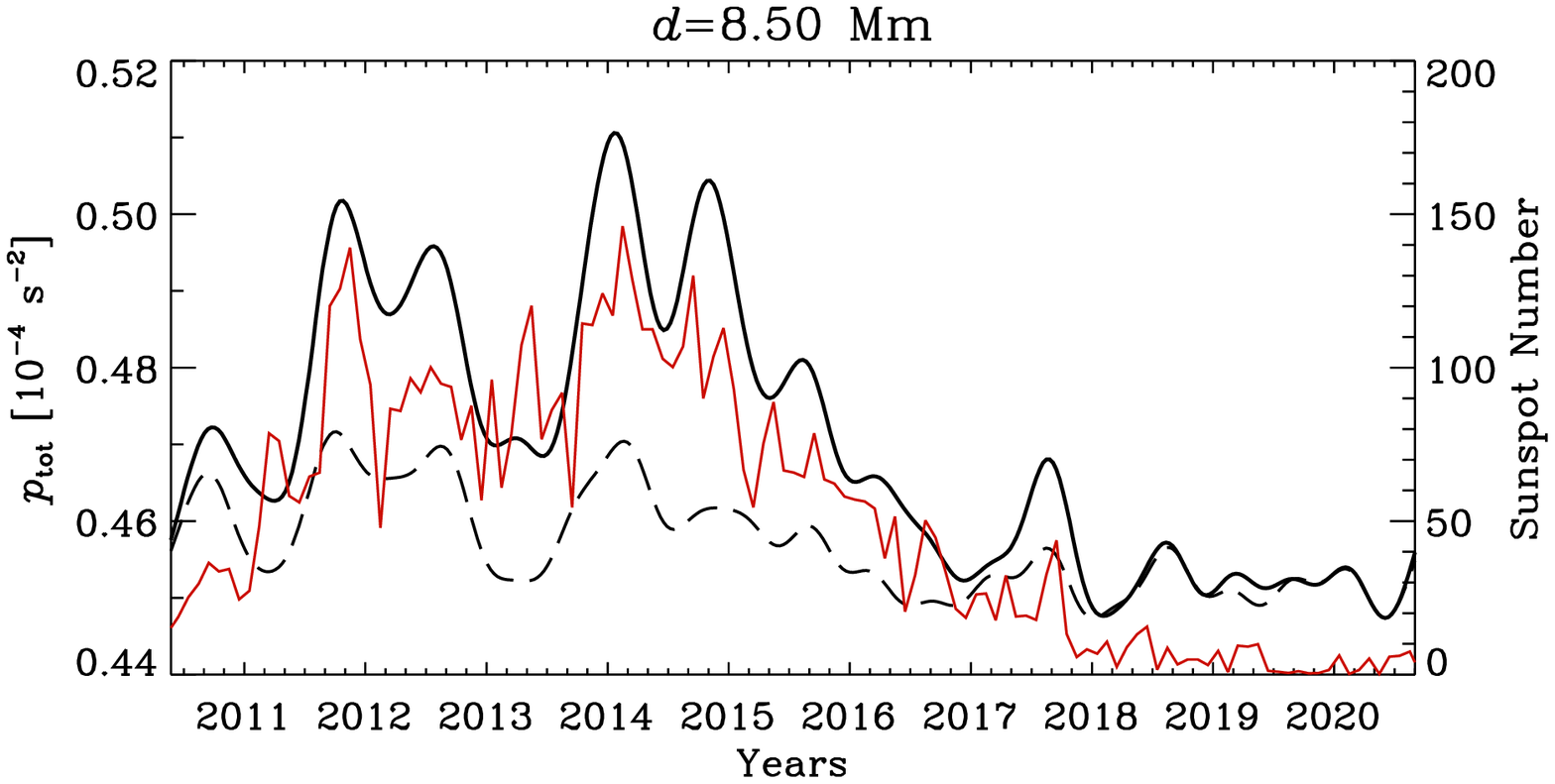}\\
    \includegraphics[width=0.31\textwidth]{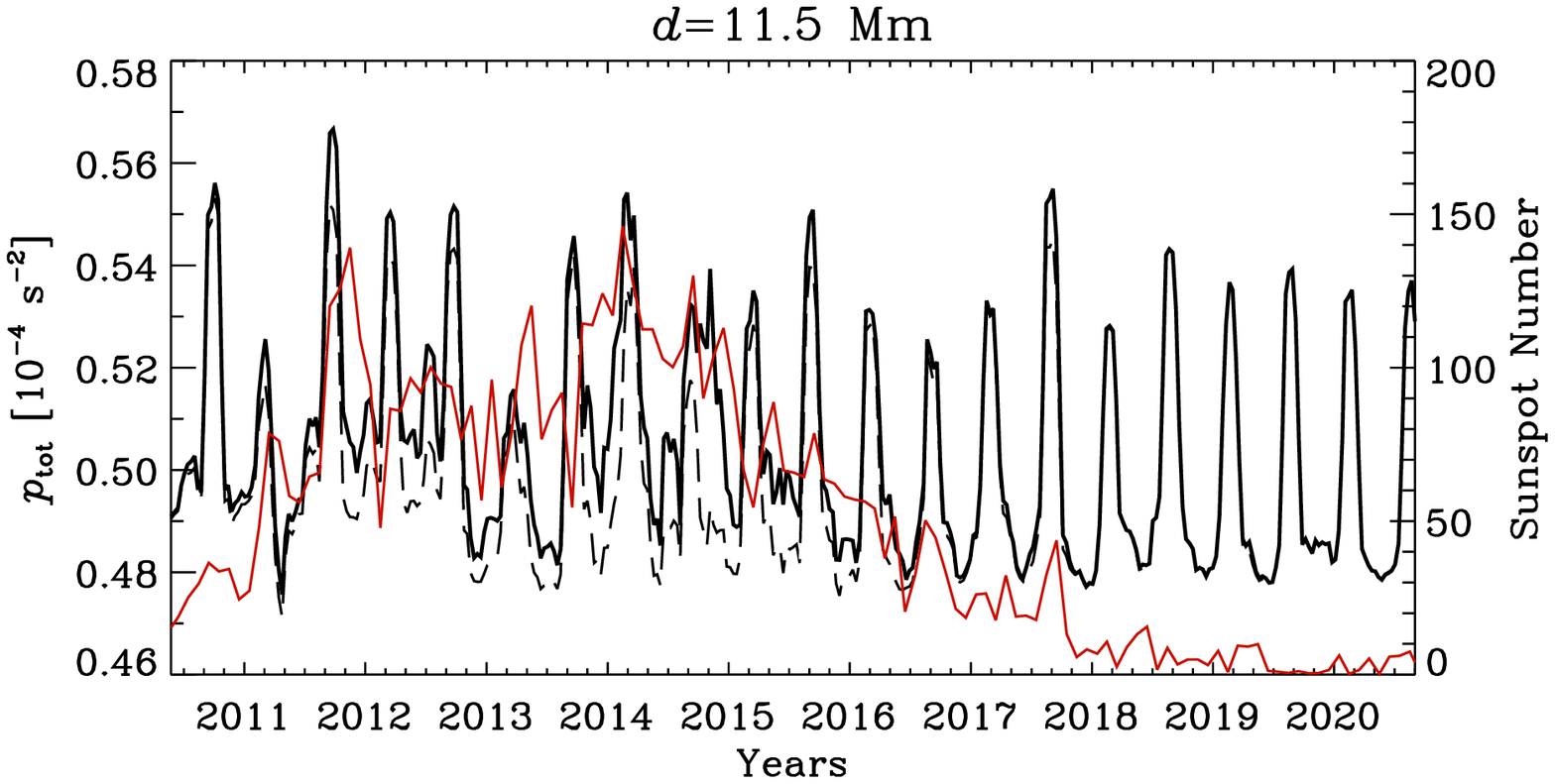}
    \includegraphics[width=0.31\textwidth]{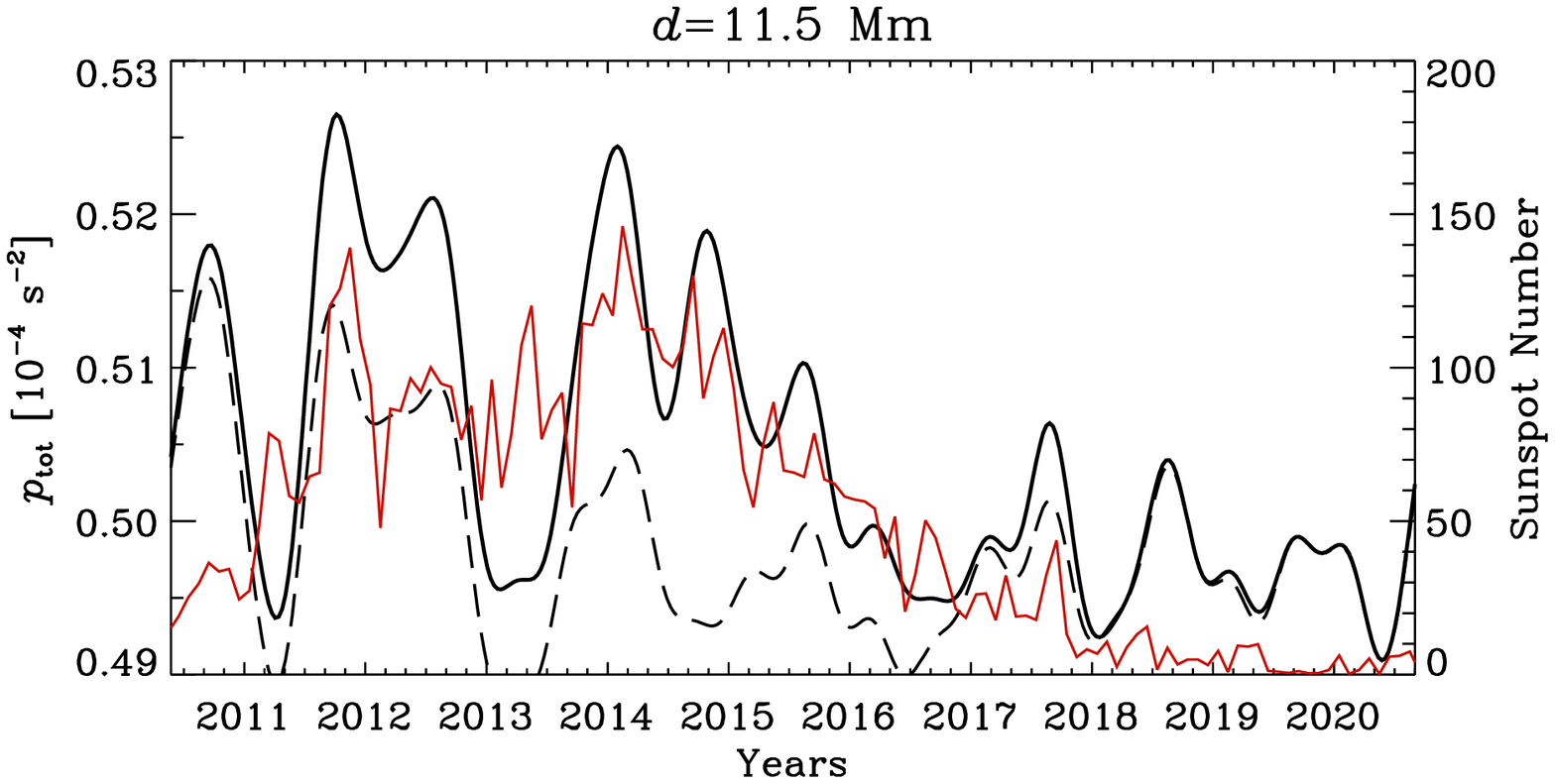}\\
    \includegraphics[width=0.31\textwidth]{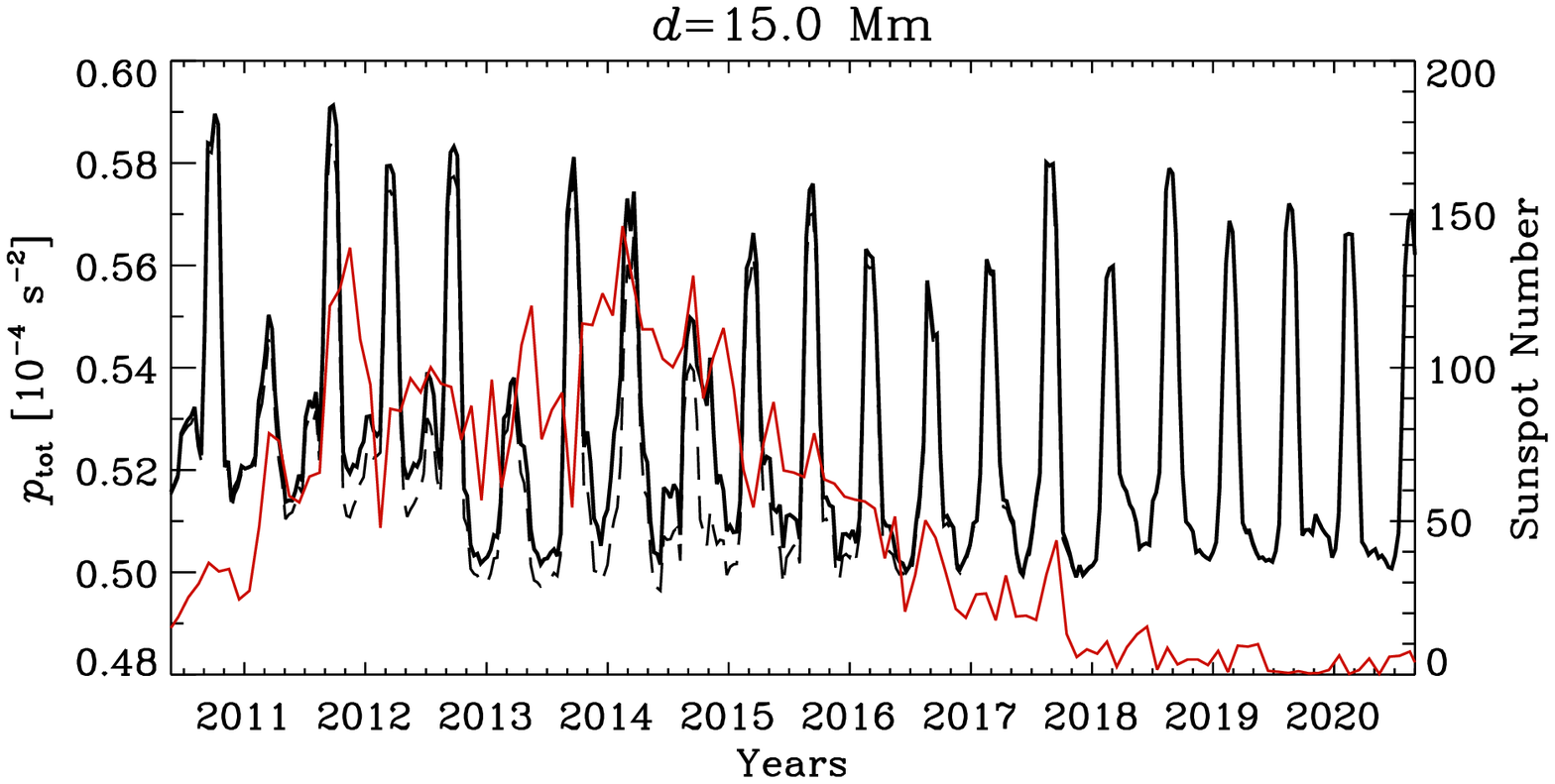}
    \includegraphics[width=0.31\textwidth]{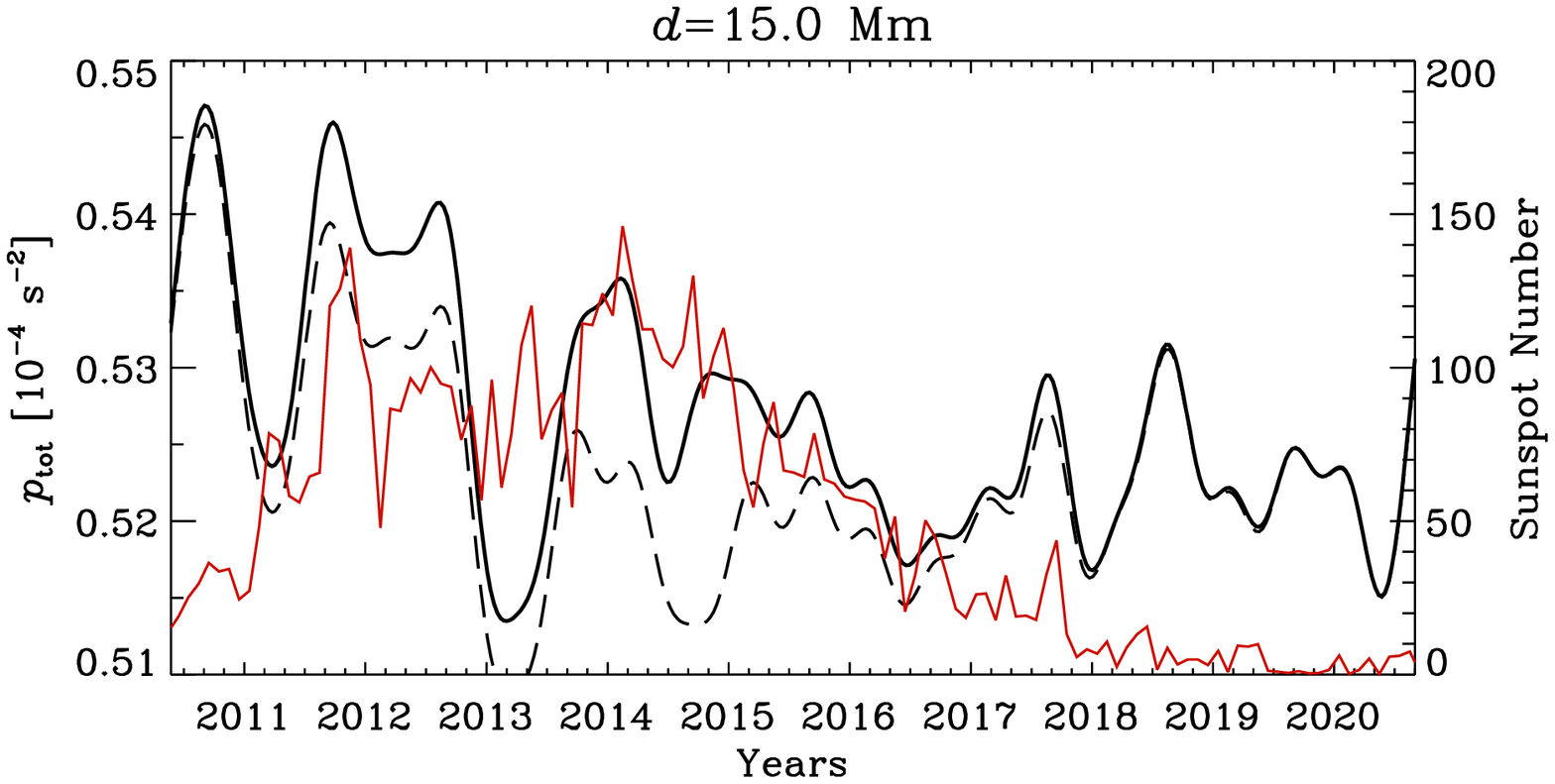}\\
    \includegraphics[width=0.31\textwidth]{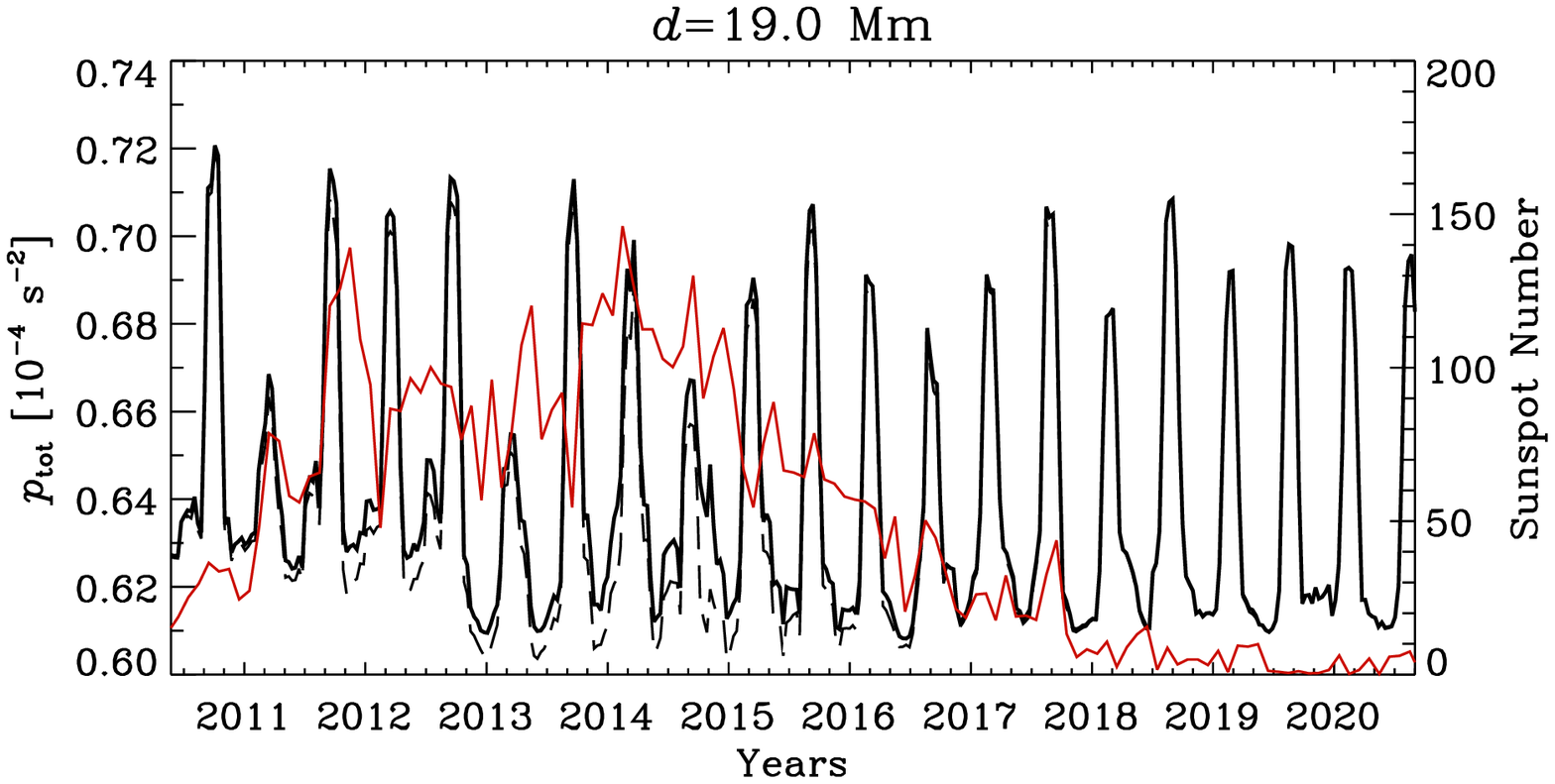}
    \includegraphics[width=0.31\textwidth]{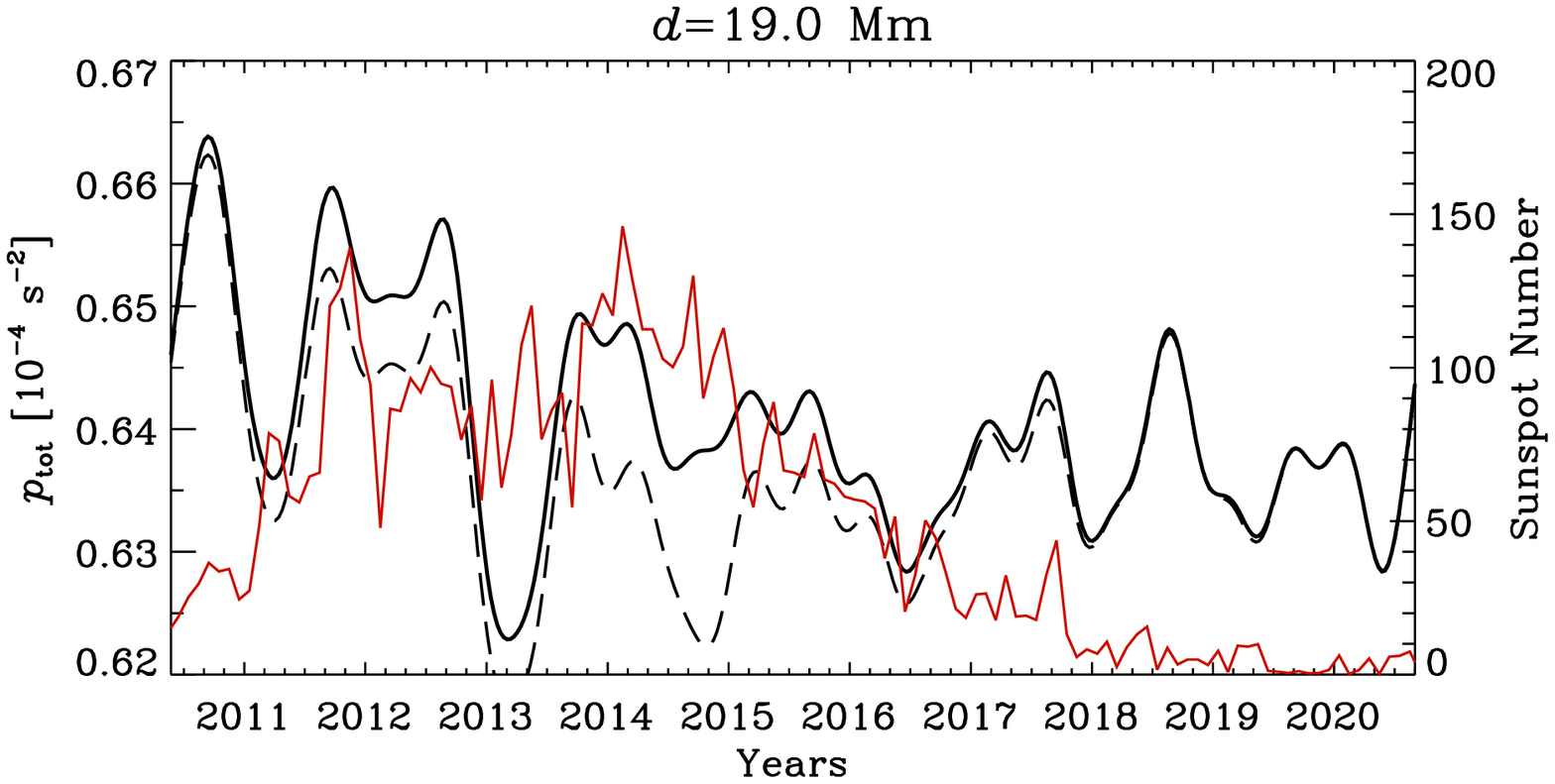}
    \caption{Time variation of the total power of all the harmonics of the unmasked (solid curves) and masked (dashed curves) fields for all the depths considered (indicated at the top of each panel). Left: unfiltered; right: filtered by applying the Butterworth filter with $f_\mathrm H=14, n=4$. The red curve in each panel represents the monthly averaged sunspot number.
	\label{time_var}}
\end{figure*}

\begin{figure} 
	\centering
	\includegraphics[width=0.5\textwidth]{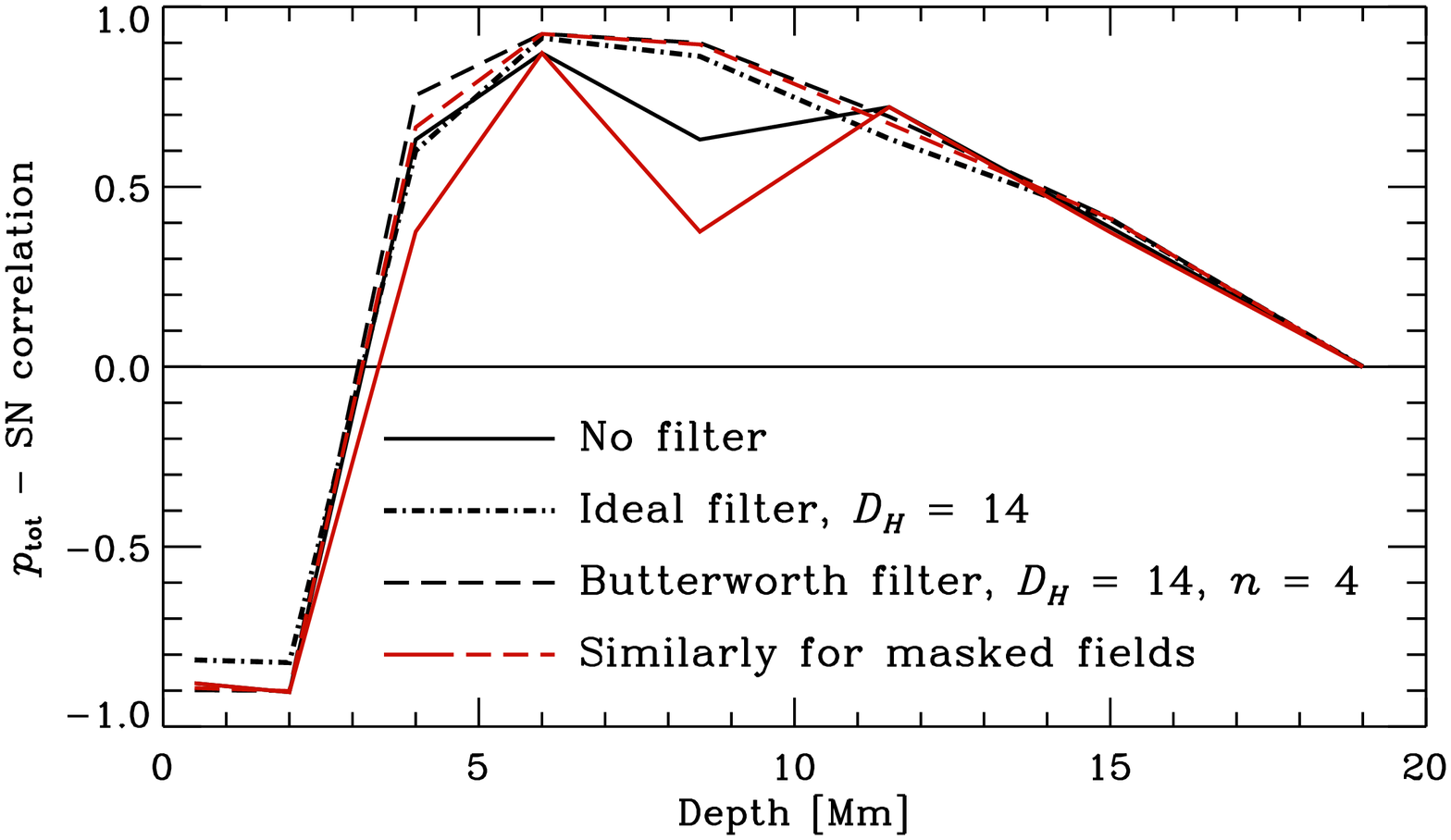}
	\caption{Correlation between the solar-cycle variations of the total power of all the flow harmonics and the monthly averaged sunspot number for unmasked (black curves) and masked (red curves) velocity-divergence fields.\label{correl}}
\end{figure}

\subsection{Time variation. The effect of magnetic fields.}\label{timevar}
	
The power spectra under consideration experience moderate variations in the course of the solar activity cycle. A careful comparison of Figures~\ref{spectra200} and \ref{spectra200HA} reveals some subtle differences. In shallow layers, $d=$0.5 and 2~Mm, the maximum spectral power, $p_{lm}$ is somewhat higher during the low-activity period, while the opposite can be noted for the deeper layers. At all depths, especially in the shallow layers, the spectrum is slightly broader in $l$ at the time of low activity.

Although the shape of the spectrum changes very little in the solar activity cycle, variations in the spectrum amplitude or in the total power of the flow, $p_\mathrm{tot}$, defined by Equation~(\ref{powertot}) are more pronounced. These variations are shown by black solid curves in the left column of Figure~\ref{time_var} for all $d$ values along with the monthly averaged sunspot number. Visually, these two quantities appear to vary nearly in antiphase at the upper levels, $d=0.5$ and 2~Mm; nearly in phase at the intermediate levels, $d=4$--8.5~Mm (where a notable oscillation with a half-year period, 0.063~$\mu$Hz, contaminates the $p_\mathrm{tot}$ variation); and without any definite correlation with the activity level at the bottom levels, $d\geqslant 11.5$~Mm. The half-year oscillation seems to stem from the variation of the inclination of the Sun's rotational axis to the line of sight. To remove this oscillation and isolate the physically conditioned $p_\mathrm{tot}$ variations on time scales of the order of the activity cycle, we apply a Fourier low-pass-filtering procedure.

Specifically, we calculate the fast Fourier transform of $p_\mathrm{tot}$ as a function of time, filter the resultant spectrum multiplying it by an appropriate filtering function, $H(\nu)$ ($\nu$ is the frequency), and perform the inverse Fourier transform of the filtered spectrum. We use two filters. The first one is an ideal filter,
$$H(\nu)=\left\{
\begin{aligned}
&1\quad \mathrm{if}\ \nu \leqslant \nu_\mathrm H,\\
&0\quad \mathrm{otherwise,}
\end{aligned}
\right.$$
and the second one is a Butterworth low-pass filter
$$H(\nu)=\frac{1}{1+(\nu/\nu_\mathrm H)^{2n}},$$
where $\nu_\mathrm H$ is the filter cutoff frequency; we assume $\nu_\mathrm H=0.05$~$\mu$Hz and $n=4$. The algorithms of discrete Fourier transform yield spectra periodic in frequency. The second half of the period corresponds to the negative frequencies decreasing in their absolute magnitude from a maximum (typically put equal to the Nyquist frequency in signal processing) to zero. Accordingly, we extend the filtering function to the second half of the spectra.
	
The time dependences of $p_\mathrm{tot}$ obtained using the Butterworth filtering are shown in the right column of Figure~\ref{time_var}. We can see that not only becomes the correlation between the sunspot number and the total power more pronounced---especially at the medium depths, $d=4$--8.5~Mm---but it emerges even at the lowest levels, $d=11.5$--19~Mm, where it was not notable without filtering. To quantify the possible solar-activity dependence of the convection-flow energy, we calculate, for each depth, the coefficient of correlation between the convection power and the sunspot number. As can be seen from the black curves in Figure~\ref{correl}, both the anticorrelation at $d=2$~Mm and positive correlation at $d=6$~Mm are especially high for filtered variations and do not significantly depend on the choice of the filter. The two temporal variations are best correlated if the Butterworth filter is used---in this case, the correlation coefficient is $-0.901$ at $d=2$~Mm, 0.925 at $d=6$~Mm, 0.695 at $d=11.5$~Mm, and 0.324 at $d=19$~Mm. The depression seen in the unfiltered variation near $d=8.5$~Mm can be attributed to the enhancement of the half-year oscillation with $d$, which disappears if a filtering procedure is applied.

A considerable contribution to the cycle dependence of $p_\mathrm{tot}$ can be made by magnetic field affecting time--distance measurements \citep[see, e.g.,][]{Liang_Chou_2015}. To assess this effect, we additionally performed our analysis for the divergence field in which its actual values in the areas with the radial magnetic field, $B$, exceeding 200~G are replaced with the mean divergence value over the remaining area. To be brief, we refer to such specially prepared fields as ``masked.'' The field mask is defined as follows:
\begin{equation}\label{mask}
    M(\theta, \varphi)=\left\{
\begin{aligned}
&0\quad \mathrm{if}\ B(\theta, \varphi) > 200\ \text G,\\
&1\quad \mathrm{otherwise.}
\end{aligned}
\right.
\end{equation}

The cycle variations of the total power, $p_\mathrm{tot}$, of the masked fields, both unfiltered and Butterworth-filtered, are shown in Figure~\ref{time_var} by dashed curves. The depth variation of the correlation between this power and the sunspot number is plotted with red curves in Figure~\ref{correl}. It can be seen that, although the replacement of the original fields with masked ones has some effect on the amplitude of the power variation, it does not change our conclusions about the anticorrelation of the power with the sunspot number in the upper layers of the convection zone and the direct correlation between these two quantities in the deeper layers.

As mentioned at the end of Subsection~\ref{scales}, the spectra of the fields differently constructed for the whole 360\degree-wide longitudinal interval differ insignificantly, only in the spectrum amplitudes, without affecting the spectrum shape and positions of the peaks. These differences do not change our findings. Similarly, the variation of the spectra during the solar activity cycle is not sensitive to the procedure of constructing the whole field from the measured-source-data sectors. Therefore, our inferences concerning the correlation of integrated power of convection with the solar-activity level remain valid.
	
\section{Conclusion and Discussion}
	
We have analyzed the running-averaged (with a 45-day window) spatial spectra of the horizontal-velocity-divergence field at depths in the solar convection zone ranging from 0 to 19~Mm. Our spectral analysis of the unsmoothed fields with angular degrees $l\leqslant l_{\max}=1000$ taken into account made it possible to separate the convection signal from the realization noise with $l$ values exceeding that of five-minute acoustic waves having their inner turning point at the given $d$. The spectra definitely reveal flow harmonics with supergranular scales at depths above $d\sim 11.5$~Mm. At depths of $d\gtrsim 6$~Mm, harmonics with giant-cell scales are prominent.

We have made additional computations to assess the possible role of the center-to-limb variations in the measured helioseismic travel times and found that the inferences of our study are not sensitive to these variations.

The spectral analysis of the divergence fields smoothed with a 17.5-Mm wide window and $l_{\max}=200$ indicates that the range of flow scales is fairly wide in shallow layers. As the depth increases, this range narrows, and the main peak shifts into the long-wavelength region. While the shortest length scales of the most energetic harmonics in the upper layers correspond to supergranular scales, $\sim 30$~Mm, the largest scales in the deepest layers are about 300~Mm, which is a giant-cell scale. The large-scale components are not clearly noticeable in the top layers because of the presence of the strong supergranulation component, but their power is of the same order of magnitude as in the deep layers. Such behavior can naturally be interpreted in terms of a superposition of differently scaled flows localized in different depth intervals within the convection zone. In addition, there is some tendency toward the emergence of meridionally elongated (banana-shaped) convection structures in the deeper layers.

We have also considered the time variation of the integrated spectral power of the flow at different levels with the solar activity cycle. To remove the half-year oscillation with a frequency of 0.063~$\mu$Hz, attributed to the variations in the inclination of the solar rotational axis to the line of sight, we applied a spectral filtering procedure with an ideal and a Butterworth low-pass filter. The results show that the time variation of the total power is anticorrelated with the sunspot number in the shallow layers, $d\lesssim 2$~Mm, and positively correlated at larger depths, $d\gtrsim 4$~Mm. To gain an idea of the possible effect of magnetic fields on the results of time--distance measurements, we additionally performed our spectral analysis replacing the actual divergence values in the areas where the radial magnetic field exceeds 200~G with the mean divergence value over the remaining area. Such masking of magnetic-field regions affects the amplitude of the power variation but does not change our conclusion about the correlation between the sunspot number and the integrated power of the flow.

The detected relationship between the solar activity and the convective-velocity power at different depths can be interpreted in terms of the depth redistribution of the convective-flow energy due to the action of magnetic fields. In particular, the large-scale converging flows around active regions, discovered by the ring-diagram \citep{Haber_etal_2004} and time-distance techniques \citep{Zhao_Kosovichev_2004} in the subsurface layers, may affect the convection spectra. In addition, it is possible that mesoscale convective motions associated with the formation of magnetic-field structures in the near-photospheric layers interact with larger-scale flows. This important issue calls for further investigation.
	
\section*{acknowledgments}
We are grateful to G.~Guerrero and A.M.~Stejko for providing the data of their numerical simulations. The helioseismological data are obtained from the SDO Joint Science Operations Center, courtesy of the NASA/SDO and HMI science teams. We also used sunspot-number data from the World Data Center for Sunspot Index, and Long-term Solar Observations (WDC-SILSO), Royal Observatory of Belgium, Brussels. The work is partially supported by NASA grants NNX14AB70G, 80NSSC20K1320, 80NSSC20K0602.

\bibliographystyle{aasjournal}
\bibliography{Scales_2}

\begin{thebibliography}{}
\expandafter\ifx\csname natexlab\endcsname\relax\def\natexlab#1{#1}\fi
\providecommand{\url}[1]{\href{#1}{#1}}
\providecommand{\dodoi}[1]{doi:~\href{http://doi.org/#1}{\nolinkurl{#1}}}
\providecommand{\doeprint}[1]{\href{http://ascl.net/#1}{\nolinkurl{http://ascl.net/#1}}}
\providecommand{\doarXiv}[1]{\href{https://arxiv.org/abs/#1}{\nolinkurl{https://arxiv.org/abs/#1}}}

\bibitem[{{Abramenko} {et~al.}(2012){Abramenko}, {Yurchyshyn}, {Goode},
  {Kitiashvili}, \& {Kosovichev}}]{Abramenko_etal_2012}
{Abramenko}, V.~I., {Yurchyshyn}, V.~B., {Goode}, P.~R., {Kitiashvili}, I.~N.,
  \& {Kosovichev}, A.~G. 2012, \apjl, 756, L27,
  \dodoi{10.1088/2041-8205/756/2/L27}

\bibitem[{{Ballot} {et~al.}(2021){Ballot}, {Roudier}, {Malherbe}, \&
  {Frank}}]{Ballot_etal_2021}
{Ballot}, J., {Roudier}, T., {Malherbe}, J.~M., \& {Frank}, Z. 2021, \aap, 652,
  A103, \dodoi{10.1051/0004-6361/202039436}

\bibitem[{{Beck} {et~al.}(1998){Beck}, {Duvall}, \& {Scherrer}}]{Beck_1998}
{Beck}, J.~G., {Duvall}, T.~L., \& {Scherrer}, P.~H. 1998, \nat, 394, 653,
  \dodoi{10.1038/29245}

\bibitem[{{Bumba} {et~al.}(1964){Bumba}, {Howard}, \&
  {Smith}}]{Bumba_etal_1964}
{Bumba}, V., {Howard}, R., \& {Smith}, S.~F. 1964, \aj, 69, 535,
  \dodoi{10.1086/109387}

\bibitem[{{Busse}(1970)}]{Busse_1970}
{Busse}, F.~H. 1970, \apj, 159, 629, \dodoi{10.1086/150337}

\bibitem[{{Busse}(2002)}]{Busse_2002}
---. 2002, Physics of Fluids, 14, 1301, \dodoi{10.1063/1.1455626}

\bibitem[{{Busse} \& {Carrigan}(1974)}]{Busse_Carrigan_1974}
{Busse}, F.~H., \& {Carrigan}, C.~R. 1974, Journal of Fluid Mechanics, 62, 579,
  \dodoi{10.1017/S0022112074000814}

\bibitem[{{Christensen-Dalsgaard} {et~al.}(1996){Christensen-Dalsgaard},
  {Dappen}, {Ajukov}, {Anderson}, {Antia}, {Basu}, {Baturin}, {Berthomieu},
  {Chaboyer}, {Chitre}, {Cox}, {Demarque}, {Donatowicz}, {Dziembowski},
  {Gabriel}, {Gough}, {Guenther}, {Guzik}, {Harvey}, {Hill}, {Houdek},
  {Iglesias}, {Kosovichev}, {Leibacher}, {Morel}, {Proffitt}, {Provost},
  {Reiter}, {Rhodes}, {Rogers}, {Roxburgh}, {Thompson}, \&
  {Ulrich}}]{Crist-Dals_etal_1996}
{Christensen-Dalsgaard}, J., {Dappen}, W., {Ajukov}, S.~V., {et~al.} 1996,
  Science, 272, 1286, \dodoi{10.1126/science.272.5266.1286}

\bibitem[{{Couvidat} {et~al.}(2005){Couvidat}, {Gizon}, {Birch}, {Larsen}, \&
  {Kosovichev}}]{Couvidat_etal_2005}
{Couvidat}, S., {Gizon}, L., {Birch}, A.~C., {Larsen}, R.~M., \& {Kosovichev},
  A.~G. 2005, \apjs, 158, 217, \dodoi{10.1086/430423}

\bibitem[{{Couvidat} {et~al.}(2012){Couvidat}, {Zhao}, {Birch}, {Kosovichev},
  {Duvall}, {Parchevsky}, \& {Scherrer}}]{Couvidat_etal_2012}
{Couvidat}, S., {Zhao}, J., {Birch}, A.~C., {et~al.} 2012, \solphys, 275, 357,
  \dodoi{10.1007/s11207-010-9652-y}

\bibitem[{{Dahlen} \& {Tromp}(1998)}]{Dahlen_Tromp_1998}
{Dahlen}, F., \& {Tromp}, J. 1998, Theoretical Global Seismiology (Princeton
  University Press)

\bibitem[{{Featherstone} \& {Hindman}(2016)}]{Featherstone_Hindman_2016}
{Featherstone}, N.~A., \& {Hindman}, B.~W. 2016, \apjl, 830, L15,
  \dodoi{10.3847/2041-8205/830/1/L15}

\bibitem[{{Frenkiel} \& {Schwarzschild}(1952)}]{Frenkiel1952}
{Frenkiel}, F.~N., \& {Schwarzschild}, M. 1952, \apj, 116, 422,
  \dodoi{10.1086/145626}

\bibitem[{{Getling} \& {Buchnev}(2010)}]{GetlingBuchnev2010}
{Getling}, A.~V., \& {Buchnev}, A.~A. 2010, Astronomy Reports, 54, 254,
  \dodoi{10.1134/S1063772910030078}

\bibitem[{{Ghizaru} {et~al.}(2010){Ghizaru}, {Charbonneau}, \&
  {Smolarkiewicz}}]{Ghizaru_etal_2010}
{Ghizaru}, M., {Charbonneau}, P., \& {Smolarkiewicz}, P.~K. 2010, \apjl, 715,
  L133, \dodoi{10.1088/2041-8205/715/2/L133}

\bibitem[{{Gizon} \& {Birch}(2004)}]{Gizon_Birch_2004}
{Gizon}, L., \& {Birch}, A.~C. 2004, \apj, 614, 472, \dodoi{10.1086/423367}

\bibitem[{{Gizon} {et~al.}(2020){Gizon}, {Cameron}, {Pourabdian}, {Liang},
  {Fournier}, {Birch}, \& {Hanson}}]{Gizon_etal_2020}
{Gizon}, L., {Cameron}, R.~H., {Pourabdian}, M., {et~al.} 2020, Science, 368,
  1469, \dodoi{10.1126/science.aaz7119}

\bibitem[{{Glatzmaier} \& {Gilman}(1981)}]{Glatzmaier_Gilman_II_1981}
{Glatzmaier}, G.~A., \& {Gilman}, P.~A. 1981, \apjs, 45, 351,
  \dodoi{10.1086/190714}

\bibitem[{{Greer} {et~al.}(2015){Greer}, {Hindman}, {Featherstone}, \&
  {Toomre}}]{Greer_etal_2015}
{Greer}, B.~J., {Hindman}, B.~W., {Featherstone}, N.~A., \& {Toomre}, J. 2015,
  \apjl, 803, L17, \dodoi{10.1088/2041-8205/803/2/L17}

\bibitem[{{Haber} {et~al.}(2004){Haber}, {Hindman}, {Toomre}, \&
  {Thompson}}]{Haber_etal_2004}
{Haber}, D.~A., {Hindman}, B.~W., {Toomre}, J., \& {Thompson}, M.~J. 2004,
  \solphys, 220, 371, \dodoi{10.1023/B:SOLA.0000031405.52911.08}

\bibitem[{{Hart}(1954)}]{Hart_1954}
{Hart}, A.~B. 1954, \mnras, 114, 17, \dodoi{10.1093/mnras/114.1.17}

\bibitem[{{Hathaway}(1987)}]{Hathaway_1987}
{Hathaway}, D.~H. 1987, \solphys, 108, 1, \dodoi{10.1007/BF00152073}

\bibitem[{{Hathaway} {et~al.}(2000){Hathaway}, {Beck}, {Bogart}, {Bachmann},
  {Khatri}, {Petitto}, {Han}, \& {Raymond}}]{Hathaway_etal_2000}
{Hathaway}, D.~H., {Beck}, J.~G., {Bogart}, R.~S., {et~al.} 2000, \solphys,
  193, 299, \dodoi{10.1023/A:1005200809766}

\bibitem[{{Hathaway} {et~al.}(2015){Hathaway}, {Teil}, {Norton}, \&
  {Kitiashvili}}]{Hathaway_etal_2015}
{Hathaway}, D.~H., {Teil}, T., {Norton}, A.~A., \& {Kitiashvili}, I. 2015,
  \apj, 811, 105, \dodoi{10.1088/0004-637X/811/2/105}

\bibitem[{{Hathaway} {et~al.}(2013){Hathaway}, {Upton}, \&
  {Colegrove}}]{Hathaway_etal_2013}
{Hathaway}, D.~H., {Upton}, L., \& {Colegrove}, O. 2013, Science, 342, 1217,
  \dodoi{10.1126/science.1244682}

\bibitem[{Hathaway \& Upton(2021)}]{Hathaway_2021}
Hathaway, D.~H., \& Upton, L.~A. 2021, The Astrophysical Journal, 908, 160,
  \dodoi{10.3847/1538-4357/abcbfa}

\bibitem[{{Herschel}(1800)}]{Herschel1800}
{Herschel}, W. 1800, Proceedings of the Royal Society of London Series I, 1, 20

\bibitem[{{Jeans}(1923)}]{Jeans_1923}
{Jeans}, J.~H. 1923, Proceedings of the Royal Society of London Series A, 102,
  554, \dodoi{10.1098/rspa.1923.0015}

\bibitem[{{Lefebvre} {et~al.}(2008){Lefebvre}, {Garc{\'\i}a},
  {Jim{\'e}nez-Reyes}, {Turck-Chi{\`e}ze}, \& {Mathur}}]{Lefebvre_etal_2008}
{Lefebvre}, S., {Garc{\'\i}a}, R.~A., {Jim{\'e}nez-Reyes}, S.~J.,
  {Turck-Chi{\`e}ze}, S., \& {Mathur}, S. 2008, \aap, 490, 1143,
  \dodoi{10.1051/0004-6361:200810344}

\bibitem[{{Leighton} {et~al.}(1962){Leighton}, {Noyes}, \&
  {Simon}}]{Leighton_etal_1962}
{Leighton}, R.~B., {Noyes}, R.~W., \& {Simon}, G.~W. 1962, \apj, 135, 474,
  \dodoi{10.1086/147285}

\bibitem[{{Liang} \& {Chou}(2015)}]{Liang_Chou_2015}
{Liang}, Z.-C., \& {Chou}, D.-Y. 2015, \apj, 805, 165,
  \dodoi{10.1088/0004-637X/805/2/165}

\bibitem[{{McIntosh} {et~al.}(2011){McIntosh}, {Leamon}, {Hock}, {Rast}, \&
  {Ulrich}}]{McIntosh_etal_2011}
{McIntosh}, S.~W., {Leamon}, R.~J., {Hock}, R.~A., {Rast}, M.~P., \& {Ulrich},
  R.~K. 2011, \apjl, 730, L3, \dodoi{10.1088/2041-8205/730/1/L3}

\bibitem[{{Muller} {et~al.}(1992){Muller}, {Auffret}, {Roudier}, {Vigneau},
  {Simon}, {Frank}, {Shine}, \& {Title}}]{Muller_etal_1992}
{Muller}, R., {Auffret}, H., {Roudier}, T., {et~al.} 1992, \nat, 356, 322,
  \dodoi{10.1038/356322a0}

\bibitem[{{Muller} {et~al.}(2018){Muller}, {Hanslmeier}, {Utz}, \&
  {Ichimoto}}]{Muller_etal_2018}
{Muller}, R., {Hanslmeier}, A., {Utz}, D., \& {Ichimoto}, K. 2018, \aap, 616,
  A87, \dodoi{10.1051/0004-6361/201732085}

\bibitem[{{November} {et~al.}(1981){November}, {Toomre}, {Gebbie}, \&
  {Simon}}]{November_etal_1981}
{November}, L.~J., {Toomre}, J., {Gebbie}, K.~B., \& {Simon}, G.~W. 1981,
  \apjl, 245, L123, \dodoi{10.1086/183539}

\bibitem[{{Rieutord} {et~al.}(2001){Rieutord}, {Roudier}, {Ludwig}, {Nordlund},
  \& {Stein}}]{Rieutord_etal_2001}
{Rieutord}, M., {Roudier}, T., {Ludwig}, H.~G., {Nordlund}, {\r{A}}., \&
  {Stein}, R. 2001, \aap, 377, L14, \dodoi{10.1051/0004-6361:20011160}

\bibitem[{{Rincon} \& {Rieutord}(2018)}]{Rincon2018}
{Rincon}, F., \& {Rieutord}, M. 2018, Living Reviews in Solar Physics, 15, 6,
  \dodoi{10.1007/s41116-018-0013-5}

\bibitem[{{Roudier} \& {Reardon}(1998)}]{Roudier_Reardon_1998}
{Roudier}, T., \& {Reardon}, K. 1998, in {ASP Conference Series}, Vol. 140,
  {Synoptic Solar Physics}, ed. K.~S. {Balasubramaniam}, J.~{Harvey}, \&
  D.~{Rabin}, 455

\bibitem[{{Scherrer} {et~al.}(2012){Scherrer}, {Schou}, {Bush}, {Kosovichev},
  {Bogart}, {Hoeksema}, {Liu}, {Duvall}, {Zhao}, {Title}, {Schrijver},
  {Tarbell}, \& {Tomczyk}}]{Scherrer_2012}
{Scherrer}, P.~H., {Schou}, J., {Bush}, R.~I., {et~al.} 2012, \solphys, 275,
  207, \dodoi{10.1007/s11207-011-9834-2}

\bibitem[{{Shcheritsa} {et~al.}(2018){Shcheritsa}, {Getling}, \&
  {Mazhorova}}]{Shcheritsa_etal_2018}
{Shcheritsa}, O.~V., {Getling}, A.~V., \& {Mazhorova}, O.~S. 2018, Physics
  Letters A, 382, 639, \dodoi{10.1016/j.physleta.2018.01.009}

\bibitem[{{Simon} \& {Weiss}(1968)}]{Simon_Weiss_1968}
{Simon}, G.~W., \& {Weiss}, N.~O. 1968, \zap, 69, 435

\bibitem[{{Zhao} \& {Kosovichev}(2004)}]{Zhao_Kosovichev_2004}
{Zhao}, J., \& {Kosovichev}, A.~G. 2004, \apj, 603, 776, \dodoi{10.1086/381489}

\bibitem[{{Zhao} {et~al.}(2012{\natexlab{a}}){Zhao}, {Nagashima}, {Bogart},
  {Kosovichev}, \& {Duvall}}]{Zhao_etal_CLS_2012}
{Zhao}, J., {Nagashima}, K., {Bogart}, R.~S., {Kosovichev}, A.~G., \& {Duvall},
  T.~L., J. 2012{\natexlab{a}}, \apjl, 749, L5,
  \dodoi{10.1088/2041-8205/749/1/L5}

\bibitem[{{Zhao} {et~al.}(2012{\natexlab{b}}){Zhao}, {Couvidat}, {Bogart},
  {Parchevsky}, {Birch}, {Duvall}, {Beck}, {Kosovichev}, \&
  {Scherrer}}]{Zhao_etal_2012}
{Zhao}, J., {Couvidat}, S., {Bogart}, R.~S., {et~al.} 2012{\natexlab{b}},
  \solphys, 275, 375, \dodoi{10.1007/s11207-011-9757-y}

\end{thebibliography}

\end{document}